# Numerical Modelling of VLF Radio Wave Propagation through Earth-Ionosphere Waveguide and its application to Sudden Ionospheric Disturbances

Thesis submitted for the degree of
Doctor of Philosophy (Science)
in Physics (Theoretical) of the
University of Calcutta

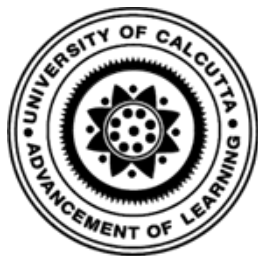

## Sujay Pal

**May** 8, **2013**

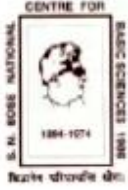 **SATYENDRA NATH BOSE NATIONAL CENTRE FOR BASIC SCIENCES**

*[Funded by the Department of Science & Technology, Government of India]*

BLOCK JD, SECTOR III, SALT LAKE, KOLKATA- 700 098
PHONE: +91-(0) 33-2335 5706-08, 2335 3057/61, 2335 0312/1313
FAX: +91-(0) 33-2335 3477/9176/1364

# CERTIFICATE FROM THE SUPERVISOR

This is to certify that the thesis entitled **"Numerical Modelling of VLF Radio Wave Propagation through Earth-Ionosphere waveguide and its application to Sudden Ionospheric Disturbances"**, submitted by Mr. Sujay Pal who got his name registered on **02.03.2011** for the award of **Ph.D. (Science)** degree of the **University of Calcutta,** absolutely based upon his own work under the supervision of **Professor Sandip K. Chakrabarti** and that neither this thesis nor any part of it has been submitted for any degree/diploma or any other academic award anywhere before.

**Prof. Sandip K. Chakrabarti**

Senior Professor & Head

Department of Astrophysics & Cosmology

S. N. Bose National Centre for Basic Sciences

JD Block, Sector-III, Salt Lake, Kolkata 700098, India

*TO*

*My PARENTS*

# ABSTRACT

Very Low Frequency (VLF) radio waves with frequency in the range 3∼30 kHz propagate within the Earth-ionosphere waveguide (EIWG) formed by the Earth as the lower boundary and the lower ionosphere (50∼100 km) as the upper boundary of the waveguide. These waves are generated from man-made transmitters as well as from lightnings or other natural sources. Study of these waves is very important since they are the only tool to diagnose the lower ionosphere.

Lower part of the Earth's ionosphere ranging $50 \sim 90$ km is known as the D-region of the ionosphere. Solar Lyman-$\alpha$ radiation at 121.8 nm and EUV radiation in $80 \sim 111.5$ nm are mainly responsible for forming the D-region through the ionization of NO, $N_2$, $O_2$ during day time. The VLF propagation takes place between the Earth's surface and the D-region at the day time. During the night hours, the D-region disappears and the VLF waves reflect from a much higher region in the ionosphere known as the E-region ($90 \sim 120$ km). The cosmic rays are the only sources for regular ionization of the ionosphere during the night time.

As a result, there is a diurnal (day-night) variation of ionization. Instead of these diurnal fluctuations, activity on the Sun or other Cosmic source can cause dramatic sudden changes to the ionosphere. When energy from solar flares or any other disturbances (e.g., Gamma Ray Bursts) reaches the Earth's upper atmosphere, the ionization in the ionosphere increases suddenly. Hence, the electron-ion density and the height of the layer change. The term Sudden Ionospheric Disturbances (SIDs) is commonly applied to such perturbations which affect the ionosphere in a non-trivial way and the ionosphere takes time (few seconds to several minutes) to recover itself to normal condition.

In this thesis, we theoretically predict the normal characteristics of VLF wave propagation through EIWG corresponding to normal behavior of the D-region ionosphere. We took the VLF narrow band data from the receivers of Indian Centre for Space Physics (ICSP) to validate our model. ICSP-VLF receivers continuously monitor the equatorial low latitude ionosphere through the recording of narrow band data from several number of VLF transmitters and also the broadband data for natural VLF sources. Detection of SIDs is common to all the measurements. We apply our theoretical models to infer the D-region characteristics and to reproduce the observed VLF signal behavior corresponding to such SIDs.

We develop a code based on ray theory to simulate the diurnal behavior of VLF signals over short propagation paths (2000∼3000 km). We applied it for the propagation of VTX (18.2 kHz) transmitter signal over the Indian sub-continent.

The day-night variation from this code are comparable to the variation obtained from a more general Long Wave Propagation Capability (LWPC) code which is based on mode theory approach. We applied the LWPC code to model the electron density variation in the D-region during a M2.0 solar flare simultaneously detected on two VLF paths (VTX-Kolkata and NWC-Kolkata). We find that the short VLF path acts as a good detector for such solar disturbances.

We simulate the observational results obtained during the Total Solar Eclipse of July 22, 2009 in India using the LWPC code. As a first order approximation, the ionospheric parameters were assumed to vary according to the degree of solar obscuration on the way to the receivers. We find that an assumption of 4 km increase of lower ionospheric height for places going through totality in the propagation path reproduces the observations very well at Kathmandu and Raiganj. We find an increase of the VLF reflection height parameter by $h' = 3.0$ km for the VTX-Malda path and $h' = 1.8$ km for the VTX-Kolkata path respectively. We also show the altitude variation of electron number density throughout the eclipse time at Raiganj.

We report and analyze a historic event, namely, the lunar occultation of a solar flare during the annular solar eclipse of $15^{th}$ January, 2010. We use the data from a multiple number of satellite and ground based observations, such as the GOES-14 (both hard and soft X-ray light curves), GONG Project (Magnetogram data), HIN-ODE (images of the flare) RHESSI (for X-ray light curves and image) for analysis of the event. We extract the time variation of the electron density profile in the D-region of the ionosphere due to the occulted solar flare from the combined effect of the eclipse and the flare.

# ACKNOWLEDGMENTS

I am very much happy to write this page finally at the completion of my thesis. It is a very pleasant task to express my heartiest thanks to all those people who made this thesis possible and an unforgettable experience for me.

First of all, I would like to express my sincere gratitude to my thesis advisor Prof. Sandip K. Chakrabarti for the continuous support during this research, for his patience, great power of motivation, enthusiasm and immense knowledge. I feel proud to have the opportunity of having a supervisor like him. His guidance helped me in all the time of research.

I would like to thank all the academic and non-academic staff of the S. N. Bose National Centre for Basic Sciences (SNBNCBS) for their support during my studies. It is great pleasure for me to work there and I could not forget the four years I have spent at SNBNCBS. It is great pleasure to thank all the colleagues and my friends of SNBNCBS, Kolkata. I would like to mention the name of Mr. Tamal Basak, Dr. Himadri Ghosh, Mr. Sudip Garain, Mr. Kinsuk Giri, Mr. Abhijit Chakrabarti for their helpful support during Ph.D. days. I can not forget the moments shared with them at SNBNCBS.

Most importantly, I would like to thank the President, all the members and my friends of Indian Centre for Space Physics (ICSP). I am attached with ICSP since my initial days of Ph.D. I have attended so many VLF workshops at ICSP and learnt the details about the VLF experimentals from here. The VLF data that have been used in this thesis are from ICSP-VLF network. I deeply appreciate all the group members of the ICSP-VLF network for sharing so many experiences. I gratefully acknowledge the assistance of Mr. Sudipta Sasmal, Mr. Sushanta Mondal, Mr. Debashish Bhowmick, Mr. Suman Ray, Mr. Sourav Palit, Mr. Surya K. Maji of ICSP during my Ph.D. period. Collaborative works with them greatly helped me to carry out this research.

I would also like to thank Dr. Kenneth J.W. Lynn of Australia and Dr. Desanka Sulic of Serbia for their generous academic help through some successful discussions.

I take the opportunity to extend my heartiest gratitude to all of my teachers specially Mr. Gurupada Nandi and Mr. Bimal Naha for providing constant support and motivation throughout my life.

Words are short to express my gratefulness to my parents, brother, sister and my close friends for providing me a great emotional support.

Financial supports for this research were provided by SNBNCBS and partly by the Council for Scientific and Industrial Research (CSIR) through fellowship. Final year of my thesis work was supported by MOES through a grant to ICSP.

# PUBLICATIONS IN REFEREED JOURNALS

1. Palit S., Basak T., Mondal S. K., **Pal S.** and Chakrabarti S. K., "Modeling of the Very Low Frequency (VLF) radio wave signal profile due to solar flares using the GEANT4 Monte Carlo simulation coupled with ionospheric chemistry", 2013, Atmos. Chem. Phys. Discuss., 13, 6007–6033, doi:10.5194/acpd-13-6007-2013.

2. **S. Pal**, S. K. Maji and S. K. Chakrabarti, "First Ever VLF Monitoring of Lunar Occultation of a Solar Flare during the 2010 Annular Solar Eclipse and its effects on the D-region Electron Density Profile", 2012, Planetary and Space Science, doi:10.1016/j.pss.2012.08.016.

3. S. K. Chakrabarti, **S. Pal**, S. Sasmal et al., "VLF campaign during the total eclipse of 22nd July, 2009: Observations and interpretations", 2012, Journal of Atmospheric and Solar-Terrestrial Physics, http://dx.doi.org/10.1016/j.jastp.2012.06.006.

4. **S. Pal**, S. K. Chakrabarti and S. K. Mondal, "Modeling of sub-ionospheric VLF signal perturbations associated with total solar eclipse 2009, in Indian subcontinent", 2012, Advances in Space Research, 50, 196–204.

5. S. K. Chakrabarti, S. K. Mondal, S. Sasmal, **S. Pal** et al., "VLF signals in summer and winter in the Indian sub-continent using multi-station campaigns", 2012, Indian Journal of Physics, Volume 86, Number 5, 323–334.

6. **S. Pal**, T. Basak and S. K. Chakrabarti, "Results of computing amplitude and phase of the VLF wave using wave hop theory", Advances in Geosciences, 2011, Solar Terrestrial (ST), Vol. 27, World Scientific, 1–11.

# PUBLICATIONS IN CONFERENCE PROCEEDINGS

1. **S. Pal** and S. K. Chakrabarti, "Theoretical models for computing VLF wave amplitude and phase and their applications", 2010, AIP Conference Proceedings, Vol-1286, 42–60.

2. S.K. Chakrabarti, S. Sasmal, **S. Pal** and S.K. Mondal, "Results of VLF campaigns in Summer, Winter and during Solar Eclipse in Indian Subcontinent and Beyond, 2010, AIP Conference Proceedings, Vol-1286, 61–76.

3. T. Basak, S. K. Chakrabarti and **S. Pal**, "Global effects on ionospheric weather over the Indian subcontinent at sunrise and sunset", 2010, AIP Conference Proceedings, Vol-1286, 137–149.

4. **S. Pal**, T. Basak, S. K. Chakrabarti, "Modeling VLF signal amplitudes over Indian sub-continent during the total solar eclipse," General Assembly and Scientific Symposium, 2011 XXXth URSI, doi: 10.1109/URSIGASS.2011.6051008.

5. **S. Pal**, S. K. Chakrabarti, "Computation of amplitude and phase of VLF radio waves: Results from comparative study between wave-hop and waveguide mode theory," General Assembly and Scientific Symposium, 2011 XXXth URSI, doi: 10.1109/URSIGASS.2011.6051014.

6. T. Basak, **S. Pal**, S. K. Chakrabarti, "VLF study of Ionospheric properties during solar flares of varied intensity for a fixed propagation path," General Assembly and Scientific Symposium, 2011 XXXth URSI , doi: 10.1109/URSIGASS.2011.6051004.

7. Chakrabarti, S.K.; **Pal, S.**; Sasmal, S.; Mondal, S.K.; Ray, S.; Basak, T.; Maji, S., "VLF observational results of total eclipse of 22nd Jul , 2009 by ICSP team", General Assembly and Scientific Symposium, 2011 XXXth URSI, doi: 10.1109/URSIGASS.2011.6051005.

8. Sasmal, S.; Chakrabarti, S.K.; **Pal, S.**; Basak, T., "A comparative study of VLF signals from several transmitters around the world as observed from Maitri station, Antarctica", General Assembly and Scientific Symposium, 2011 XXXth URSI, doi: 10.1109/URSIGASS.2011.6051002.

9. Chakrabarti, S.K., **Pal, S.**, Sasmal, S., Mondal, S.K., Ray, S., Basak, T., "Results of VLF campaigns in Summer and Winter in Indian subcontinent", General Assembly and Scientific Symposium, 2011 XXXth URSI, doi: 10.1109/URSIGASS.2011.6051007.

10. Basak, T., Chakrabarti, S.K., **Pal, S.**, "Computation of the effects of solar phenomena on Global Ionospheric Weather using waveguide mode theory of VLF propagation", General Assembly and Scientific Symposium, 2011 XXXth URSI, doi: 10.1109/URSIGASS.2011.6051009.

# Contents

# List of Figures

xi

# List of Tables

# Chapter 1

# Introduction

The research activities contained in this thesis primarily examine the theoretical and observational characteristics of Very Low Frequency (VLF) radio wave propagation mainly in the equatorial low latitude regions in terms of the modification of the Earth-ionosphere waveguide boundaries primarily due to the disturbances of the lower ionospheric region extended from 50 km to 100 km. The VLF wave acts here as the main diagnostic tool for remote sensing of this region.

The VLF group of Indian Centre for Space Physics (ICSP) has set up a network of VLF receivers in India and also in Nepal that continuously monitor the VLF transmitters signal amplitude and phase throughout 24 hours. The observational data that has been used here are mainly from this network.

In what follows, we will discuss about the atmosphere and plasma environment around the Earth, the basics of the ionosphere, its characteristics under normal and excited space weather conditions, various ionospheric disturbances, effects on the radio waves.

## 1.1   The Earth's Atmosphere and Plasma Environment

The increased dependence on space-based systems in recent days increase more and more interests on the subject of space research. The near-Earth space consisting with the upper mesosphere and the ionosphere, also the homeland of many satellites, is the most vulnerable region to outer radiation. The physical properties of this region are directly affected by the behavior of the Sun and space weather. The interaction of the solar wind with the Earth's magnetic field also controls plasma supply in the ionosphere and magnetosphere.

### 1.1.1 The atmosphere

The atmosphere of the Earth is not only just the air we breathe in but also a buffer that protects us from meteorites, harmful radiation from outer space and helps the radio waves to propagate globally for long distances. Due to all pervasive influence of gravity, the atmosphere can be approximated as horizontally stratified and organized into five principal layers by a representative temperature profile. Figure 1.1 represents a schematic diagram of Earth's atmosphere showing different regions and can be considered as the main scientific background figure for this dissertation.

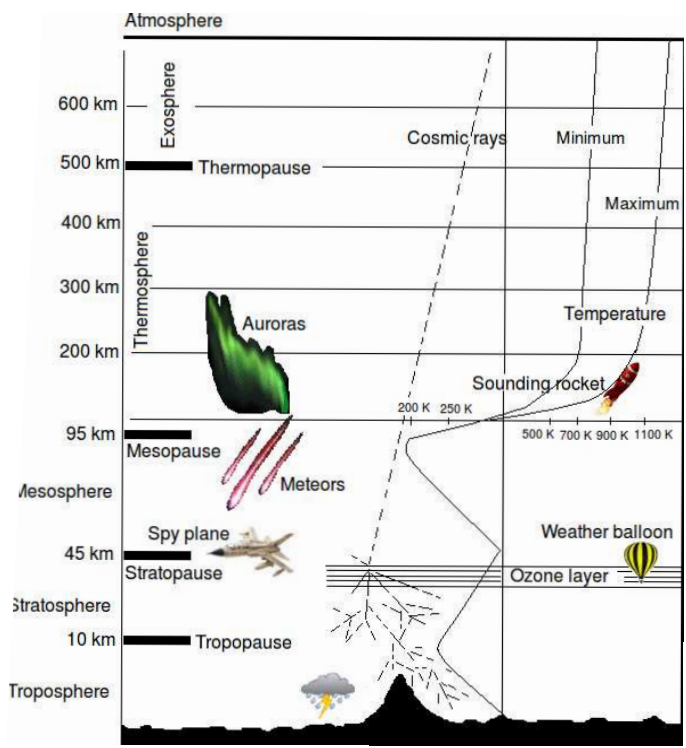

Figure 1.1: Schematic diagram of the Earths atmosphere showing different regions. Dark solid curves show atmospheric temperature profiles for solar maximum and minimum conditions (Not to scale) [adapted from Schunk & Nagy, 2009].

The closest layer to the Earth is the troposphere. It contains about four-fifths of Earth's air and is normally associated with atmospheric weather. Troposphere extends up to about 17 km at the equator and somewhat less at the poles. The atmospheric temperature in the troposphere decreases with altitude from the surface temperature, with a lapse rate of about 7 degree per km, up to a minimum value which defines it upper boundary known as the tropopause.

The stratosphere begins from tropopause and extends up to 45 km. It includes the Ozone layer which absorbs the harmful solar ultraviolet radiation and protects our life. The temperature in this region basically increases with altitude up to a local maximum which defines its top boundary known as the stratopause.

The next layer, known as the mesosphere, extends up to 95 km. The atmospheric temperature decreases again to a minimum at mesopause due to radiative cooling creating the coldest region of the atmosphere with temperature getting as low as 190 K. Meteors generally burn up in this region keeping the Earth's surface not poked with meteor craters, like Moon.

The thermosphere extends from the mesopause to 500 km. The atmospheric temperature here first increases with altitude to an overall maximum value (1000 K) and then becomes constant. The initial temperature increase in the thermosphere is explained by the absorption of solar UV and EUV photons. In the sunlit hemisphere these solar photons have sufficient energy to ionize the atmospheric constituents. Also due to the gravitational separation of the different neutral species, the heavy molecular constituents dominate at low altitudes and the atomic neutrals dominate at high altitudes. Near the thermopause i.e., at about 500 km, the atmosphere becomes so thin that collisions are negligible and, hence, the upper atmosphere can no longer be treated as a fluid. This transition altitude is called the exobase and the region above it is called the exosphere with mark of beginning outer space.

The atmospheric pressure p, mass density $\rho$ and temperature T are connected by the barometric equations as given below:

$$p = p_0 \exp(-\int_{h_0}^{h} \frac{dh}{H}),$$

$$\rho T = \rho_0 T_0 \exp(-\int_{h_0}^{h} \frac{dh}{H}),$$

where $p_0$, $\rho_0$ and $T_0$ are the values of $p$, $\rho$ and T at the reference height $h_0$ and $H = \frac{KT}{mg}$ is the scale height. Here $m$ is the mean molecular mass and $g$ is the acceleration due to gravity. For an isothermal atmosphere, the above two equations are reduced to,

$$p = p_0 \exp(-\frac{h - h_0}{H}) \ \ and \ \ \rho = \rho_0 \exp(-\frac{h - h_0}{H}).$$

## 1.1.2   The Ionosphere

The ionized region of the upper atmosphere extending from an altitude of about 60 km to more than 1000 km is defined as the ionosphere. The presence of free electrons and ions effectively makes this region an electrical conductor, which reflects radio waves over a broad range of frequencies. The existence of an ionized layer in the upper atmosphere was appreciated at the beginning of the 20th century. Marconi first described that radio waves could propagate to large distances beyond the horizon via multiple reflections between a conducting layer in the upper atmosphere and the ground. Experimentally, the existence of an ionized upper layer was proven in 1924. Appleton and Barnett [Appleton & Barnett, 1925] in England and Breit and Tuve [Breit & Tuve, 1926] in USA, conducted experiments to demonstrate the presence of the ionosphere in guiding the radio waves.

Historically, the ionosphere has been divided into different layers or regions. The first detected layer is named the E-layer (90 $\sim$ 160 km) as it reflects the electric fields. The D and F layers are below and above the E layer respectively. The D-region extends from 60 km to 90 km and is highly variable with much lower electron density. However, the electron density in all of these regions widely varies with time of the day, season, solar cycle and the level of geomagnetic and solar wind activity [Davies, 1990; Demirkol, 1999]. Typical neutral atmospheric constituents and day-night electron density profiles from MSISE-90 Atmospheric Model and IRI-2007 Model respectively are shown in Figure 1.2.

*Formation of the Ionosphere*

The formation of the ionosphere strongly depends on the Sun. Considering $O_2$ as the active constituent in the D-region ionized by its first ionization potential, the rate of ion production is given by the following two equations depending on the temperature distribution of the D-region [Mitra, 1951]. For the D-region between $\sim$50–63 km and $\sim$76–83 km, the temperature is nearly constant with height and the rate of ion production is given by the Chapman relation,

$$q = AN_0 Q exp[-z - N_0 AH \sec \chi exp(-z)],$$

where $A$ is the atomic absorption coefficient, $N_0$ is the density of neutral particles at the bottom of the region concerned, $Q$ is the available flux of solar quanta at the wavelengths concerned, $z = h/H$ is the reduced height and $h$ is the height measured from the bottom of the layer and $H = \frac{kT}{mg}$ is the scale height.

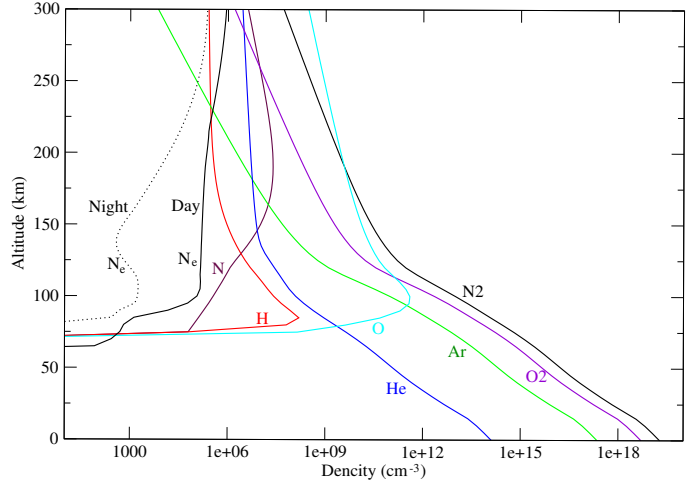

Figure 1.2: Typical neutral atmospheric constituents from MSISE-90 model and electron density profiles from IRI-2007 model.

For the region between 63-76 km and 83-100 km, the temperature vary linearly with height, the equation for the rate of ion production becomes [Mitra, 1951],

$$q = AN_0Q(1 + \gamma\frac{h - h_0}{T_0})^{-(1 + T_0/H_0\gamma)} exp[-AN_0H_0 \sec \chi(1 + \gamma\frac{h - h_0}{T_0})^{-T_0/H_0\gamma}],$$

where $\gamma$ is the temperature gradient, $h_0, T_0$ are the height and temperature of the reference level and $H_0$ is the scale height at the reference level.

Electrons are recombined with the positive molecular ions through the dissociative recombination processes. In the lower D-region, the electrons can attach themselves with neutral molecules in the absence of Sun light producing negative ions. The reactions are fast in the E- and F-regions. In the E-region, all the ions are almost molecular. So the rate of recombination of electrons is equal to the rate of recombination of molecular ions and the rate of increase of electron density is given by the continuity equation,

$$\frac{dN_e}{dt} = q - \alpha N_e^2,$$

where $\alpha$ is the mean dissociative recombination coefficient [Davies, 1990].

In the D-region, negative ions play an important role. The continuity equation in general can be written as [Rishbeth & Garriott, 1969; Zigman et al., 2007],

$$\frac{dN_e}{dt} = \frac{q}{1-\lambda} - \alpha N_e^2 - N_e \frac{d\lambda}{dt} \frac{1}{1+\lambda},$$

where $\lambda$ is the ratio of negative ion to electron number density and $\alpha$ is the effective recombination coefficient. $\alpha$ can be written as [Mitra, 1951],

$$\alpha = \alpha_e + \lambda \alpha_i + \frac{1}{N_e T} \frac{dT}{dt} + \frac{1}{N_e(1+\lambda)} \frac{d\lambda}{dt},$$

with

$$\lambda = \frac{\beta N_e N}{K N_e N + \rho N_e + \alpha_i N_e^2 + \alpha_e N_e^2}.$$

$\alpha_e$ is electronic recombination coefficient, $\lambda$ is the ratio of negative ions to electrons, $\beta$ is coefficient of attachment and $K$ is the coefficient of collisional detachment.

### *The D-region*

It is the more complex part of the ionosphere from the chemical point of view. This is because the D-region is the relatively high pressure region with a large number of species taking part in photochemical reactions and different sources contribute to ion production. Though the solar Lyman-$\alpha$ at 1215 $\mathring{A}$ is the main source of ionization in the D-region, minor contribution from EUV spectrum between 1027 $\mathring{A}$ and 1118 $\mathring{A}$ is also responsible. Soft X-rays in the range 1-8 $\mathring{A}$ are the strongest source of ionization during solar flares and they overwhelm the normal ionizations. The D-region is grouped into four sub-classes [Mitra, 1974]:

Altitude range $\geq 85$ km: This is the simplest region with electron density dominating over negative ions.

Altitude range $\sim 82 - 85$ km is characterized by a disappearance of positive water cluster ions $H^+(H_2O)_n$.

Altitude range $\sim 70 - 82$ km: Here positive ions, mainly water cluster ions $H^+(H_2O)_n$ dominate over negative ions $NO_3^-$ and $HCO_3^-$.

Altitude range $\sim 50 - 70$ km: Here negative ions like $NO^-$, dominate over the electrons.

With so many positive, negative and cluster ions and neutrals, the D-region becomes the most difficult region to observe and to model. In addition to the usual neutrals that are found in the E and lower F regions ($N_2$ , $O_2$ , O, N), there are several important minor neutral species [NO, $CO_2$ , $H_2$ O, $O_3$ , OH, $NO_2$ , $HO_2$ ,

$O2(^1\Delta g)$] in the D-region which take part in the chemical processes [Shunk & Nagy, 2009]. There are many complex chemical models developed to explain the ionizations for the D-region. One of the first successful chemical model for the D-region is the Mitra-Rowe six-ion chemical scheme [Mitra & Rowe, 1972].

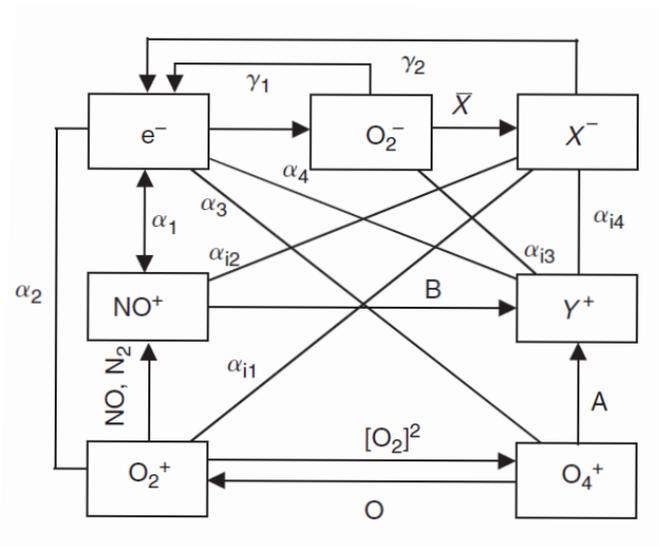

Figure 1.3: Simplified positive and negative ion chemistry for D-region [Mitra and Rowe, 1972].

### 1.1.3   The Magnetosphere

The Earth's magnetosphere is the region above the ionosphere where the motions of ionized particles are controlled by Earth's magnetic field and extends up to 10-12 $R_e$ ($R_e$: Earth radius) on the day-side and 60-100 $R_e$ on the night side. The boundary between the ionosphere and magnetosphere is not fixed and can be considered as smooth continuation of the ionosphere. The magnetosphere ends at the magnetopause boundary where it meets and interacts with the solar wind. Eventually, the Earth's magnetic field and solar wind control the shape and physical properties of the magnetosphere. The solar wind pushes the Earth's magnetic field away from the Sun and produces a comet-like structure as shown in the Figure 1.4.

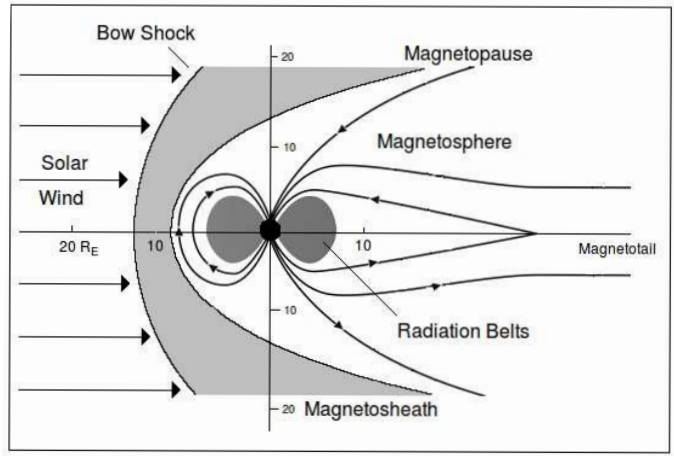

Figure 1.4: The Earth's magnetosphere. The solar wind interacts with the Earth's magnetic field to create a shock front behind which lies the magnetosphere [adapted from Zeilik & Gregory, 1998].

### *The Van Allen Radiation Belts*

The protons and electrons from solar wind which leaks into the magnetosphere are trapped by the Earth's dipolar magnetic field in toroidal belts called Van Allen radiation belt named after the discoverer James Van Allen in 1958 [Van Allen et al., 1959]. These belts are axisymmetric around the magnetic axis and compressed on the Sunward side while it is elongated on the other side. Though both the electrons and protons are found throughout the magnetosphere, the concentration of these particles are high in the inner and outer radiation belts. The high energy protons ($\sim$50 MeV) and electrons ($\geq$30 MeV) tend to reside in the inner radiation belts between $1R_e$ and $2R_e$. In the outer radiation belts ranging between $3R_e$ to $4R_e$, the less energetic protons and electrons are concentrated. The inner belt is relatively stable compared to the outer belt which varies in its number of particles by as much as a factor of 100.

The trapped charge particles in the radiation belts spiral along Earth's magnetic field lines while bouncing between the northern and southern mirror points with periods from 0.1 to 3 seconds. The particles also drift in longitude in addition to spiraling and oscillating in North-South. This is because the strength of both the magnetic and gravitational fields decrease with the increase of distance from the Earth. High energy protons drift westward around the Earth in about 0.1 s and low energy electrons drift eastward in about 1 to 10 h leading to the longitudinal uniformity of the radiation belts.

*Earth's magnetic field*

The Earth's external magnetic field plays an important role in space physics controlling the motion of charged particles. It is believed that the slowly rotating fluid motions in the electrically conducting outer core are responsible for the origin of Earth's magnetic field through some type of dynamo. The near-Earth magnetic field can closely be approximated by the familiar dipolar magnetic field with the dipole positioned at the center of the Earth and whose axis is inclined about 12° to the Earth's rotation axis. Such field can be expressed by the following simple expressions [Walt, 1994]:

$$B_r = -B_0 (\frac{R_e}{r^3})^3 sin\lambda,$$

$$B_\lambda = B_0 (\frac{R_e}{r})^3 cos\lambda,$$

where $r$ is the radial distance from the center of the Earth, $R_e$ is the mean radius of the Earth (approximately 6370 km), $\lambda$ is the geomagnetic latitude measured northwards from the equator and $B_0$ is the mean value ($\sim 3.12 \times 10^{-5}$) of the magnetic field at the magnetic equator on the Earth's surface.

The true magnetic field on or above the Earth's surface differ from the dipolar approximations and can be best represented numerically at times between 1900 A.D and at present by the International Geomagnetic Reference Field (IGRF) model developed and maintained by the International Association of Geomagnetism and Aeronomy (IAGA) working group. The IGRF model represents the geomagnetic field $\mathbf{B}(r, \theta, \phi, t)$ produced by internal sources in terms of a scalar potential $V(r, \theta, \phi, t)$ given by the following expression:

$$V(r, \theta, \phi, t) = a \sum_{n=1}^{N} \sum_{m=0}^{n} (\frac{a}{r})^{n+1} [g_n^m(t) \cos m\phi + h_n^m(t) \sin m\phi] \times P_n^m(\cos \theta).$$

Here, $r$ denotes the radial distance from the centre of the Earth in km, $a = 6371.2$ km is the magnetic reference spherical radius which is close to the mean Earth radius, $\theta$ denotes geocentric co-latitude (i.e., $90^0$ - latitude) and $\phi$ denotes East longitude. $P_n^m(\cos \theta)$ are the Schmidt semi-normalized associated Legendre functions of degree n and order m [Winch et al., 2005]. $g_n^m$ and $h_n^m$ are the Gauss coefficients.

The geomagnetic field vector, $\mathbf{B}$ can be specified by seven quantities (shown in Figure 1.5) which are called the magnetic elements. They are, the orthogonal components X (northerly intensity), Y (easterly intensity) and Z (vertical intensity, positive downwards); total intensity F; horizontal intensity H; inclination (or dip) I (the angle between the horizontal plane and the field vector, measured positive downwards) and declination (or, magnetic variation) D (the horizontal angle between true North and the field vector, measured positive eastwards). Declination, inclination and total intensity can be computed from the orthogonal components using the equations:

$$D = \arctan \frac{Y}{X},$$

$$I = \arctan \frac{Z}{H},$$

$$F = \sqrt{H^2 + Z^2}.$$

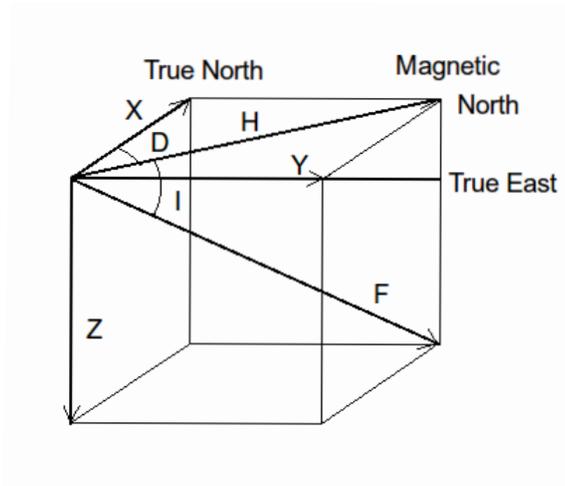

Figure 1.5: Geomagnetic field elements.

At Earth's surface, the total intensity varies from 24,000 nT (nanoTesla) to 66,000 nT. In the ionospheric height of D-region (70 km) over Kolkata, the intensity of magnetic field is about 43,970 nT while that at ground level is about 45,670 nT [IGRF-10 model]. Figure 1.6 is the model magnetic field intensity map from IGRF-10 and is useful for ionospheric radio propagation purposes.

*Geomagnetic Variations and Disturbances*

The observed magnetic field at a point on the Earth varies diurnally, seasonally and with solar activity due to the electric currents that flow above the Earth surface.

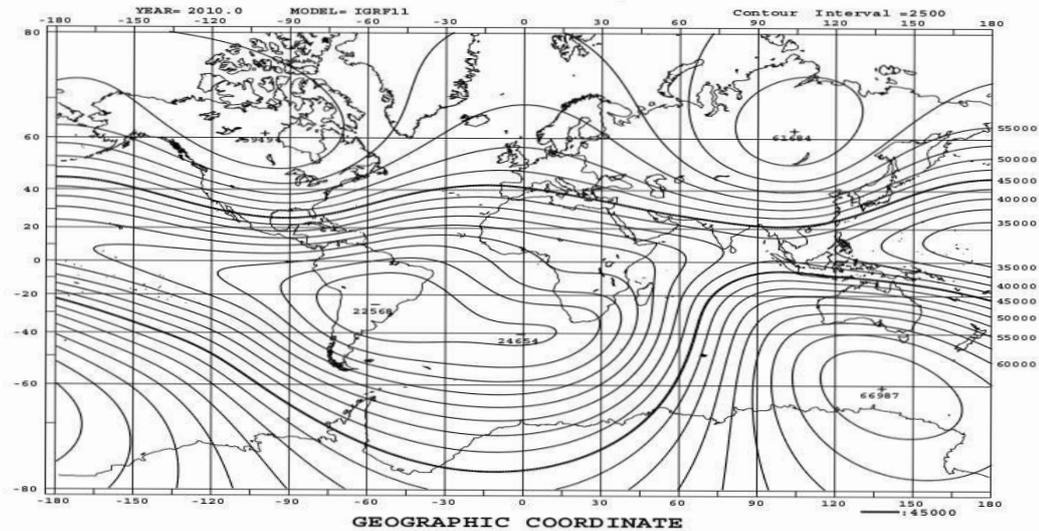

Figure 1.6: Total intensity of Earth's magnetic field (F) from International Geomagnetic Reference Field (IGRF-10) model for the year of 2010.

On quiet days, the magnetic field varies smoothly with local solar time mostly due to the current flowing in the E-layer known as solar daily ($S_q$) current [Matsushita, 1967; Davies, 1990]. Generally, the $S_q$ currents are stronger in day time than in night time, stronger in summer than in winter and approximately 50% stronger at solar cycle maximum than at solar cycle minimum.

During a geomagnetic storm, a temporary disturbance in the Earth's magnetosphere, additional currents circulate in the ionosphere causing a variation in magnetic field. Generally, this happens when high speed supersonic solar wind or coronal mass ejection (CMEs) followed by solar flares hit the Earth's magnetosphere and force to oscillate. Geomagnetic disturbances are monitored by measuring the three magnetic field components from ground-based magnetic observatories.

The short-term variation of the geomagnetic field is measured with the K-index. The K-index indicates disturbances in the horizontal component of Earth's magnetic field in 3 hour UT interval. K-index is numbered with an integer in the range 0-9 with 1 being quiet and 5 or more indicating a geomagnetic storm. The global planetary index $K_p$ is obtained as the mean value of the disturbance levels in the two horizontal field components, observed at 13 selected, sub-auroral stations.

The Dst index, calculated hourly, estimates the globally averaged change of the horizontal component of the Earth's magnetic field at the magnetic equator based on measurements from a few magnetometer stations. During quiet times, Dst is between +20 and -20 nano-Tesla (nT). The real-time information about the geomagnetic indices can be obtained from the National Geophysical Data Center (http://www.ngdc.noaa.gov).

## 1.2   Radio waves and the Ionosphere

Most of the knowledge about the ionosphere is obtained from the remote sensing by radio waves as various frequencies which are reflected from different regions of the ionosphere. Therefore some basic understanding of the radio refractive index is necessary to interpret the reflection and transmission mechanism of the radio waves from the ionosphere. The Appleton-Hartree formula determines the refractive index of electromagnetic waves propagating through the ionized medium in the presence of an external magnetic field [Budden, 1966]. The expression for the refractive index is given by the following equation:

$$n^2 = 1 - \frac{X}{1 - iZ - \frac{Y_T^2}{2(1-X-iZ)} \pm [\frac{Y_T^4}{4(1-X-iZ)^2} + Y_L^2]^{1/2}}, \tag{1-1}$$

where the dimensionless quantities are defined as follows,

$$X = \frac{\omega_p^2}{\omega^2},$$

$$Y = \frac{\omega_B}{\omega},$$

$$Y_L = \frac{\omega_B \cos\theta}{\omega},$$

$$Y_T = \frac{\omega_B \sin\theta}{\omega},$$

$$Z = \frac{\nu}{\omega},$$

where $\omega_p$ is the plasma frequency, $\omega_H$ is the gyro frequency, $\nu$ is the effective electron collision frequency, $\omega$ is the angular frequency of the wave and $\theta$ is the angle between the propagation direction and ambient magnetic field.

In general, the refractive index is complex. But at higher frequencies (greater than 8 MHz), the collisional term becomes negligible and correspondingly, the refractive index becomes purely real or imaginary.

There are two special classes of the Appleton-Hartree equation. The first one is the **Quasi-Longitudinal** (QL), where the wave propagates almost parallel to the magnetic field vector ($\theta \sim 0°$ and $Y_T \sim 0$), and the second one is the **Quasi-Transverse** (QT), where the wave propagation occurs along perpendicular direction to the magnetic field vector ($\theta \sim 90°$ and $Y_L \sim 0$).

For quasi-longitudinal propagation, equation (1) reduces to

$$n_L^2 = 1 - \frac{X}{1 \pm Y_L},\tag{1-2}$$

where the positive sign represents the ordinary wave (left hand circularly polarized) and the negative sign denotes the extraordinary wave (right hand circularly polarized). Similarly, for quasi-transverse propagation, the Appleton-Hartree formula reduces to the following form for the ordinary wave,

$$n_T^2 = 1 - X,\tag{1-3}$$

and for the extraordinary wave,

$$n_T^2 = 1 - \frac{X(1-X)}{1 - X - Y_T^2}.\tag{1-4}$$

Thus, in this case the propagation of ordinary wave is independent of the magnetic field whereas the refractive index for the extraordinary wave depends on the magnetic field.

## 1.2.1   Properties of the Earth-Ionosphere Waveguide

Extremely Low Frequency (30 Hz to 3 kHz) and Very Low Frequency (3-30 kHz) waves provide an alternative and convenient means of communication over the high frequency band due to its slow fading rates [Davies, 1990]. The propagation of VLF signal is characterized by relatively low attenuation rates of about 2–3 dB per megameter (Mm) allowing the propagation possible over long distances (5000–20 000 km) within the Earth-ionosphere waveguide. This waveguide is formed by the surface of the Earth as lower boundary and lower ionosphere (mainly the D-region) as the upper boundary.

An electromagnetic wave totally reflects from a medium with varying dielectric properties at the point at which the refractive index becomes zero [Jackson, 1999]. For an isotropic loss-less plasma, this condition is satisfied when $\omega \simeq \omega_p$, where $\omega_p$ is the plasma frequency of the medium given by

$$\omega_p^2 = \frac{4\pi N_e q_e^2}{\varepsilon_0 m_e},$$

where $N_e$ is the number density of the electrons, $e$ is the charge of an electron, $\varepsilon_0$ is the permittivity of free space and $m$ is the mass of an electron. But this condition does not generally apply to VLF signals. The value of refractive index for VLF signals in the ionosphere is complex, containing both a refractive and an absorptive term. It depends upon local properties of the medium and wave direction with respect to the Earth's magnetic field in a complicated manner. Due to the effect of magnetic field and the presence of absorption due to collisions, the refractive index never reaches to zero at any D-region altitude, so total reflection never occurs, but a substantial partial reflection is possible. Partial reflection occurs from the region where the refractive index changes very rapidly over distances comparable to a wavelength and this region acts like a sharp boundary between two different media. Quantitatively, this occurs in the region where the following expression holds [Wait & Spies, 1964]:

$$\omega \simeq \frac{\omega_p^2}{\nu},$$

where $\omega_p$ is the plasma frequency as defined above, $\omega$ is the wave frequency and $\nu$ is the effective collision frequency of electrons with heavy particles. Thus the reflection conditions depend on the number density of electrons of the ionosphere. A typical D-region electron density profile, frequently used in numerical modeling, is given by a two parameter exponential profile [Wait & Spies, 1964],

$$N_e(h) = 1.43 \times 10^{13} exp(-0.15h') exp[(\beta - 0.15)(h - h')],$$

where $N_e$ is electron density $(cm^{-3})$, $h'$ is the effective reflection height in km, and $\beta$ is the sharpness (slope) of the profile in $km^{-1}$. There are several other electron density profiles which are based on rocket measurements and various types of soundings, pulse cross modulation, incoherent radar and are not simply mathematical relationships between the electron density and altitude.

Absorption of radio waves in the ionosphere occurs both during the process of reflection and during the passage through the lower ionosphere below the reflection region. It is the collisions between electrons, ions and neutral molecules of the ionosphere which transfer the propagating wave energy to the thermal energy and thus reduce the wave intensity. There exist various collision frequency altitude profiles for electron-neutral particle $\nu_e(z)$, positive ion - neutral particle $\nu_+(z)$ and negative ion - neutral particle $\nu_-(z)$ collision frequencies [Wait & Spies, 1964; Morfitt & Shellman, 1976]. A typical example of these is the set of exponential profiles given in Wait & Spies [1964],

$$\nu_e(z) = 1.816 \times 10^{11} e^{-0.15z},$$

$$\nu_+(z) = 4.540 \times 10^9 e^{-0.15z},$$

$$\nu_-(z) = 4.540 \times 10^9 e^{-0.15z},$$

where $z$ is the altitude measured in km and $\nu$ is in units of $s^{-1}$. This strong exponential decay of collision frequency with altitude does not continue above 100-110 km. Kelly's collision frequency profiles [Kelley, 2009] can be used thereafter.

## 1.2.2   Natural ELF/VLF Waves in the Ionosphere

Naturally-occurring electromagnetic (radio) signals emanating from lightning storms, aurora, and Earth's magnetosphere are very common phenomena and caught the interest of space-weather researchers for the past few decades. The majority of Earth's natural radio emissions are due to tropospheric lightnings and are audible with ground-based radio receivers. Lightning discharge from thunderstorm produces a spectrum of electromagnetic energy ranging from a few Hz to tens of MHz, with a maximum energy concentrated in the ELF/VLF band [Weidman et al., 1986; Burke et al., 1992; Cummer et al., 1998]. The electromagnetic pulses in the ELF/VLF band are called radio atmospherics or sferics. The signals from sferics propagate within the Earth-ionosphere waveguide with very low attenuation and can be detected at very large distances from their source location. Sferics sound like sharp cracking noise when heard.

Spherics signals disperse while propagating considerable distances through the Earth-ionosphere waveguide. As a result, low frequencies travel slower than high frequencies and produce tweeks. Tweeks don't sound like sferics due to the dispersion. They sound like a quick musical ricochet. A fraction of the wave energy transmitted through the ionosphere couples into the magnetosphere where they are guided approximately along the geomagnetic field and can propagate one hemisphere to another. While propagating they are dispersed even more than tweeks. As a result, a gliding tone is heard when played through a loudspeaker of ELF/VLF receiver. These signals are called whistlers.

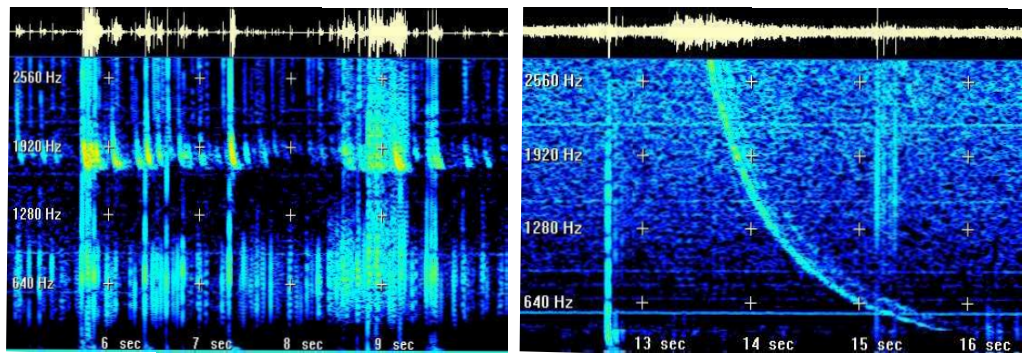

Figure 1.7: Natuaral ELF/VLF waves: example of spherics, tweeks and whistlers (Image coutesy: NASA).

## 1.3   Remote sensing of D-region by VLF Radio waves

During ionospheric disturbances, the D-region and F-region are mostly affected. This dissertation studies the VLF waves propagation within the Earth-ionosphere Waveguide (EIWG) and thus tries to understand the properties of the normal and perturbed D-region using VLF waves as one of the main diagnostic tools. The usual technique of ionospheric measurements such as Ionosonde and Incoherent Scatter Radar are not usable for continuous D-region measurements due to lower electron concentrations at the D-region. Also, the D-region is too low for satellites and too high for balloons. As a matter of fact, the little knowledge about the D-region that we have is insufficient and comes from indirect experimental sources. One of the most fruitful methods of monitoring the D-region continuously is to study the propagation of Very Low Frequency (VLF) waves in the EIWG. The monitored VLF waves can be of natural origin or man-made VLF transmitters. The properties of the lower waveguide boundary i.e., the conductivity and permitivity of the Earth's surface do not change with time, while the properties of the ionosphere highly depend on number density of ions and electrons which change from time to time with space-weather conditions. Thus continuous monitoring of the lower ionosphere is a valuable tool to study the lower ionosphere by the ground based VLF recording systems.

## 1.4   The Sudden Ionospheric Disturbances (SIDs)

The Earth's ionosphere, in general, is not quiet. Rather it is always fluctuating. Quiet Sun and cosmic rays are the regular sources of ionization. The ionization in the ionosphere is dominated by the Sun during the day time and by the cosmic rays during the night hours. As a result, there is a diurnal (day-night) variation of ionization. Instead of these diurnal fluctuations, activity on the Sun or other Cosmic sources can cause sudden changes to the ionosphere. When energy from solar flares or any other disturbances (e.g., Gamma Ray Bursts) reaches the Earth's upper atmosphere, the ionization in the ionosphere increases suddenly. Hence, the electron-ion density and location of the layers change. The term Sudden Ionospheric Disturbances (SID) is commonly applied to such perturbations which affect the ionosphere in a non-trivial way and ionosphere takes time (few seconds to several minutes) to recover itself to normal condition. SIDs are generally observed as the short wave fadeout (SWF), sudden phase anomaly (SPA), sudden frequency deviation (SFD), sudden cosmic noise absorption (SCNA), sudden enhancement/decrease of atmospherics (SEA) [Davies, 1990; Stonehocker, 1970; Liu et al., 1996; Mitra, 1974].

### Short Wave Fadeout (SWF)

Short Wave Fadeout is the absorption of short wave radio waves (in the HF range) by the increased ionization in the lower ionosphere. SWF was the first detected sudden ionospheric disturbance associated with the solar flare. Onset of SWF is manifested with a sudden decrease in HF signal followed by a relatively slow recovery. It may last from a few minutes to several hours, sometimes causing complete radio black out on certain HF band. The magnitude of the SWF events depend on the solar X-ray flux, the solar zenith angle and also on the background ionospheric conditions.

### Sudden Phase Anomaly (SPA)

Sudden Phase Anomaly is a sudden phase advancement of the VLF signals caused by the changes in reflection height. SPA is usually characterized by a rapid increase in phase (in 1-5 min) and slow recovery (in 30 min to 3 h) and phase shifts of 30 - 60 degree are common for moderate flares [Davies, 1990]. The phase variations roughly follow the time profile of the solar X-ray flare. In addition to positive phase enhancement, negative SPA is also reported by Ohshio, [1971] on the Fort-Collins (WWVL 20 kHz) to Japan path.

### Sudden Frequency Deviation (SFD)

Sudden Frequency Deviation is a sudden increase in received frequency of a high-frequency radio wave reflected from the E- or F-region of the ionosphere followed by a decay to the transmitted frequency. The frequency deviation may have several peaks and may take a negative value during the decaying portions. The start to maximum time is about 1 minute and peak frequency deviation is usually less than 1/2 Hz [Donnelly, 1971]. Generally, SFDs are the results of an almost time-coincident increase in E- and F-region electron densities.

*Sudden Cosmic Noise Absorption (SCNA)*

Sudden cosmic noise absorption (SCNA) is a sudden decrease in the strength of cosmic noise followed by a gradual recovery mainly associated with solar disturbances. Generally, three types of SCNA has been observed:

1. small change in the intensity of absorption and short duration,

2. moderate intensity change and relatively long duration,

3. large intensity change and long duration.

The main contribution towards total noise absorption comes from the D-region and the F2 region [Mitra & Shain, 1953].

*Sudden Enhancement/Decrease of Atmospherics (SEA)*

Sudden enhancement of atmospherics (SEAs) occur when there is a sudden increase in the number of atmospherics detected caused by the solar X-ray flare which enhance the reflectivity of the ionosphere. SEAs occur approximately between 10 kHz and 75 kHz [Sao et al., 1970] and also at ELF below 1 kHz. SEAs allow us to detect signals from more distant lightning discharges. Instead of enhancement, sudden decrease of atmospherics in the frequency range between 1 kHz and 10 kHz has also been observed.

*Sudden Increase in Total Electron Content (SITEC)*

Sudden increase in total electron content indicates sudden enhancement of electron density in the ionospheric F region, where the electron density is produced mainly by the extreme ultraviolet (EUV) flares [Tsurutani et al., 2005]. Nowadays, SITEC studies have been done using networks of worldwide GPS receivers which allow for two dimensional imaging of the ionosphere using TEC data [Afraimovich et al., 2001; Liu & Lin, 2004]. These studies have also observed a solar zenith angle dependence of SITEC [Zhang & Xiao, 2003; Wan et al., 2005].

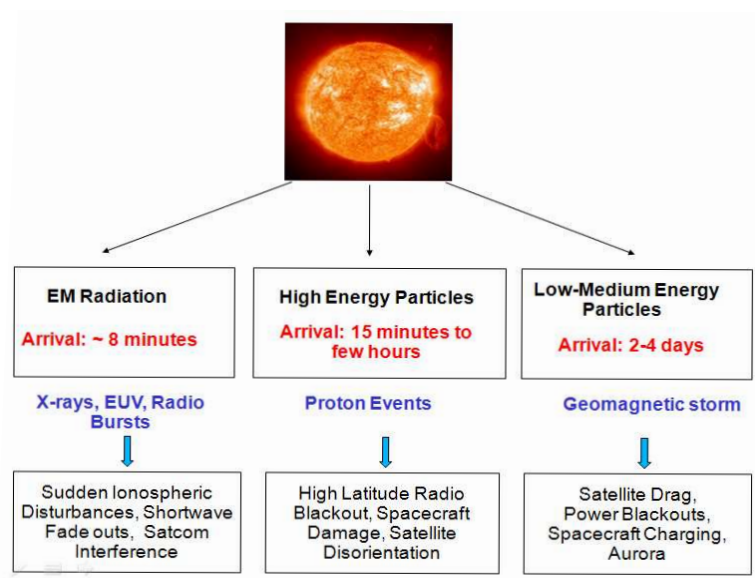

Figure 1.8: Solar eruption of energy and particles and corresponding effects on Earth.

### 1.4.1   Different Sources of Ionospheric Disturbances

The main causes of SIDs are described below.

*(i) Solar flares*

Solar flares are the sudden burst of electromagnetic energy and particles from coronal magnetic loops in the active regions near the sunspots area of the solar disk. The generated electromagnetic energy emissions contain the X-rays, Gamma rays, UV, microwaves, optical and radio waves. The total energy output of a typical solar flare could be of the order of $10^{30}$ ergs. The duration of solar flares vary from about 3 min to 2 h. Solar flares are associated with the sunspot activity and are more frequent around the peaks of the 11 year sunspot cycles.

*Causes of Solar flares*

Recent theory states that the magnetic reconnection of the field lines in the corona releases magnetic energy (Figure 1.9) which accelerates particles to relativistic speed. These particles travel down the magnetic loops emitting microwave radiations. When they reach the foot points of the loops in the chromosphere, the result is a electron-ion breaking radiation (Bremsstrahlung radiation) in the hard X-ray range. During this time the loops begin to gradually fill with a hot plasma, emitting thermal soft X-ray radiation. Thus there are three phases of solar flares:

(a) Pre-flare phase: Slight brightening of soft X-ray and microwave emissions.

(b) Impulsive phase: Particles are accelerated to relativistic speeds and travel down the magnetic loops into the chromosphere where non-thermal electron-ion bremsstrahlung is emitted in the hard X-ray range.

(c) Gradual phase: Magnetic flare loops gradually fill with plasma and heat up, emitting thermal soft X-ray radiation.

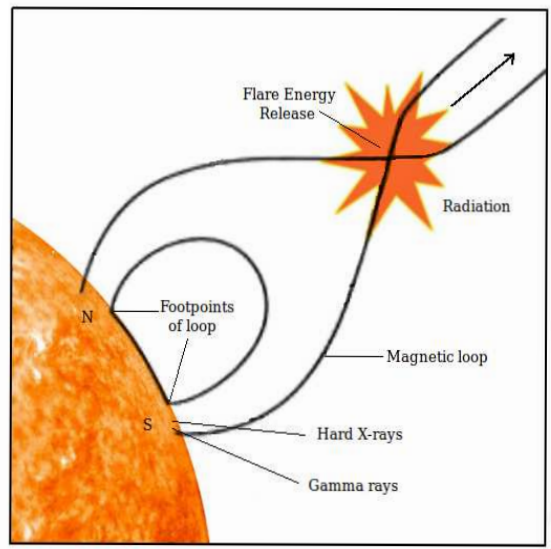

Figure 1.9: Model of a solar flare. Magnetic reconnection of the field lines in the corona is responsible for flare energy (adapted from Masuda, 1994).

*Ionospheric effects*

X-rays and UV radiation emitted during solar flares affect the Earth's ionosphere and thereby disturb the radio communication systems on the Earth and satellites operations in space. In general, a solar flare affects the whole upper atmosphere and ionosphere depositing different amount of energy at different altitudes and producing variation of ionization density with altitudes. Enhancement of UV flux during a flare depends on both the flare intensity and position on the solar disk while X-ray flux enhancement depends only on the flare intensity [Manju et al., 2012]. Thus it has been seen that the response of the E-region to flares is directly related to the X-ray flux enhancement whereas the F-region response depends on the central meridian distance (CMD) of flares on the Sun. This is known as the limb effect and it indicates that the UV flux controls the F-region. The effects of solar flares on the D-region of the ionosphere will be described in detail in Chapter 3. Figure 1.8 gives an illustration of the effects caused by solar flares on Earth.

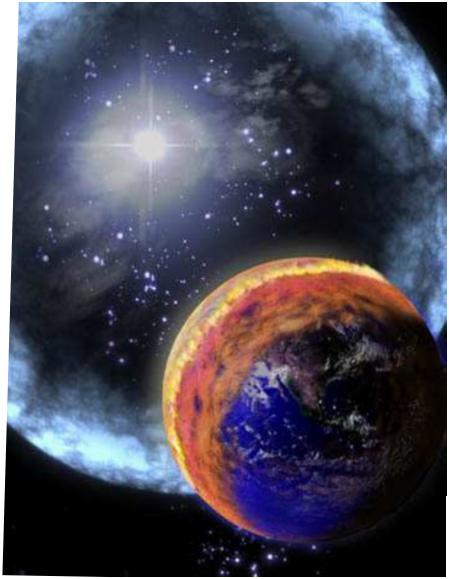

Figure 1.10: Gamma rays (invisible) from GRB hit the Earth's atmosphere initiating changes in the atmosphere that deplete ozone and create a brown smog of NO2 (Image credit: NASA).

*(ii) Gamma Ray Bursts*

Gamma Ray Bursts (GRBs) are the outburst of gamma ray photons associated with extremely energetic short-lived explosions in distant galaxies. Typical life-time of GRBs vary from milliseconds to several seconds. Most observed GRBs are believed to happen due to supernova events, as a rapidly rotating high mass star collapses to form a black hole, or due to the merger of binary neutron stars. The inverse Compton process is the most accepted model in the scientific community by which GRBs convert energy into radiation. In this model, pre-existing low energy photons are scattered by the relativistic electrons within the explosion which augment their energy by a large factor and transform them into gamma rays.
*Ionospheric effects*

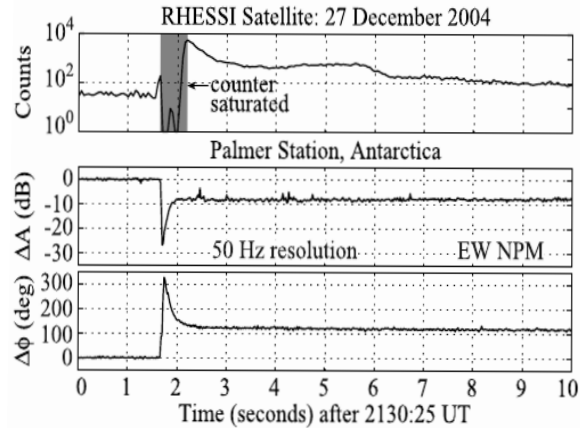

Figure 1.11: RHESSI and VLF observations, 10 s after the flash, due to SGR1806-20 [Inan et al., 2007].

The powerful rays from GRBs affect the Earth's ionosphere modifying the electrical conductivity of the ionosphere and produce detectable SIDs on the amplitude and phase of the VLF signals. In fact, calculations show that if gamma ray photons from a relatively nearby star explosion hit the Earth for only 10 seconds, it could destroy half of the Ozone layer which would in turn trigger mass extinctions [Russell, 1979; Thorsett, 1995].

Since GRBs occur in random directions, GRB-facing part of the Earth's atmosphere will be primarily affected. For VLF paths between a transmitter and receiver, it is necessary that the total or most of the path lies in the hemisphere illuminated by the gamma ray photons. The first ionospheric response of a GRB using VLF was detected on 1983 [Fishman & Inan, 1988]. They observed simultaneous amplitude changes in three VLF paths (England-Palmer, Maryland-Palmer, Hawaii-Palmer) due to the GRB (GRB830801). Since then there are many more detections using VLF signals as the main diagnostic tool [Inan et al., 1999; Fishman et al., 2002; Schnoor et al., 2003, GCN-2176; Tanaka et al., 2008].

The largest gamma ray burst, till now, was detected on December 27, 2004 on Hawaii-palmer VLF path in daytime solar illuminated ionosphere. The GRB came from a soft gamma ray repeater (SGR1806-20), a rotating neutron star with an enormous magnetic field, 50,000 light years away and caused a decrease in VLF signal amplitude by 20 dB (shown in Figure 1.11). It was so intense that it ionized the atmosphere down to an altitude of 20 kilometers, though lasted for few seconds [Inan et al., 2007].

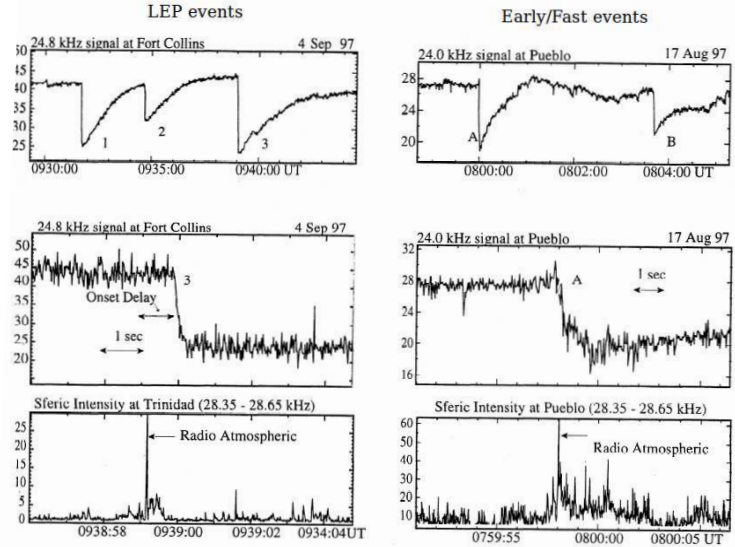

Figure 1.12: Difference between LEP event and early/fast event with associative spherics [Sampath et al., 2000].

### (iii) Lightning

Lightning can perturb the sub-ionospheric VLF propagation directly or indirectly. Direct perturbations in the sub-ionospheric VLF signals occur within 20 ms of (i.e., early) the associated lightning discharges and typically are rapid (i.e., fast) and are called "Early/Fast" VLF perturbations [Inan et al., 1988, Sampath et al., 2000]. The onset durations of these events are less than 100 ms. They are attributed to heating and ionization of the lower ionosphere by intense electromagnetic pulses from lightning discharge which implies a direct energetic coupling process between lightning in the troposphere and the lower ionosphere. Early VLF events have been extensively studied mostly at Stanford University [Inan et al., 1995; Johnson & Inan, 2000; Barrington-Leigh et al., 2001; Moore et al., 2003; Marshall et al., 2006] and also from many others [Dowden et al., 1996; Molchanov et al., 1998; Rodger, 2003; Mika et al., 2008].

Lightning generated VLF impulses which produce whistlers in the magnetosphere, undergo cyclotron resonance with the trapped electrons in the magnetosphere. The electrons which are pitch angle scattered into the loss cone, precipitate into the ionosphere producing secondary ionization in the lower ionosphere. This process produce VLF perturbations known as lightning induced electron precipitation (LEP) events. The onset of LEP event occurs after a characteristic time delay of $\sim$ 1s following the sferic occurrence while the early/fast events occur within 20 ms after the associated sferic (Figure 1.12).

*(iv) Meteor Showers*

Meteors while entering the Earth's atmosphere from outer space produce ionized trails in the height range 80–120 km. Billions of these ionized trails are produced daily. They usually diffuse very rapidly and disappear within a few seconds [Davies, 1990]. Due to collisions with the neutral molecules, the kinetic energy of the meteors converted into potential energy of ionization and produce extra ionization in the ionosphere. VLF phase perturbation associated with the Lyrid, *delta*-Aquarid and Perseid meteor showers was first reported by Chilton (1961) on the 16 kHz transmission path from Rugby, England, to Boulder, Colorado. It was also reported that meteor shower produces ELF-VLF waves which propagate and reach the ground at the same instance as the optical signal [Keay 1995; Beech et al., 1995; Zgrablic et al., 2002]. Many workers have also detected perturbations in the amplitude of sub-ionospheric VLF signals [Chakrabarti et al., 2002; De et al., 2012].

## 1.5 Other Ionospheric Disturbances

There are many other ionospheric disturbances which are associated either directly or indirectly with the events on the Sun.

### 1.5.1 Polar Cap Disturbances

Polar Cap is the upper atmosphere cum magnetosphere region which is enclosed by the poleward boundary of the auroral oval characterized by open geomagnetic field lines [Sivjee et al., 2003]. High energy particles, mainly protons, emitted with the solar flares or CMEs enter the polar ionosphere through the polar cap and ionize the upper atmosphere. Protons may have energy in the range 1 to 400 MeV. Protons with energies grater than 30 MeV can reach 50 km and are responsible for polar cap radio absorption (PCA). While the extremely energetic protons ($\sim$ 500 MeV) can reach the ground where they are detected by the cosmic ray detectors producing ground-level events (GLE).

PCAs are studied by riometers, VHF forward scatter, VLF reflections and ionosondes. Amplitude and phase anomalies occur on transpolar VLF wave propagation path during PCA events. Generally much larger signal losses and phase advances are reported on transpolar paths during PCA events [Bailey, 1964; Burgess & Jones, 1967; Crombie, 1965; Mitra, 1974; Reder & Westerlund, 1970; Westerlund et al., 1969]. Relativistic electron precipitations (REP) at auroral latitudes (60° - 75°) also disturbe the VLF radio wave propagation (Figure 1.13) [Westerlund, 1973; Larsen, 1979; Schmitter, 2011]. REP events have a large latitudinal extent ($\sim$ 2000 km) with electron energies about 500 keV and most events generally last from 1 to 6 hours.

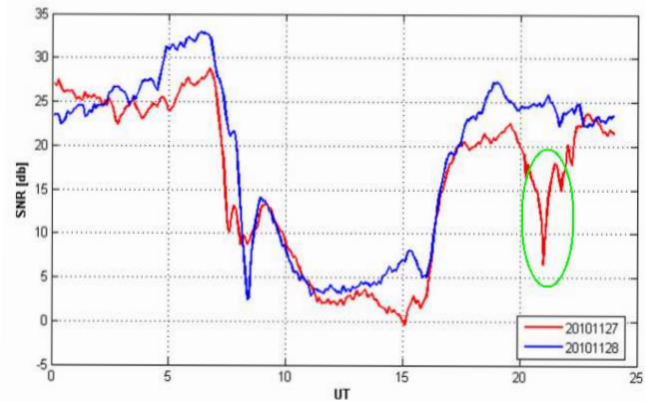

Figure 1.13: Sub-auroral particle precipitation event on the propagation path from the NRK transmitter (Iceland, 37.5 kHz) to (52° N 8° E) displayed by the red curve over quite day data (blue curve) [Schmitter, 2011].

## 1.5.2   Earthquakes

In addition to the solar or extra-terrestrial forcing from above, the upper atmosphere can also be affected by volcanic eruptions, earthquake phenomena, different meteorological conditions which excite atmospheric gravity waves in the troposphere or stratosphere (forcing from below). Among these, earthquakes are reported to produce large perturbations in the upper atmospheric plasma states. Recently, it has been established [Parrot et al., 1993, 2006 and references therein] that the ionospheric plasma was affected over the epicenters of future earthquakes and various anomalies have been detected by satellites and ground-based facilities. Scientists are trying to use this phenomena for short-term earthquake prediction.

There are many reports on the pre-seismic anomalous states in the atmosphere and ionosphere as well as those in the near-Earth (telluric) currents and ultra-low frequency electromagnetic variations prior to earthquakes [Hayakawa & Fujinawa, 1994; Hayakawa, 1999; Hayakawa & Molchanov, 2002]. Among different electromagnetic effects, the detection of seismo-ionospheric perturbations in VLF/LF phase and amplitude have attracted a lot of interests in the scientific community and have been reported to occur before large earthquakes [Hayakawa & Fujinawa, 1994; Parrot, 1995].

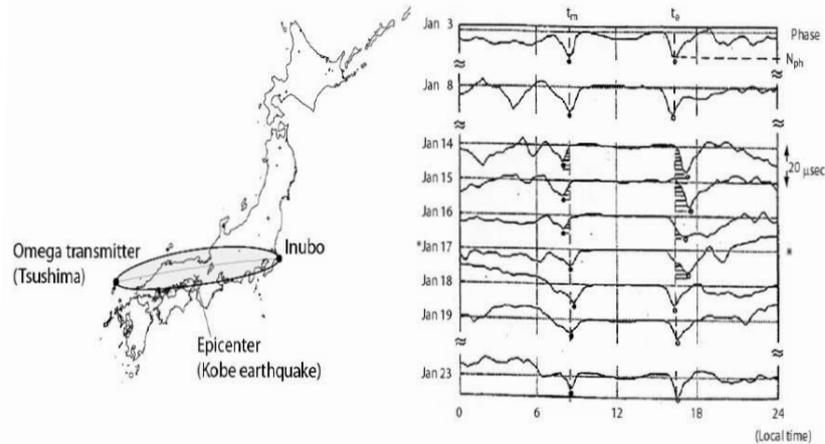

Figure 1.14: (Left) Location of the VLF transmitter (Omega, Tsushima) and receiver (Inubo) and (right) plot of diurnal variation of phase measurement (f = 10.2 kHz) around the Kobe earthquake. Significant change in the TTs has been observed just before the EQ (January 17) [Hayakawa et al., 1996].

Significant shifts in both morning and evening terminator times of the VLF signal from Omega Japan received at Inubo ∼1000 km away have been observed (Figure 1.14) before the Kobe earthquake (M=7.2) on 17 January 1995 [Hayakawa et al., 1996]. Terminator times (TT) are defined as the times of minimum in amplitude (or phase) around sunrise and sunset. They interpreted the shift in terminator time in terms of the lowering of lower ionosphere by using the full-wave mode theory. Since then there are many reports about the shifts in terminator times of the subionospheric VLF signals before the Earthquakes [Molchanov & Hayakawa, 1998; Clilverd, Rodger & Thomson, 1999; Hayakawa & Molchanov, 2000; Hayakawa et al., 2003; Chakrabarti et al., 2005, 2007]. Enhancement of fluctuations in the night time amplitude and phase of the VLF signals possibly associated with seismic activities have also been observed [Maekawa et al., 2006; Kasahara et al., 2010; Ray et al., 2011]. Figure 1.15 shows the anomalous VLF signal behavior prior to earthquakes monitored from ICSP-VLF network [Chakrabarti et al., 2010b].

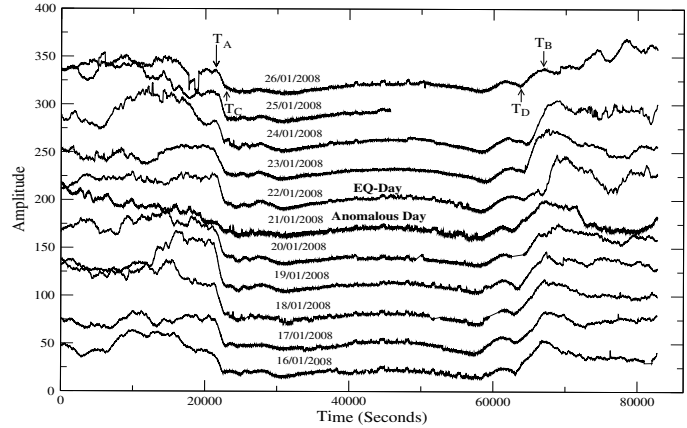

Figure 1.15: Anomolous diurnal behavior of the VTX (18.2 kHz) signal just before the day of earthquake [Chakrabarti et al., 2010b].

### 1.5.3 Traveling Ionospheric Disturbances

Traveling ionospheric disturbances (TIDs) are relatively large scale, wavelike disturbances in the ionospheric plasma. Generally the horizontal wavelengths of TIDs vary from 100 km - 1000 km with periods ranging from a few minutes to hour. Their propagation speed ranges from 50 m/s to 1000 m/s. TIDs are classified into very-large-scale and medium-scale disturbances. Very-large-scale disturbances originate in the auroral zone associated with the geomagnetic storm and travel to great distances with wavelengths of the order of 1000 km and periods of 1 h or more. Medium-scale disturbances (MSTIDs) originate locally in the troposphere or stratosphere and can not propagate to large distances. MSTIDs have wavelengths ranging from 100 - 200 km with periods 5 to 45 minutes [Davies, 1990].

One of the main reasons behind the traveling ionospheric disturbances to occur is the atmospheric gravity waves (GWs). Such gravity waves transfer energy and momentum between the lower, middle and upper atmosphere while propagating upward and could manifest themselves as TIDs in the ionosphere. Strong evidence for gravity wave seeding of ionospheric plasma irregularities has been found by the Arecibo incoherent scatter radar [Nicolls & Kelley, 2005]. The causative gravity waves, generated by the intense geomagnetic activity at auroral latitudes, observed to propagate primarily in the meridional direction as a train of large-scale traveling ionospheric disturbances (TIDs).

## 1.6    Major Scientific Achievements Reported in this Thesis

1. We developed a wave-hop code to simulate the short-path VLF signal propagation characteristics. We applied this code for the propagation of VLF signal from the Indian Naval transmitter named VTX to anywhere within the Indian sub-continent.

2. We used the Long Wave Propagation Capability (LWPC) code to predict and explain the propagation characteristics of VLF signal in Indian sub-continent. We simulated the normal diurnal behavior of VLF signal from the coupled IRI and LWPC model.

3. We estimated the effects of solar flare on EIWG. We showed the variation of electron density profile of the D-region during a M2.0 solar flare via the effects on two different VLF frequencies along a short and long VLF paths. The short VLF path is found to be a good detector for solar flares.

4. We simulated the VLF signal perturbations for several propagation paths associated with the Total Solar Eclipse of July 22, 2009. Both wave-hop and LWPC codes have been used. The enhancement of the VLF reflection heights along the different paths and the variation of D-region electron density profile have been shown quantitatively.

5. We have modeled and derived the D-region parameters from the perturbation on VLF signal during the Annular Solar Eclipse of January, 2010. During the eclipse, a C1.3 type solar flare occurred which was partly blocked by the Moon and left its trace on the VLF signal. The effect of this occulted flare on D-region electron density profile along with the solar eclipse have been deduced. This was a first report where the effects of two simultaneous events on the D-region electron density profile are seen clearly from our VLF data.

# Chapter 2

# Simulation of VLF Wave Propagation

The goal of this thesis is to simulate the VLF wave propagation characteristics under various realistic conditions of the EIWG boundaries. The simulated results are compared with the experimental measurements. For these, we have taken data from the VLF receivers of ICSP-VLF network and also from the receiver of other networks that are available publicly. ICSP-VLF network monitors several worldwide VLF transmitters. These transmitters are mainly used for Naval communication by the Navy people. Before going into the simulation part, first we will briefly describe the recording of VLF signals and their general diurnal and seasonal behavior. Then, we will discuss the propagation mechanism, simulation of VLF signal behavior using ray-theory and waveguide mode theory.

## 2.1   Recording of VLF Signals: Experimental Set-up

The VLF recording system consists of mainly three parts; one receiving antenna, one receiver/amplifier, one computer (Figure 2.1). The receiving antenna can be of two types, namely, the magnetic loop or B-field antenna which receives the magnetic component of the VLF signal and the electric field or E-field antenna which receives the electric component of the VLF signal. The E-field antenna picks more electrical noise and also noise picking ability depends on the local meteorological conditions, while the loop antenna is much less dependent on the local environment. Most of the VLF data used in this thesis are received by loop antennas. The VLF waves have huge wavelengths ranging from 10–100 km. Since the dimension of the magnetic loop antenna is very small as compared to the wavelength of VLF waves, the time varying magnetic field can be considered as uniform across the area of the loop inducing an open circuit voltage at the loop terminals. Generally, the axis perpendicular to the loop is oriented orthogonal to the propagation path from a particular transmitter to get the maximum signal strength from that transmitter. The output from the antenna is fed to the receiver where significant amplification of the signal amplitude is done by using suitable amplifiers. The receiver output is finally fed into the sound-card of the computer where the analog signal is converted into the digital signals and the data acquisition software records the amplitude and phase of the signals with time.

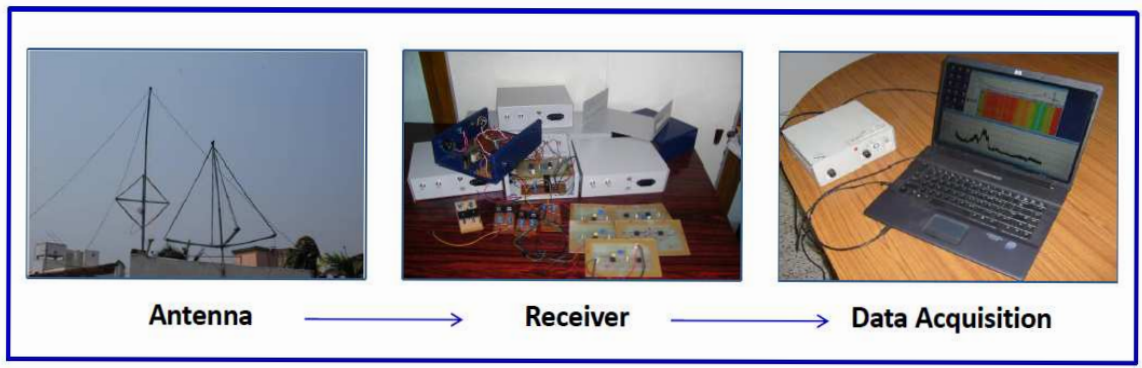

Figure 2.1: Typical VLF recording system of ICSP-VLF network.

In addition to taking data form ICSP-VLF network, one VLF receiving system with a loop antenna and a SUPERSID VLF receiver [http://solar-center.stanford.edu /SID/sidmonitor/] is also installed in the SNBNCBS. Figure 2.2(a) and Figure 2.2(b) show a square loop antenna and the SUPERSID VLF receiver installed there. Typical spectrum received at the Centre is shown in Figure 2.2(c), where the vertical spikes standing on the natural noise floor represent the signal from VLF transmitters. The receiver monitors these steady spikes while the noise floor may rise and drops from time to time. Generally signals from the NWC (19.8 kHz) and VTX (18.2 kHz) and sometimes from the DHO (22.1 kHz) were stronger at this site. In Figure 2.3(d), we have shown a typical variation of signal amplitude from the NWC transmitter (green curve) for quiet solar conditions and (red curve) for active solar conditions with an M1.8 solar flare detected on the top of the diurnal variations.

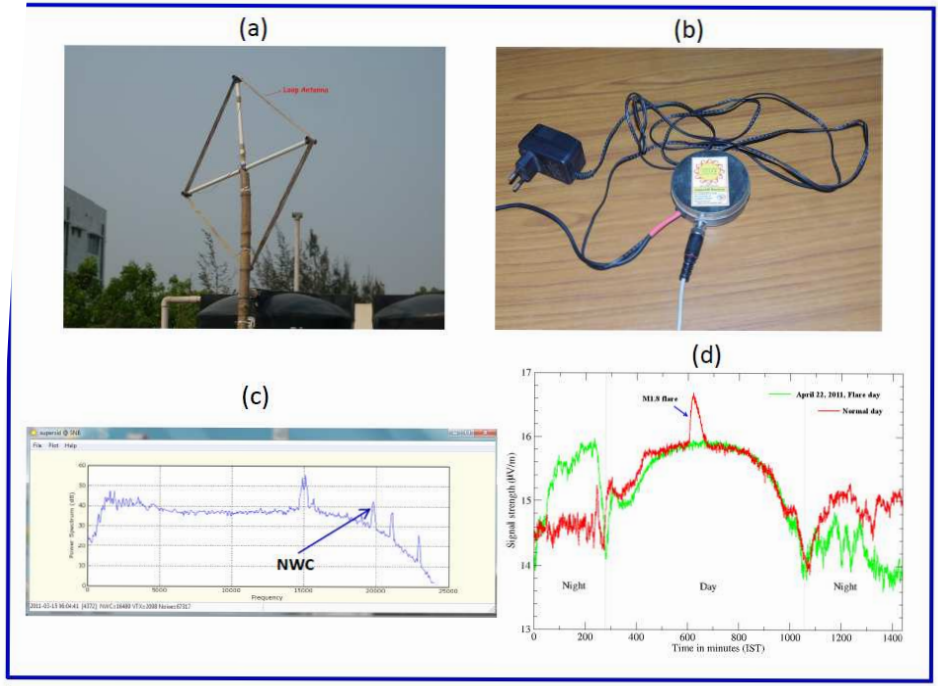

Figure 2.2: Square loop magnetic antenna (a) installed at SNBNCBS with a SU-PERSID VLF Receiver (b). (c) Typical spectrum where the vertical spikes standing on the natural noise floor represent the signal from VLF transmitters. (d) Diurnal variations of NWC (19.8 kHz) signal amplitude for a solar quiet (green curve) and active (red curve) day.

## 2.2   Typical Characteristics of Recorded VLF Signals

In Figure 2.3, we present the standard and solar-quiet diurnal variations of the amplitude (and phase for phase stable transmitters) of NWC, VTX and JJI VLF transmitters recorded at IERC/ICSP. The phase of the VLF signal is also recorded for phase stable transmitter (like, NWC). In Figure 2.3(a), the blue curve is the phase variation while the red curve is the amplitude variation for the NWC (19.8 kHz) signal. Figure 2.3(b) and 2.3(c) are the amplitude variations for the VTX (18.2 kHz) and the JJI (22.1 kHz) signals respectively. The great circle path distances between the transmitters and the receiver are shown in Figure 2.4.

Each diurnal variation is characterized by the presence of two signal minima around the local ground sunrise and sunset time of the receiver. The minimum around the sunrise is called the sunrise terminator time (SRT) and the minimum around the sunset is called the sunset terminator time (SST). The D-region disappears at night and also the E-region becomes weak with less uniform electron density. The fluctuations in the night time VLF signal amplitude are attributed to the fact that the VLF signal reflects randomly from the upper E-or lower F-region with no non-deviated attenuation at the lower D-region. Also, the atmospheric convections and oscillations may result in such fluctuations. The appearance of the sunrise terminator time (SRT) in the diurnal behavior actually signifies the formation of the D-region, while the sunset terminator time represents the start of disappearance of the D-region [Sasmal et al., 2009]. During the day time between the SRT and SST, the signals vary with the solar zenith angle with the maximum amplitude around the noon when the electron production rate in the ionosphere reached its maximum level.

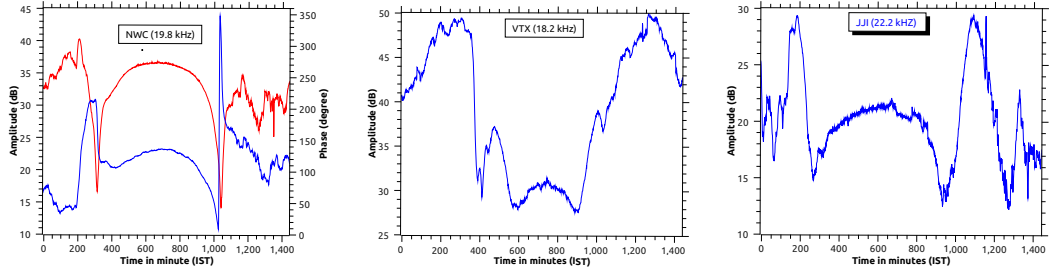

Figure 2.3: Amplitude and Phase of NWC (19.8 kHz) signal (panel a), amplitude of VTX (18.2 kHz) signal (panel, b), amplitude of JJI (22.4 kHz) signal (panel, c).

## 2.3   VLF Propagation Theory

The theory of propagation of VLF radio waves via multiple reflections between the Earth and lower ionosphere has been known to us since 1930s. There are two methods available in the literature to calculate and predict the field strength of very low frequency (VLF) radio signals.
a) Ray theory.
b) Waveguide mode theory.
The choice of representation depends on the frequency of the signal and the distance of propagation [Wait, 1962; 1998].

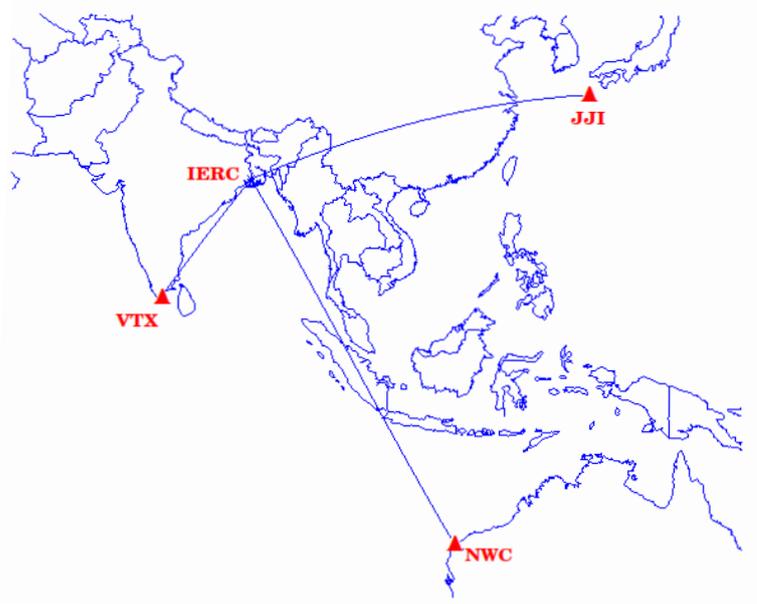

Figure 2.4: Great circle path distances between the receiver at IERC and the transmitters VTX, NWC and JJI.

### 2.3.1   Ray Theory

The ray theory is generally easier to visualize and analyze the propagation mechanism using the principles of geometric optics. Thus it is assumed that the ionization density is not changing very appreciably within the distance of one wavelength of the propagation wave. This method simply traces the different discrete ray paths (sky waves) that complete an integral number of reflections between the transmitter and the receiver (Figure 2.5). The number of complete reflections of the wave between the Earth and ionosphere before it reaches the receiver is called the order of the ray path. Theoretically, there are infinite number of ray paths between a transmitter and a receiver. Since the absorption and transmission rates are higher for higher order ray paths, the net effect of these rays at the receiver becomes less and less significant. Thus it is sufficient to consider only a finite number of low order rays [Davies, 1990; Wolf, 1990; Poulsen, 1991].

Near the transmitter ($d < 300$ km), most of the energy is propagated by the direct and the ground reflected waves (over line-of-sight distances) and by the surface wave. The combination of these three is referred to as the ground wave. As the distance increases, another component of the wave which is reflected from the ionosphere, plays a dominant role. This ionosphere-reflected components are referred to as the sky-waves. Phase differences between the sky-waves and the ground wave can cause constructive and destructive interferences at any given receiving point. As a result, the composite field can vary significantly from point to point over a propagation path. Within the ray theory, the total field is represented in terms of the sum of the ground wave and the discrete sky-waves.

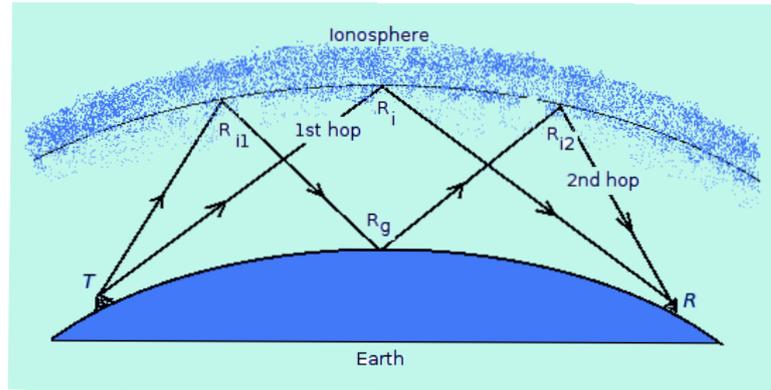

Figure 2.5: Ray path geometry within the EIWG between the transmitter (T) and the receiver (R). The points from where the wave is reflected at the ionosphere are marked.

## 2.3.2   Mode Theory

ELF and VLF radio waves propagate globally without much attenuation ($\sim 2 - 3$ dB/Mm) since the reflection coefficient is high. That is why the study of ELF and VLF radio waves is very significant. The ray theory is not very useful for larger propagation distances. In such a case, it is useful to represent the propagation mechanism in terms of the waveguide modes: where the waveguide is formed by the Earth surface as the lower boundary and the lower ionosphere as the upper boundary. The total field at any point in the guide is obtained in terms of sum of an infinite series of discrete "waveguide modes" [Morfitt & Shellman, 1976]. These "waveguide modes" are found by solving the Maxwell's equations with proper boundary conditions. Consequently, the waveguide modes can be defined as "a form of propagation of waves that is characterized by a particular field pattern in a plane transverse to the direction of propagation, while the field pattern is independent of position along the axis of the guide" [Davies, 1990].

The order of a mode actually represents the number of maxima and minima in the transverse vertical field pattern. For a VLF signal propagation at long distances it is sufficient to consider only a few low order modes as the higher order modes attenuate at a faster rate, thus the mode theory becomes an automatic choice for relatively larger propagation distances. Figure 2.6 shows a simple picture of mode configuration within an ideal EIWG with ionospheric and ground reflection coefficient of $-1$ and $+1$ respectively. The cut-off frequency of the waveguide is given by,

$$f_n = \frac{nc}{2h},$$

where n is the order of the mode. The modes with a frequency greater than $f_n$ propagate with group velocity,

$$v_{gn} = \frac{c}{\sqrt{(1 - \frac{f_n^2}{f^2})}},$$

which approaches zero as $f$ approaches $f_n$. Below a cut-off frequency, the wave will attenuate strongly within the waveguide and is called an evanescent mode. The normal day time cut-off frequency of the first order mode for EIWG is $\sim$ 2 kHz.

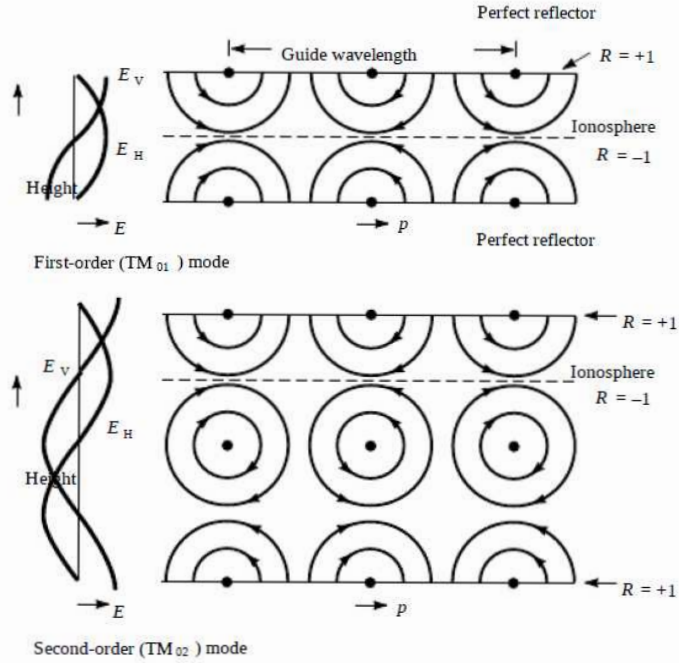

Figure 2.6: Electric field pattern in an ideal EIWG [following Davies, 1990].

*Brief Review of VLF Waveguide Mode Theory*

There are three primary mathematical formulations of waveguide mode theory for VLF wave propagation in the EIWG. These are pioneered by J. Galejs (1972), J. R. Wait (1962) and K. G. Budden (1966). The main difference between these formulations lies in the treatment of the ionospheric and ground boundaries. Galejs (1972) described the ionosphere as sharply stratified with a free space below the interface and a homogeneous ionosphere above. Since the real ionosphere is smoothly varying, the Galejs formulation is inadequate for those works which require accurate modeling of the ionospheric reflection process [Cummer, 1997].

The comprehensive derivation of waveguide mode theory of VLF propagation was published by Wait in a number of papers in the 1950's and 1960's and was summarized in his book [Wait, 1962]. In this formulation the upper and lower boundary are specified by the conductivity, permittivity, and permeability of the boundary regions explicitly. The method also applies for smoothly stratified ionosphere. However, Wait assumes that the upper boundary regions are isotropic, which is not the case for the magnetized plasma comprising the ionosphere. On the other hand, Budden waveguide theory is the most general one, because it describes the upper and lower waveguide boundaries in terms of completely general reflection coefficients, so that they can be comprised of any medium, sharply bounded or stratified, or even anisotropic.

After the development of the basic theory, the focus was towards the solution of modal equations under realistic conditions of the ionosphere and Earth surface. The first computation of VLF propagation using a computer program was made by Pappert et al., (1967) with exponential ionospheric profiles as defined by Wait & Spies (1964). Then the significant development in VLF mode theory came through the inclusion of the effects of modal conversion resulting from the inhomogeneity of the upper or lower boundary of the waveguide across the day night terminator and the land-sea boundary [Wait, 1970; Smith, 1974; 1977; Pappert & Snyder, 1972; Pappert & Morfitt, 1975; Barr, 1987]. As a result of all these developments it was possible to solve the modal equation for an inhomogeneous, anisotropic waveguide with finite conductivity of the ground and then provide the modal summations along a path over which the ionosphere and ground parameters vary with propagation distance [Barr et al., 2000].

## 2.4    Simulations Using Ray or Wave-hop Theory

Here we calculated the received field strength of VLF waves using ray theory approach. As discussed earlier, the resultant field strength is the sum of all sky waves and ground wave for short propagation paths. The paths between a given transmitter and the receiver are shown geometrically in Figure 2.5. Depending on the distance between the transmitter and the receiver, these propagating waves may reflect one or more times by the reflector (i.e., the lower ionosphere). The paths are named as 1-hop, 2-hop, etc. according to the number of reflections taking place in the ionosphere. The total field at R, the receiving point, is then the vector sum of the fields due to each path including the ground wave.

### 2.4.1    Basic Model and Calculations

*Sky-waves*

Consider a short vertical dipole (L$\leq 0.1\lambda$) radiating $P_t$ kW of power. The ground wave component propagates along the surface of the Earth and depends on the property of the ground. The other component, i.e., the sky-waves propagate along the EIWG and their properties are highly dependent on the electron concentration in the ionosphere as well as on the height of the ionosphere. The incoming sky-waves beyond the skip-zone can be received by a short vertical electric antenna or by a plane loop antenna. If the reception is made by a plane loop antenna located on the surface of the Earth, the signal strength is given by:

$$E_1 = 600\sqrt{P_t}Cos\Psi R_i F_i F_t F_r/L \qquad \text{mV/m},$$

where
L: sky-wave path length in km
$\Psi$: angle of departure and arrival of the sky-wave relative to the horizontal
$R_i$: the ionospheric reflection coefficient which is the ratio of the reflected electric field component to that of the incident field component parallel to the plane of incidence
$F_i$: ionospheric focusing factor
$F_t$: transmitting antenna factor
$F_r$: receiving antenna factor

On the other hand, if the reception is made using a short vertical antenna then the field strength is given by:

$$E_1 = 600\sqrt{P_t}(Cos\Psi)^2 R_i F_i F_t F_r/L \qquad \text{mV/m.}$$

For a 2-hop sky-wave mode, the field strength $E_2$ received by a loop antenna is given by:

$$E_2 = 600\sqrt{P_t} Cos\Psi R_{i1} R_{i2} F_{i1} F_{i2} R_g F_t F_r D_g/L_2 \qquad \text{mV/m,}$$

where, $R_{i1}$ and $R_{i2}$ are the reflection coefficients for the first and second reflections
$D_g$ is the divergence factor caused by spherical Earth, approximately equal to $1/F_i$
$R_g$ is the effective reflection coefficient of the finitely conducting Earth
$F_{i1}$ and $F_{i2}$ are the focusing factors for first and second reflection
$L_2$ is the total radio path length of 2-hop sky-wave

Assuming an uniform ionosphere over short distances, we can say $F_{i1}$ $F_{i2}$, (Rec. ITU-R P.684-3, 2002; Wakai et. al., 2004) so that

$$E_2 = 600\sqrt{P_t} Cos\Psi R_i^2 F_i R_g F_t F_r/L_2 \qquad \text{mV/m}$$

Similarly, the field strength for higher order hop sky-waves can be obtained.

### Ground-wave

Propagation curves for the ground wave are obtained from the GRWAVE computer program. GRWAVE calculates the ground wave field strength in an exponential atmosphere as a function of frequency, antenna heights and ground constants; approximate frequency range 10 kHz-10 GHz [Rec. ITU-R P.368-7, 1992]. Here we present the ground wave propagation curves of the VTX (18.2 kHz) transmitter (Figure 2.7) for three ground conditions of sea water ($\epsilon$:70, $\sigma$:5 S/m), medium ground ($\epsilon$: 15, $\sigma$: 0.002 S/m) and dry ground($\epsilon$: 7,$\sigma$: 0.0003).

### Resultant field strength and phase

The resultant field strength due to superposition of 1-hop sky-wave and ground wave is calculated as following [Yoshida et al., 2008: Wakai, 2004]:

$$I^2 = E_1^2 + G^2 + 2E_1 G Cos\phi$$

where $\phi$ is phase delay of 1-hop sky-wave with respect to the ground wave. Similarly, to calculate the resultant field due to a superposition of 1,2-hop sky-waves and

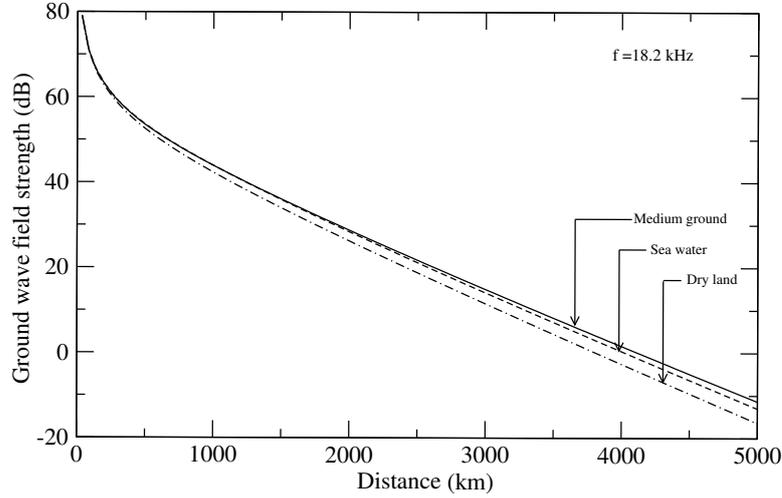

Figure 2.7: Variation of the field strength of the ground wave with the propagation distance from the transmitter for three ground conditions with frequency 18.2 kHz.

ground wave the following formula is used:

$$I^2 = E_1^2 + E_2^2 + G^2 + 2E_1GCos\phi_1 + 2E_2GCos\phi_2 + 2E_1E_2Cos\phi.$$

Here $\phi_1$ and $\phi_2$ are the phase delays of 1-hop and 2-hop sky-waves with respect to the ground wave and $\phi$ is the phase delay of the 2-hop sky-wave with respect to 1-hop sky wave. Following the same procedure one can find the resultant filed as a superposition of more number of sky-waves and ground wave as required.

*Reflection height model*

Effective ionospheric reflection height varies as a function of solar zenith angle ($\chi$) and season. The diurnal variation of VLF field strength is very sensitive to the variation of reflection height. One has to know this variation accurately in order to reproduce the diurnal variation of VLF signals. We have introduced two parameters $h$ and $\delta h$ in the reflection height model, where $h$ is the effective reflection height of the day time lower ionosphere and $\delta h$ is such that,

$$h + \delta h = h_n,$$

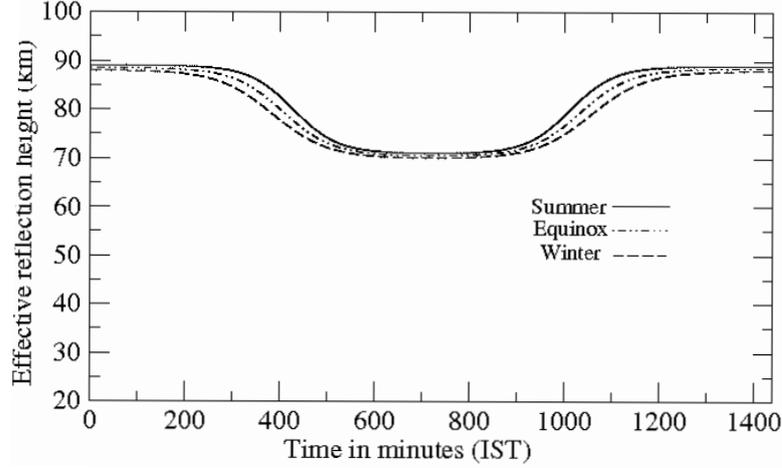

Figure 2.8: Variation of the effective ionospheric reflection height (km) with time of the day used for simulating the diurnal variation of VLF field strength [Pal & Chakrabarti, 2010].

where $h_n$ is the night time reflection height. Figure 2.8 shows the variation of $h(\chi)$ for three seasons (i.e., summer, winter and equinox).

### *Focusing and Antenna Factors*

The ionospheric focusing factors for day and night conditions and the transmitting and receiving antenna factors for three ground conditions of sea water, land and dry-ground are adapted from Rec. ITU-R P.684-3 (2002).

### *Ground reflection coefficient*

The ground reflection coefficient for vertical polarization is calculated as a function of frequency $f$ in kHz, elevation angle $\Psi$, dielectric constant $\epsilon$ and conductivity $\sigma$ in S/m from the equation given in Wakai (2004):

$$R_g = (n^2 sin\Psi - (n^2 - (cos\Psi)^2)^{1/2})/(n^2 sin\Psi + (n^2 - (cos\Psi)^2)^{1/2}),$$

where $n^2 = \epsilon - j18\sigma10^6/f$.

Figure 2.9 shows the variation of ground reflection coefficient of 18.2 kHz as a function of elevation angle for three different ground conditions.

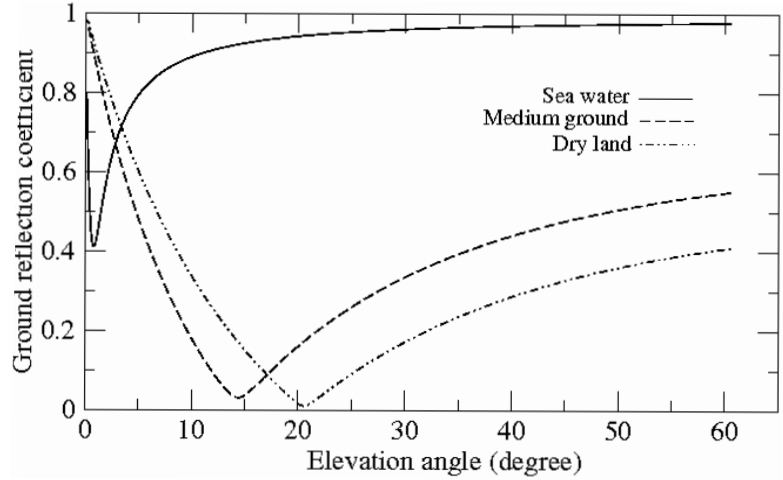

Figure 2.9: Variation of ground reflection coefficient with elevation angle for three ground conditions of sea water ($\epsilon$: 70, $\sigma$: 5 S/m), medium ground ($\epsilon$: 15, $\sigma$: 0.002 S/m) and dry ground($\epsilon$: 7,$\sigma$: 0.0003).

*Ionospheric reflection coefficient*

The ionospheric reflection coefficient for a West-East path is set to 0.6 at night and 0.3 at day and for the simulation of diurnal variation of VLF field strength, the reflection coefficient at day time is set to vary according to the solar zenith angle.

## 2.4.2   Results

Here we present the results of our computations. In Figure 2.10 we present the diurnal variation of the resultant electric field strength of sky-waves and the ground wave for the VTX-Kolkata path in Summer (chosen to be on June 21 for concreteness). The green curve is obtained considering the contribution from 1, 2-and 3-hop sky-waves and the ground wave, whereas the red curve is obtained only by considering 1-and 2-hop sky-waves and the ground wave. Addition of more hops produces additional constructive and destructive interferences.

In Figure 2.11 we show the diurnal variation similar to Figure 2.10, however we assume a winter (December 21) condition. The curves have the same meaning as above. Note that the length of the 'VLF' day as obtained from the time of the first fall of the signal and the final rise of the signal is shorter in winter.

Figure 2.12 shows the typical spatial variation of signal strength along the same path (VTX-Kolkata) with bearing 34.6° with respect to the transmitter. The black curve represents the typical day time variation while the blue curve represents the typical night time variation. Note that the day time path is quite stable with less number of minima and maxima in the signal strength. Here the value of day time and night time ionospheric reflection coefficients are 0.3 and 0.6 respectively. The wave-hop theory can also be used to compute the phase of the received wave under ideal conditions. In Figure 2.13, we present the variation of the wave phase, the dashed curve being for the summer and the solid curve being for the winter.

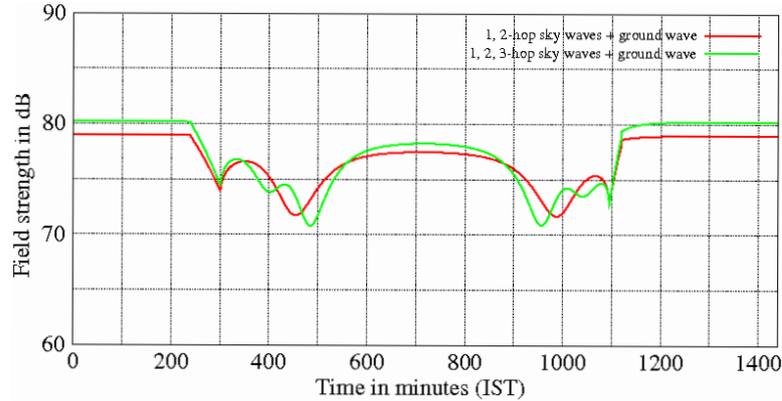

Figure 2.10: Diurnal variation of the resultant electric field strength obtained from the sky-waves and the ground wave for the VTX-Kolkata path in Summer. See text for details.

In order to understand the variation of the ionospheric parameters from day to day and from season to season, we can check the variation of the diurnal signal as the parameters are changed. In Figure 2.14, we give an example of how the signal changes with height (km) of the D-region in the day time. We also vary $dh$, the differential change in height of the ionosphere between the day time and night time. This is drawn for VTX-Kolkata propagation path at 18.2 kHz. We can compare the observational results with this chart to obtain $h$ and $dh$.

In Figure 2.15, we show how the diurnal behavior of the signal changes with the month of the year. We note that the interference patterns around the local sunrise or sunset time gradually die out from June to December. These are to be compared with actual observations.

To check the wave-hop model, one needs to compare the simulation results with the observational data. Here in Figure 2.16, we compare typical variations of VTX signal (18.2 kHz) in summer (blue curve) and winter (red curve) as recorded at ICSP, Kolkata. The weak signal in winter day time shows a strong absorption of the wave amplitude in the D-region ionosphere as compared to summer. This phenomenon is called the D-region winter anomaly which may not be directly related to the solar activity, but sometimes directly correlated to the planetary wave activity in the lower troposphere-mesosphere region [Schmitter, 2011]. In the second panel of Figure 2.16, we show our simulation results for a typical summer and a winter day.

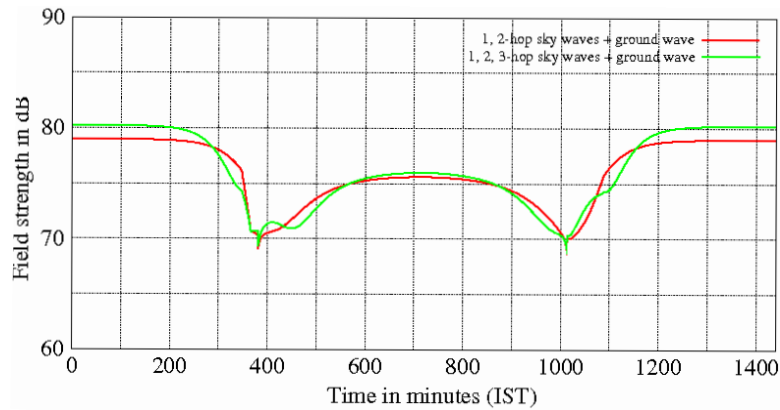

Figure 2.11: Same as Figure 2.10 but the day was chosen to be December 21. Note that the length of the 'VLF' day as obtained from the time of the first fall of the signal and the final rise of the signal is shorter in winter.

### 2.4.3   Limitations of Ray Theory

The ray theory gives an easy way to calculate the total field as a sum of ground wave and sky waves. For shorter VLF propagation distances and also for higher frequency (LF and higher) ray theory can be applied with sufficient confidence and becomes the method of choice. However, it is difficult to use ray theory for distances greater the 3000–4000 km as the number of sky-waves will be large and mathematical expression will become very complex. The effect of the geo-magnetic field on VLF propagation does not arise explicitly in the wave-hop theory, which affects the ionospheric reflection coefficient for different direction of propagation. Also, for inhomogeneous waveguides such as across day-night ionospheric gradients, the geometry is ill-defined and depends on the approximations made about the waveguide cross-section.

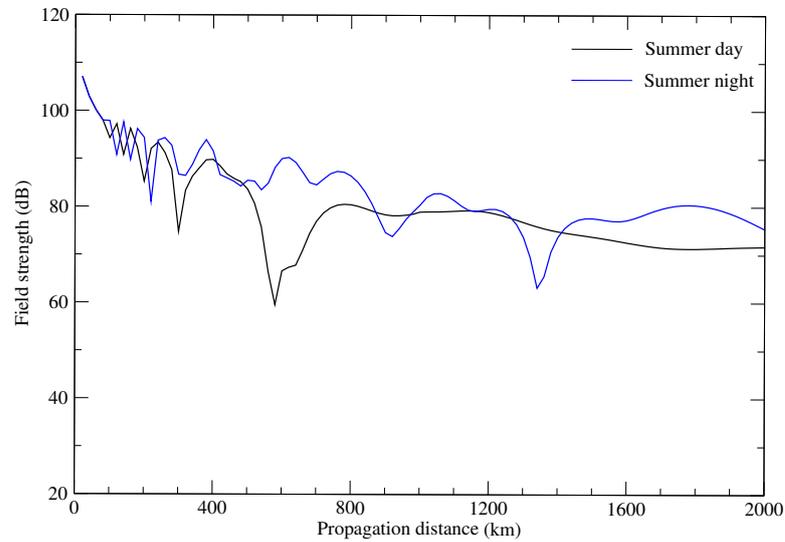

Figure 2.12: Variation of signal strength along the VTX-Kolkata propagation path. Black curve represents the typical day time variation while blue curve represents the typical night time variation.

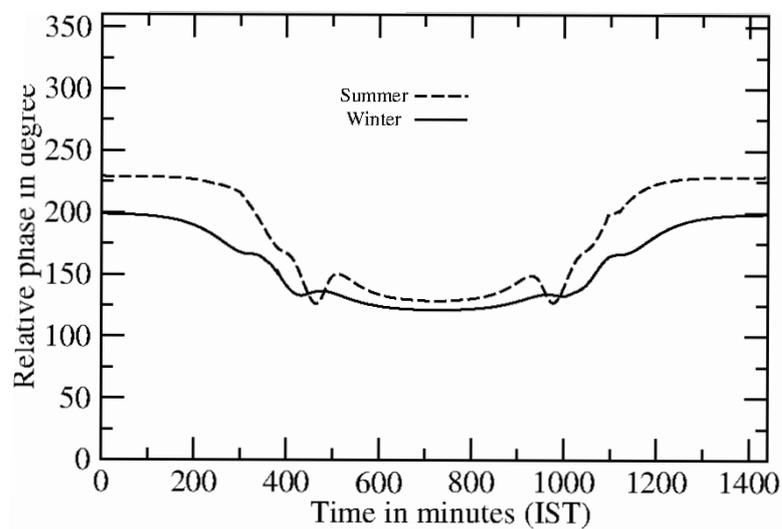

Figure 2.13: Typical phase variation of VTX (18.2 kHz) signal for summer and winter [Pal et al., 2011].

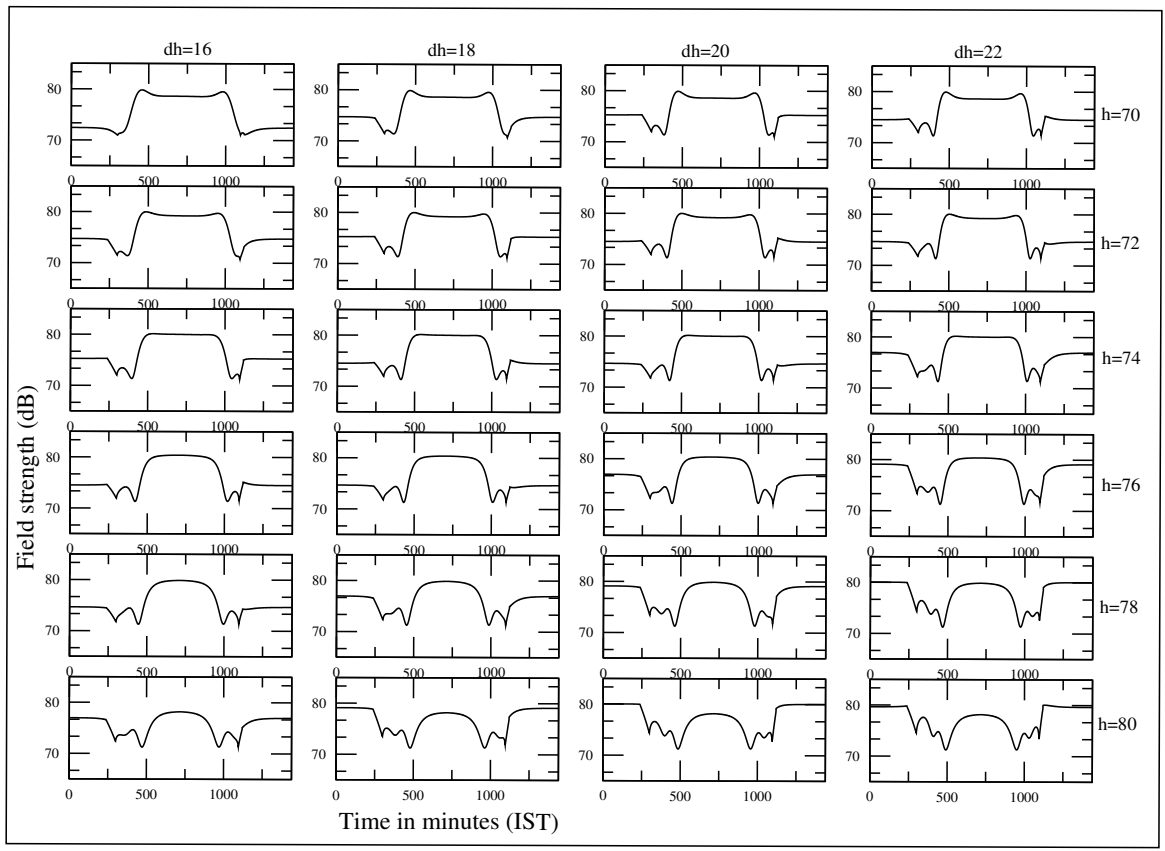

Figure 2.14: Variation of diurnal signal strength of 18.2 kHz VTX wave at Kolkata with the D-region height ($h$) and its increase at night ($dh$) as parameters [Pal & Chakrabarti, 2010].

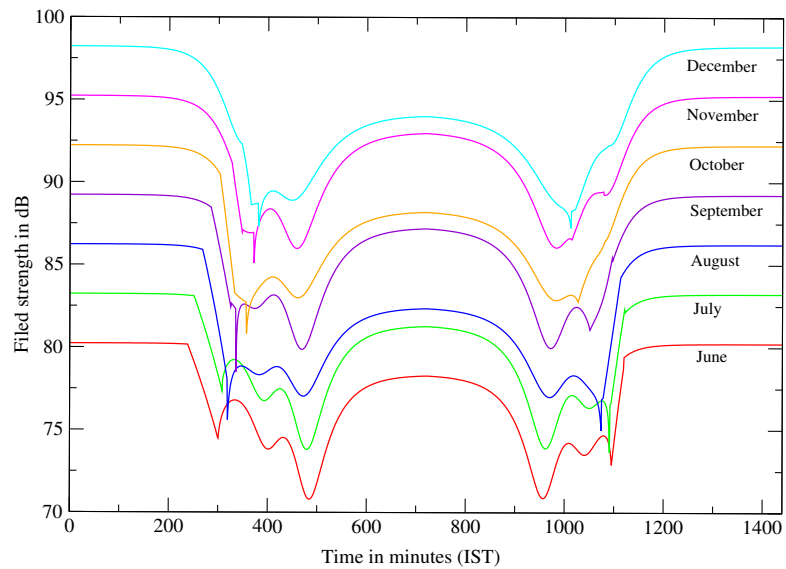

Figure 2.15: Seasonal change of diurnal variation for VTX (18.2 kHz) signal at Kolkata [Pal et al., 2011].

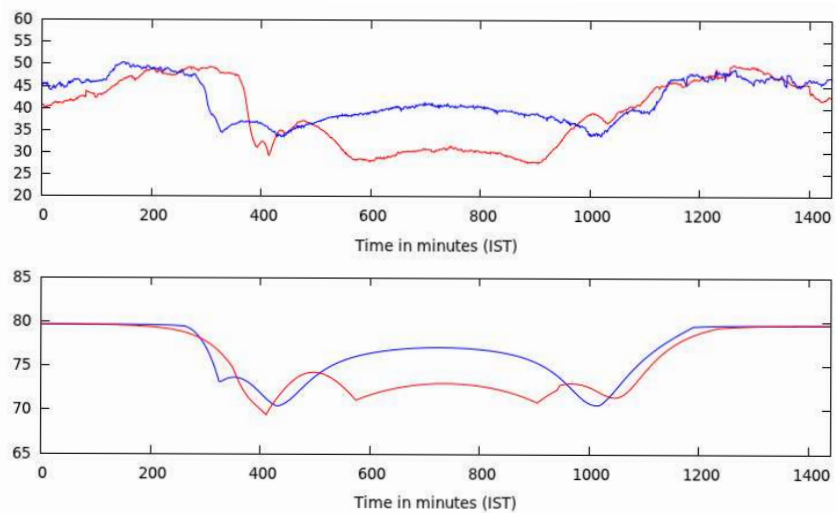

Figure 2.16: First panel shows the observed diurnal behavior of VTX signal in summer (blue curve) and winter (red curve) as recorded in Kolkata. Second panel shows the simulated results for the same obtained from the ray theory code.

## 2.5   Simulation Using Waveguide Mode Theory

Mode theory assumes that the energy within the waveguide is distributed among a series of modes. Here each mode is characterized by a discrete set of angles of incidence of the waves on the ionosphere for which the wave energy will propagate away from the source. These complex angles are called eigen angles or modes and are obtained by solving the determinantal equation -

$$F(\theta) = [R_g(\theta)R_i(\theta) - I] = 0,$$

which is called the modal equation. Here, $R_g$ and $R_i$ are the $2 \times 2$ reflection matrices of the ground and the ionosphere respectively:

$$R_i = \begin{pmatrix} {}_\parallel R_\parallel & {}_\parallel R_\perp \\ {}_\perp R_\parallel & {}_\perp R_\perp \end{pmatrix},$$

$$R_g = \begin{pmatrix} {}_\parallel R'_\parallel & 0 \\ 0 & {}_\perp R'_\perp \end{pmatrix},$$

and I is the unit matrix.

The left subscript on the matrix elements denotes the incident wave polarization and the right subscript denotes the reflected wave polarization. $R$ denotes a downward reflection coefficient and $R'$ denotes an upward reflection coefficient. The cross terms are also called the conversion coefficient and are zero in $R_g(\theta)$ because the ground is not anisotropic.

The ionospheric reflection coefficient matrix $R_i$ is numerically obtained by the wave-admittance method, which is one of the full wave methods suggested by Budden. The ground reflection matrix $R_g$ is calculated in terms of the solutions to Stokes equation and their derivatives [Rec. ITU-R P.684-3, 2002]. Once the values of reflection matrices are obtained, the modal equation is solved for as many modes ($\theta_n$) as consistent with the imposed boundary conditions. From the set of $\theta$'s, the propagation parameters such as - the phase velocity, the attenuation rate, the magnitude and phase of the excitation factors, height gain function are calculated. Using these parameters in a proper mode summing procedure the total field amplitude and phase are computed at some distant point.

## 2.5.1   Long Wave Propagation Capability Code

The Long Wave Propagation Capability (LWPC) code developed by the Naval Ocean Systems Center [Ferguson et al., 1989; 1998] computes field values for VLF signal propagation using the two-dimensional waveguide propagation formulation for an arbitrary propagation path and ionospheric conditions along the path, including the effects of the Earth's magnetic field. This program has been used by many workers and is quiet successful for modeling of long-range propagation of VLF signals also in the presence of ionospheric anomalies. The working of the entire code can be described briefly by describing its three main parts: PRESEG, MODEFNDR/MODESRCH and FASTMC.

PRESEG automatically determines the necessary waveguide parameters along the given propagation path. The path parameters are distance in km, geomagnetic azimuth in degrees East of North, geomagnetic dip in degrees from the horizontal, strength of the geomagnetic field in Webers/$m^2$, ground conductivity in Seimens, ratio of the dielectric constant of the ground to that of free space, the slope of an exponential ionosphere ($\beta$) in $km^{-1}$, and its reference height ($h'$) in km. To include any inhomogeneity in the lower or upper waveguide boundary, one can modify or override the default path parameters.

Earlier developed MODESRCH/MODEFNDR program for obtaining ELF/VLF/ LF mode constants in an EIWG [Morfitt & Shellmann, 1976] is the workhorse of the LWPC model. It takes the waveguide parameters from PRESEG and finds the eigenangles and the necessary mode constants which satisfy the modal equation. MODEFNDR also computes the excitation factors of each mode, which are needed to calculate the final field strengths.

FASTMC performs the mode conversion calculations using the mode conversion theory [Ferguson & Snyder, 1980] along the discontinuities in the inhomogeneous waveguide. The outputs from the MODEFNDR are used as the inputs of the FASTMC.

The default propagation model in LWPC is Long-Wave Propagation Model (LWPM) which treats the ionosphere as having exponential increase in conductivity with height. A log-linear slope ($\beta$ in $km^{-1}$) and a reference height ($h'$) define this exponential model. This default model defines an average value of the slope and reference height that depends on frequency and diurnal conditions. For day time ionosphere $\beta = 0.3$ and $h' = 74$ are constant for all frequencies between 10 kHz -60 kHz. In case of night time ionosphere $h' = 87$ km is constant for all frequencies but $\beta$ varies with frequency in the range 0.3 $km^{-1}$ at 10 kHz to 0.8 $km^{-1}$ at 60 kHz.

## 2.5.2    Results from LWPC

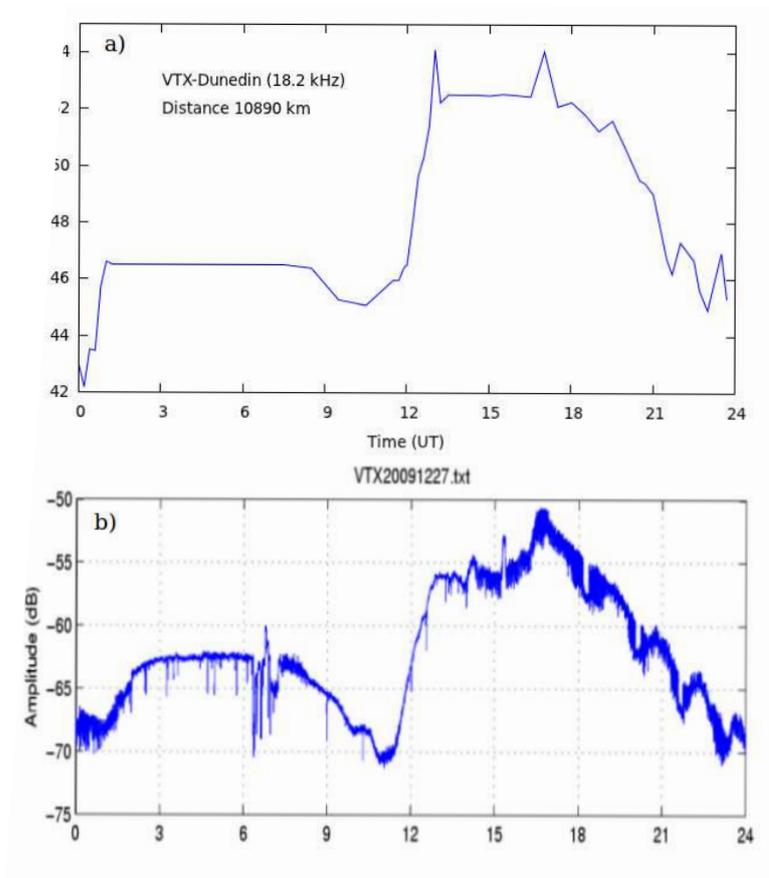

Figure 2.17: a) LWPC predicted diurnal behavior for VTX (18.2 kHz) signal at Dunedin. b) Diurnal behavior of VTX signal received at Dunedin, New Zealand by UltraMSK network.

While the LWPC is generally used for signal coverage analysis, the default propagation model in LWPC is quite usable for approximate prediction of diurnal behavior of VLF signals at any place. As an example of such a calculation, we present Figure 2.17. Here, Figure 2.17(b) represents the diurnal behavior of VTX (18.2 kHz) signal received at Dunedin, New Zealand (great circle distance ~10890 km) by UltraMSK network [image taken from http://ultramsk.com]. Time is in UT hour. Figure 2.17(a) represents the calculated diurnal behavior of the same using the default LWPC model. We can see that the diurnal shapes in this case are roughly the same.

*Application in VLF Campaign*

Indian Centre for Space Physics (ICSP) conducted VLF campaigns in which data from more than a dozen places in India and Nepal were collected [Chakrabarti et al., 2010a, 2012b]. Figure 2.18 shows the map of India showing all the receiving stations employed in the VLF campaigns. Based on campaign results and VLF diurnal shapes, the Indian sub-continent was roughly divided into four regions. In the lower most region close to the transmitter, the signal is ground wave dominated. In the Eastern region, the signal is E-type and in the Western region the signal is W-type. In the middle, the signals in day and night are of roughly equal strength (see Chakrabarti et al., 2010a; 2012b for diurnal shapes). The geomagnetic equator and the magnetic longitude passing close to the VTX transmitter station are plotted for reference.

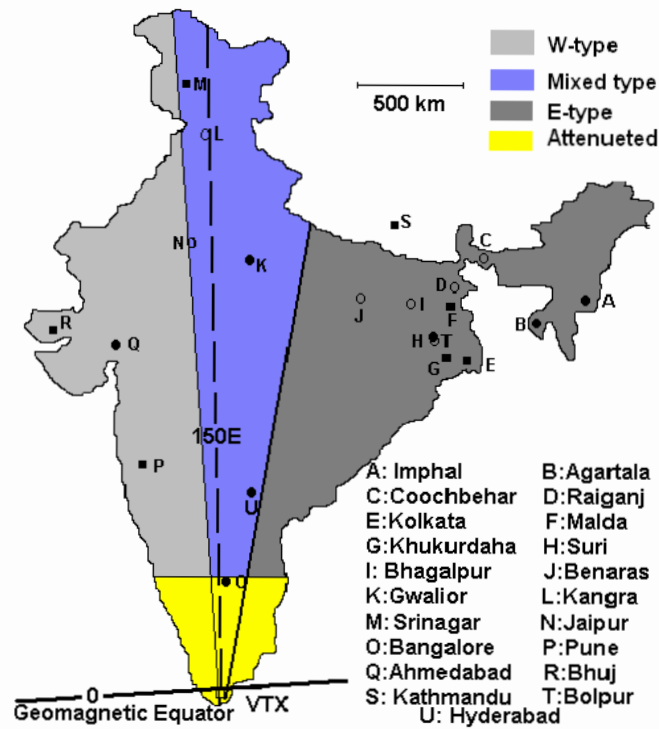

Figure 2.18: The division of the Indian sub-continent as a results of VLF campaigns. In the lower most region close to the transmitter, the signal is ground wave dominated. In the Eastern region the signal is E-type and in the Western region the signal is W-type. In the middle, the signals in day and night are of roughly equal strength [Chakrabarti et al., 2010a].

It has been shown in Chakrabarti et al., (2012b) that the predicted shape of diurnal behavior from LWPC code roughly agrees with that of observations of VTX (18.2 kHz) signal at different places in India during summer and winter VLF campaign. In Figure 2.19, we present the attenuation co-efficients of the first two waveguide modes during the day time and night time propagations. The color bars represent the attenuation in units of the dB/Mm. Contours of constant attenuation are superposed on the plot. Along the X-axis we present the bearing angle with respect to VTX transmitter and along the Y-axis we present the distance from the VTX station which is located at the origin of the plot. All the stations where receivers were kept are represented by filled circles. The circles having smaller positive bearings are to the East of the geomagnetic meridian and those having larger bearings are to the West of the geomagnetic meridian. The top left and top right panels show attenuations of the first and the second modes at 12.00 IST. The bottom left and the bottom right panels show attenuations of the first and second modes at 0.00 IST.

To show the signal behavior more clearly over the Indian sub-continent, Figure 2.20 is plotted for attenuation coefficient as a function of latitude and longitude. Basically, both Figures 2.19 and 2.20 represent the same thing. It is interesting to note that the schematic division of Indian sub-continent with respect to VTX signal behavior (in Figure 2.18) is quantified in Figure 2.20 since it represents the distribution attenuation coefficient.

### *A Comparison Between the LWPC and Wave-hop Codes*

It is instructive to compare the results from our wave-hop model with those obtained from the Long Wave Propagation Capability (LWPC) code. In Figure 2.21, we present the day time spatial variation of the field strength of VTX (18.2 kHz) transmitter as a function of propagation distance from the transmitter with a fixed bearing of $34.6°$ which corresponds to VTX-Kolkata base line. The green curve shows the day time (06:30 UT) variation of the field strength as obtained from the LWPC code. In this case, the so-called Wait's exponential parameters of the lower ionosphere are $\beta = 0.3$ and $h' = 74$ km respectively. The red curve shows the day time field strength variation as obtained from the wave-hop theory with parameters $h = 80$ km $R_i = 0.3$. In this case, the field strength up to 1000 km is the resultant of 1-and 2-hop sky-waves and the ground wave. Beyond 1000 km, the field strength is calculated as a resultant of 1,2-and 3-hop sky-waves and ground wave. The black curve shows the variation of field strength of the ground wave as a

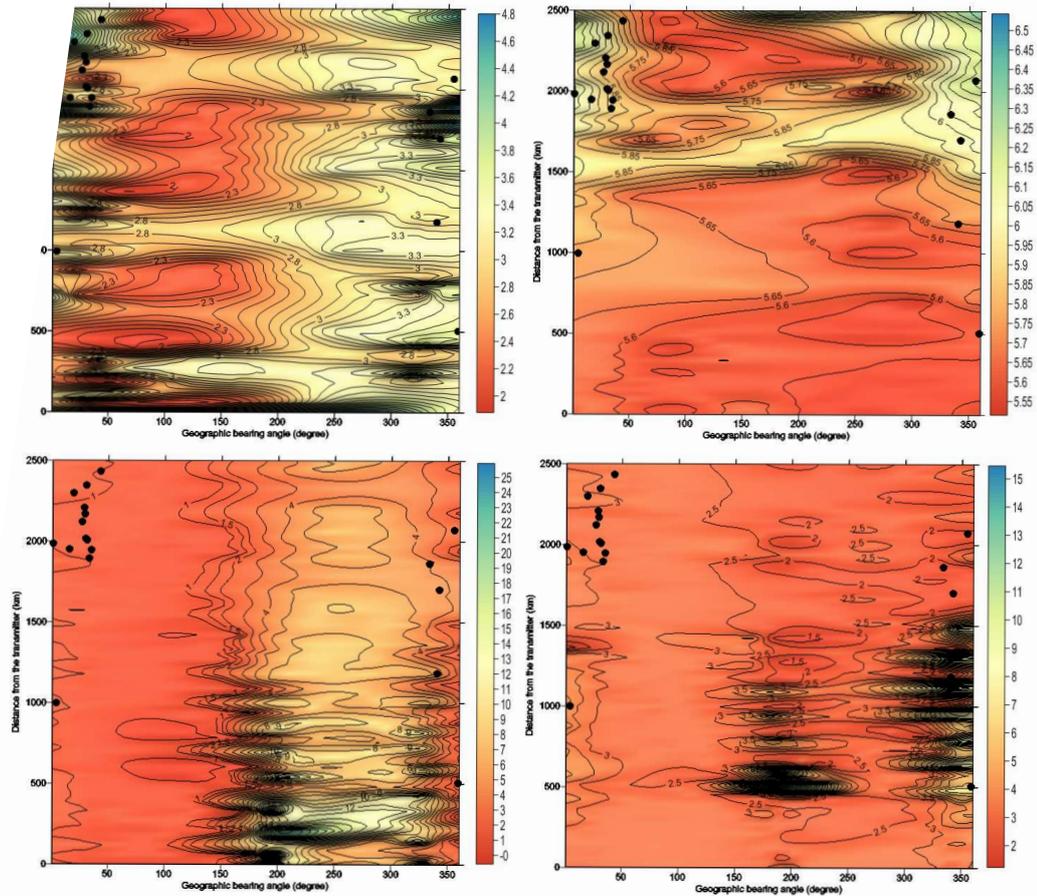

Figure 2.19: Variation of the signal attenuations as a function of the bearing angle (X-axis) and distance from the transmitter (Y-axis). Superimposed are the contours of constant attenuation. VTX is located at the origin of this plot. The locations of all the stations are denoted by filled circles. a Mode 1 (daytime). b Mode 2 (daytime). c. Mode 1 (nighttime). d Mode 2 (nighttime) [Chakrabarti et al., 2012b]

function of distance over sea water with $\epsilon = 70$ and $\sigma = 5$ S/m.

We clearly see the evidence of the alternating destructive and constructive interference patterns and both the theories seem to be predicting their locations at roughly the same places. Note that below about 750 km, the ground wave is dominant. However, with distance, the ground wave weakens and the sky wave dominates.

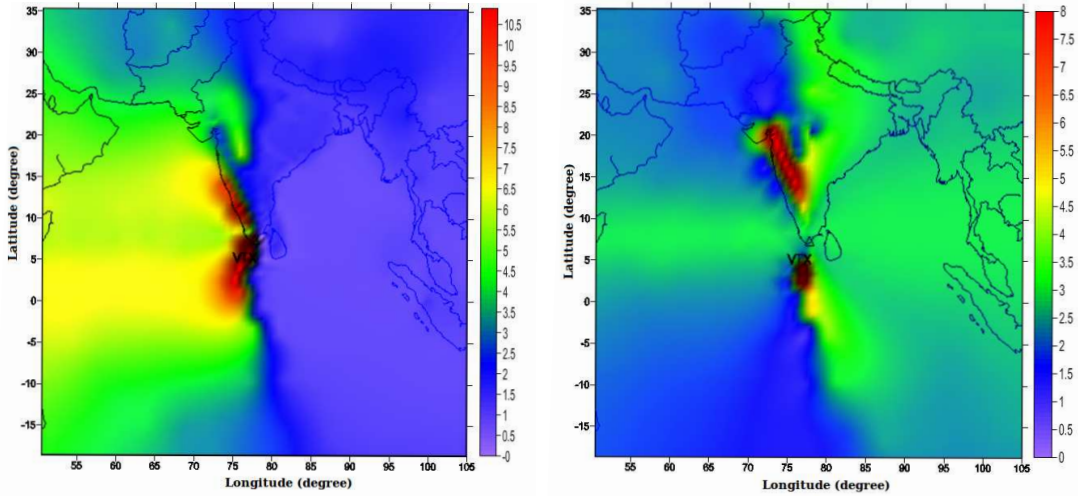

Figure 2.20: Variation of attenuation coefficients of the VTX (18.2 kHz) signal as a function of latitude and longitude. The location of VTX transmitter is indicated in the figure. The left and right plots represent the night time attenuation coefficient of mode 1 and mode 2 respectively.

We can repeat the same exercise for the night time signals. In Figure 2.22, we show the signal amplitude as computed by our code (wave-hop) and LWPC. The results are compared with the ground wave result (black curve). In this case the bearing is fixed at 34.6° which corresponds to VTX-Kolkata base line. The Wait's exponential parameters are $\beta = 0.38$ and $h' = 87$ km (green curve). The wave-hop parameters are $h = 97$ km and $R_i = 0.6$ (red curve).

*Coupled IRI and LWPC*

It is difficult to obtain the exact diurnal shape from the LWPC code, specially, the day time variation of signal amplitude as a function of solar flux. It is necessary to model the variation of ionospheric parameters as a function of solar zenith angle. To avoid such difficulties we couple the FORTRAN code of the International Reference Ionosphere model [Bilitza et al., 2008] with the LWPC code. As a result, the electron-ion density profiles from the IRI-2007 model are computed along the propagation path. These electron-ion density profiles are then used as the inputs in the LWPC code. In Figure 2.23, we compare the day-night variation of the amplitude of VTX signal at 18.2 kHz as observed in Kolkata with the simulated variation as reproduced from the LWPC code. The dotted curve is obtained assuming the electron-neutral collision frequency as described by Wait [Wait & Spies, 1964] which is also the default collision frequency as set in LWPC code. The dashed curve is obtained assuming the collision frequency as described by Kelley [Kelley, 2009].

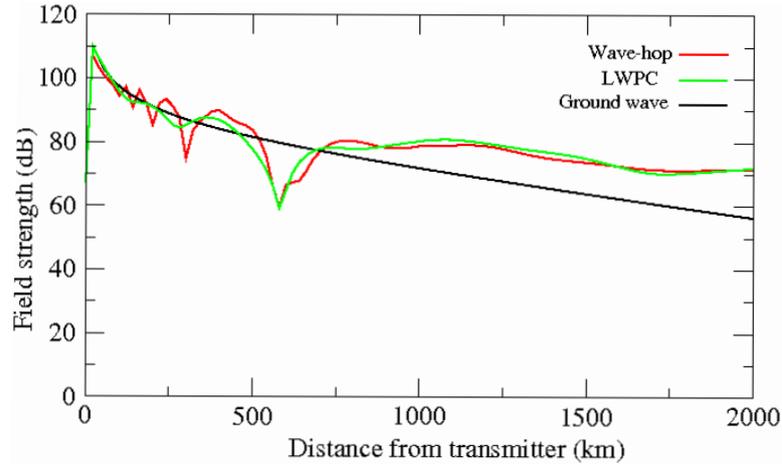

Figure 2.21: Comparison of the field strength with distance (from the transmitter) as obtained by the our wave-hop theory (red) and the LWPC code (green). The ground wave (black) amplitude decays with distance and at a long distance, the sky wave becomes dominant [Pal et al., 2011].

## 2.6    Non-Reciprocity of VLF Propagation Theory

The propagation characteristics of VLF waves depend on the direction of propagation. There exists a difference in phase velocity and attenuation rate between the East-to-West propagation and the West-to-East propagation. In 1925, Round et al. suggested for the first time from their VLF signal strength measurements that the West-East propagating signals were less attenuated than the East-West propagating signals. Until, 1950, the existence of such an event was not certain. In 1954, Budden showed theoretically that propagation between two similar antennas along the Earth's magnetic field must be reciprocal. For certain angles of incidence of the waves, the reciprocity condition was not satisfied if the direction of propagation was across the Earth's field. The non-reciprocity has been explained by suggesting an interaction between the horizontal component of the Earths magnetic field transverse to the direction of propagation and a component of the electric vector which rotates in the vertical plane of propagation of the wave. Thus there exists a difference in electron orbits in the plane of propagation [Barber & Crombie, 1959; Davies, 1990] for the East-West and the West-East paths. Figure 2.24 shows the variation of magnitude of the reflection coefficient with ionospheric collision frequency parameter ($\beta$) for 30 kHz signal. Red curve is the variation of reflection coefficient with collision frequency parameter in the absence of a magnetic field, while the green and blue curves are the same but in the presence of the magnetic field, for the East-West and West-East paths respectively.

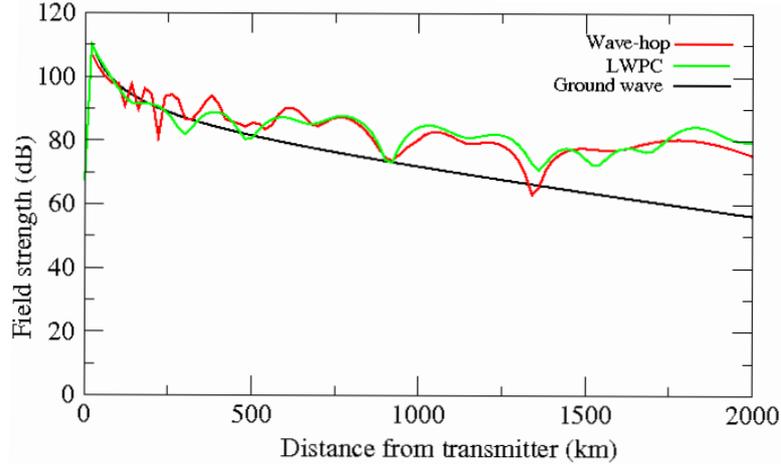

Figure 2.22: A figure similar to Figure 2.21, however, the night time parameters were used. In this case, the Wait's exponential parameters are $\beta = 0.38$ and $h' = 87$ km (green curve) and the wave-hop parameters are $h = 97$ km and $R_i = 0.6$ (red curve) [Pal et al., 2011].

In Figure 2.25, we present a classic example of non-reciprocity between eastern path (Kolkata) and western path (Pune) for the VTX (18.2 kHz) signal observed by ICSP-VLF network in India. A careful observation shows that the sunrise pattern of Pune is similar to the sunset pattern of Kolkata and vice-versa. This is due to the fact that Pune is in West while Kolkata in East with respect to the VTX transmitter. Also day time signal strength is much stronger for Pune.

## 2.7 Variation of Sunrise-Interference Pattern of VLF Signal

Figure 2.26 shows the variation of VTX (18.2 kHz) signal amplitude as it changes from Winter (January) to Summer (June). It is clear that the VLF day-length becomes broader from Winter to Summer as is obvious. The most noticeable fact is that the shape of the interference pattern after the sunrise terminator time (SRT) changes with season and there is a maximum and minimum followed by the SRT in each diurnal shape. Denoting this bump between the two minima as post-sunrise signal fluctuation (PSF) (as shown an inset in Figure 2.26), we found a relation between the PSF and the angle ($\alpha$) made by the sunrise terminator path with the propagation path which is plotted in Figure 2.27. Thus we can see that the larger PSF occurs when the sunrise terminator makes a small angle with with the propagation path and vice-versa. The PSF may be related to the origin of another lower ionospheric region called the C-region (60 - 75 km) [Raulin et al., 2010]. The intensity of ionization in the C-region depends on the angle between the propagation path and sunrise terminator line.

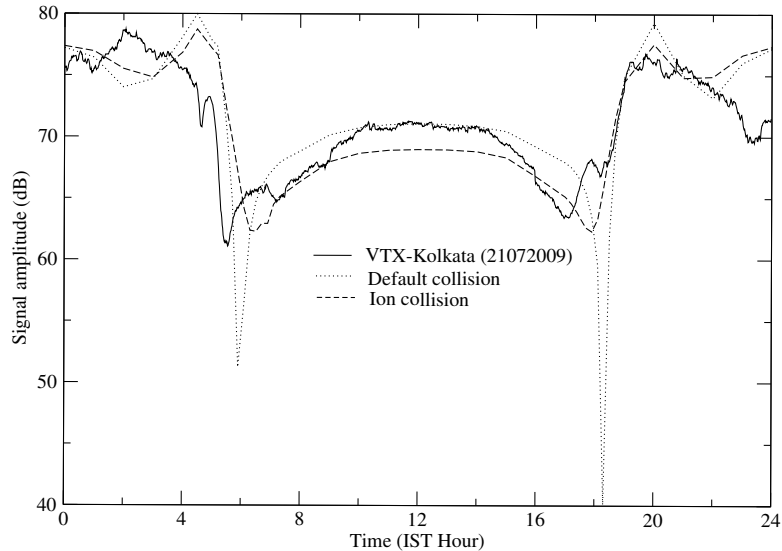

Figure 2.23: Diurnal variation of VTX (18.2 kHz) signal amplitude received at Kolkata. Solid line represents the observed data of July 21, 2009. Dotted curve represents the reproduced diurnal variation obtained using the electron-neutral collision frequency as defined by Wait, which is the default in LWPC code. Dashed curve represents the reproduced diurnal variation using the electron-collision frequency as defined by Kelley (2009).

The nature of signal interference pattern around the ground sunrise time of the receiver depends in general on the frequency of the signal, the position of the receiver along the propagation path as well as on the interference pattern between the ground wave and sky waves. In Figure 2.28, we have shown the behavior of the VTX (18.2

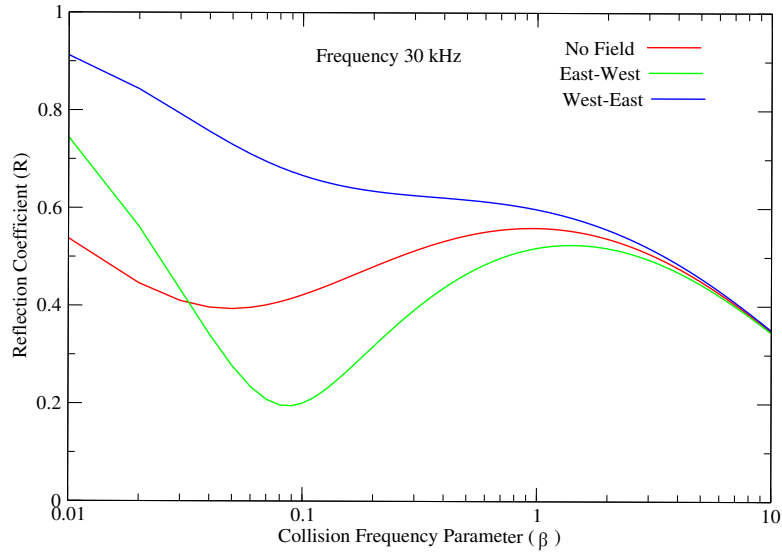

Figure 2.24: Variation of magnitude of the reflection coefficient with ionospheric collision frequency parameter ($\beta$) for 30 kHz signal [following the model in Barber and Crombie, 1959].

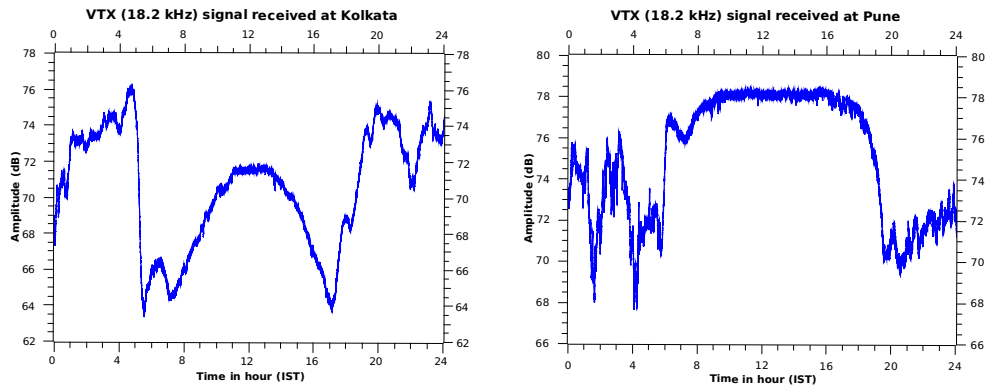

Figure 2.25: Example of non-reciprocity between the eastern path (Kolkata) and western path (Pune) for the VTX (18.2 kHz) signal observed by ICSP-VLF network.

kHz) signal around the sunrise time of the receiver at different places received during the summer VLF campaign of ICSP [Chakrabarti et al., 2010a]. The vertical arrows represent the time of ground-sunrise time at the corresponding places. The proper

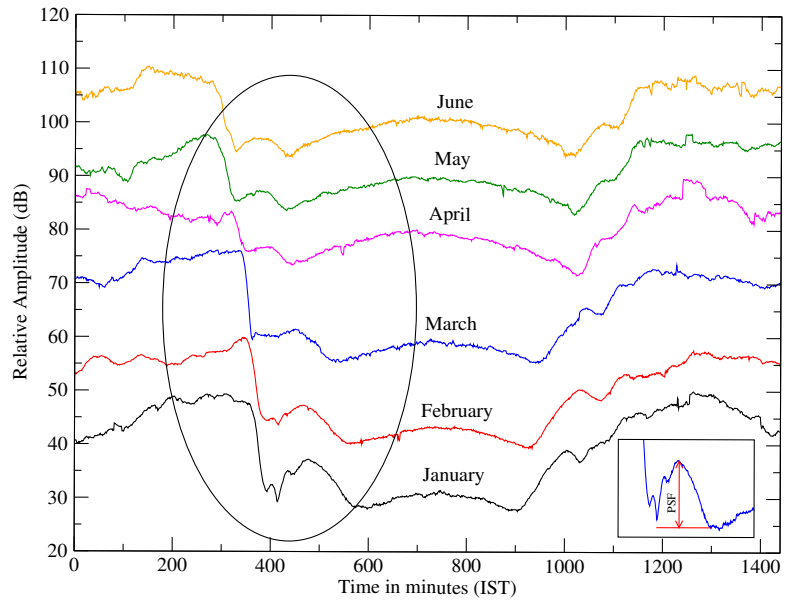

Figure 2.26: Observed seasonal variation of VTX signal in terms of diurnal variation.

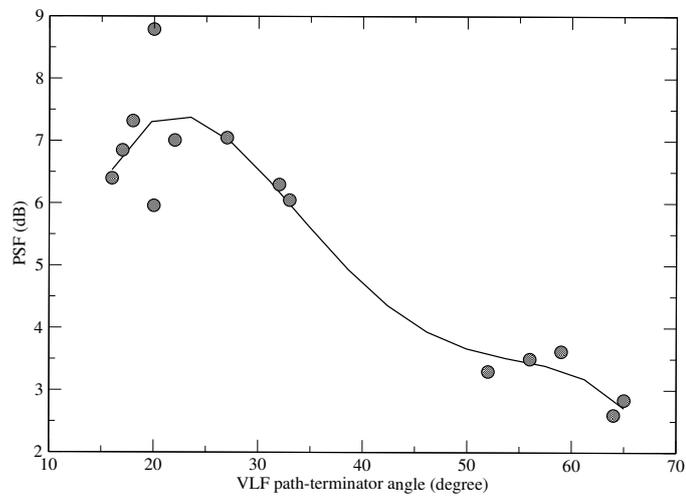

Figure 2.27: Variation of the post-sunrise signal fluctuation (PSF) versus the angle between the propagation path and sunrise terminator $\alpha$.

simulation of such behavior at different places requires a complete general VLF propagation model.

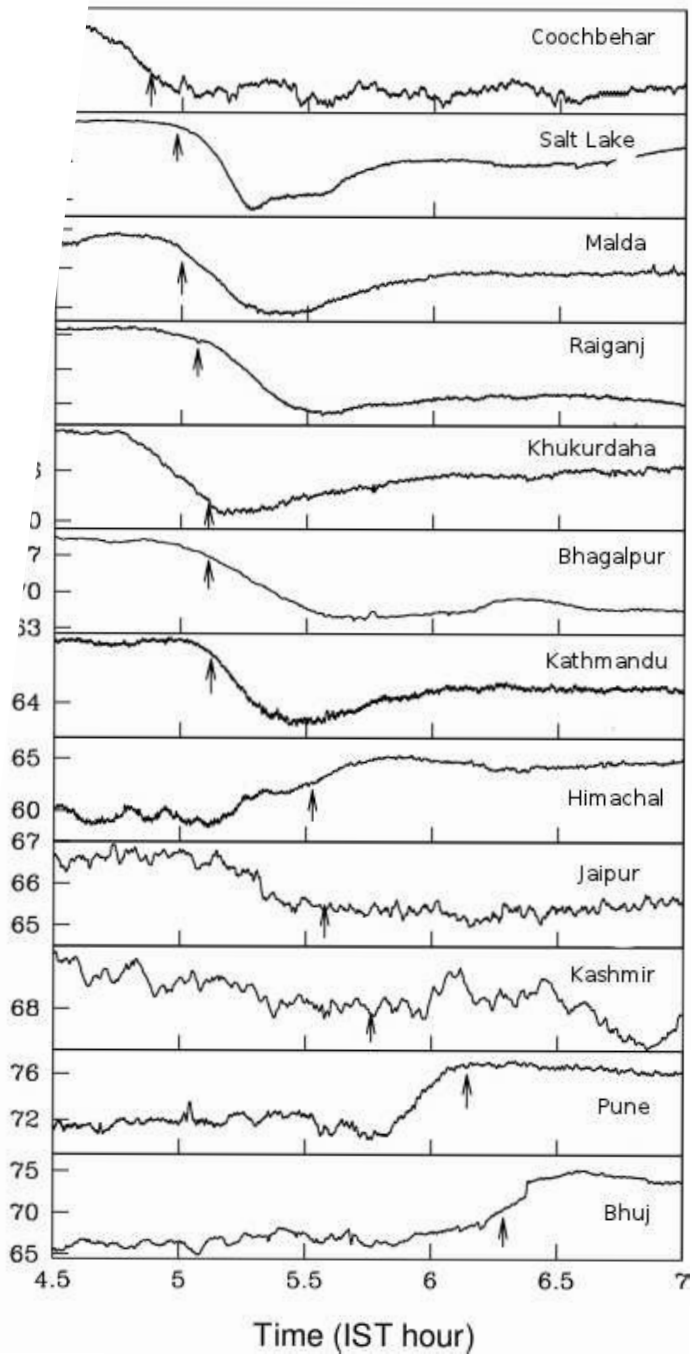

Figure 2.28: The inference pattern around the sunrise time at different places in India for VTX signal. Arrows represent the local sunrise time of the places.

# Chapter 3

# Effects of Solar Flares in the Earth-Ionosphere Waveguide

## 3.1 Introduction

Solar flares have strong hard UV and X-ray components which increase the conductivity in the D-region resulting significant perturbations in the received amplitude and phase of VLF radio signals propagating in the Earth-ionosphere waveguide (Mitra, 1974; Thomson et al., 2001; Todoroki et al., 2007; Zigman et al., 2007). Soft X-rays (below 1 nm) penetrate down to the D-region where they produce extra ionization by (partially) ionizing all the neutral constituents, particularly the dominant ones such as Nitrogen and Oxygen. Though Lyman-$\alpha$ is the main source of ionization in this region, there is no significant change in Lyman-$\alpha$ flux during solar flare (Swift, 1961). The enhancement in ionization in the D-region ionosphere due to flares increases the VLF phase velocity and lowers the VLF reflection height. Apart from the general effect of solar flares on ionosphere regarding the chemical reactions, nowadays the research activity is concentrated towards modeling of the VLF propagation in the presence of flares and estimation of electron density and finding relations between the flares and its effects on VLF propagation.

In this Chapter, we have shown the effects of a solar flare on the D-region using two VLF paths. Using LWPC code we computed the time variation of the VLF reflection heights and the electron number densities during the flare event. We have also shown the dependence of VLF perturbations due to solar flares on frequency, propagation path and on the solar zenith angle.

## 3.2   Observational Data

The VLF receivers of ICSP-VLF network are continuously monitoring several world-wide VLF transmitter signals which include NWC (19.8 kHz), VTX (18.2 kHz) and JJI (22.2 kHz) transmitters, as described in Chapter 2. In Figure 3.1, we present a typical examples of NWC (19.8 kHz) signal as perturbed by solar flare events observed from a ground based VLF receiver at IERC/ICSP. Here, the blue curve represents the diurnal variation of NWC signal on a normal day (November 19, 2010) when there is no flaring activity on the Sun. The red curve shows the same variation but with the presence of solar flares on an active day (February 16, 2011). There is a sudden rise of the amplitude followed by a slow decay associated with each type of solar flare.

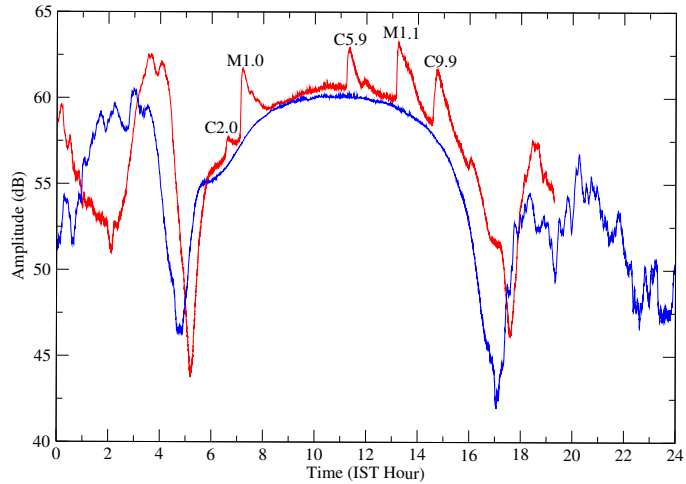

Figure 3.1: Detection of solar flares by monitoring the NWC signal at 19.8 kHz by ICSP receiver. Blue curve represents the diurnal variation of NWC signal on a normal day when there is no flaring activity on the Sun. Red curve shows the perturbation of the signal due to solar flares on an active day.

Figure 3.2 shows the GOES-15 satellite detection of solar flares on February 16, 2011 (IST day) in two different X-ray bands. The red curve is for the soft X-ray flux (in $W/m^2$) between 1.0–8.0 $\dot{A}$ and the blue curve is for the hard X-ray flux (in $W/m^2$) between 0.5–4.0 $\dot{A}$. Solar flares are classified as B, C, M and X depending on the peak flux in the soft X-ray band, with each class having a peak flux ten times greater than the preceding one. .

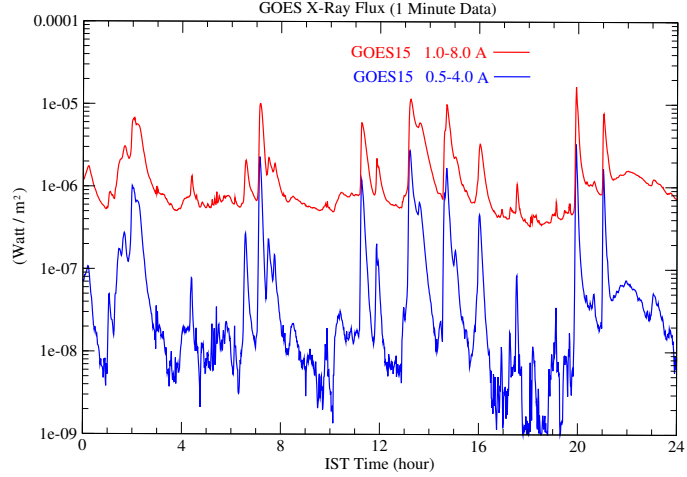

Figure 3.2: GOES-15 soft X-ray (1.0–8.0 $\dot{A}$) and hard X-ray (0.5–4.0 $\dot{A}$) flux on February 16, 2011 (IST).

## 3.3    Effects of a Solar Flare on Two VLF Frequencies

An M2.0 class solar flare has been detected simultaneously on two different VLF paths (VTX-Kolkata and NWC-Kolkata) by a VLF receiver at Kolkata (lat: 22° 34′ N, long: 88° 24′ E) on 12th June, 2010. In case of VTX-Kolkata and NWC-Kolkata, the flare produces a maximum amplitude deviation of 5.6 dB and 3.5 dB respectively.

In Figure 3.3, we present the GOES satellite data (solid curve) of the X-ray flare and the corresponding deviation of the VTX signal amplitude (dashed curve) due to the flare as received from Kolkata. The VTX signal amplitude deviation in dB was obtained by subtracting the data of the previous day from the data of the flare-day. The time is given in Indian Standard time (IST=UT+5:30 h). The VLF signal is behind the GOES signal by about 2.5 minutes. In Figure 3.4, we compare the same X-ray data (solid curve) with the NWC (19.8 kHz) signal amplitude (dashed curve) of the flare as received from Kolkata. The VLF signal lags behind the GOES signal by about 1.5 minutes.

This time-delay between the peak of the satellite detection and VLF detection is an important parameter for the ionosphere. It represents the "sluggishness" of the ionosphere (Valnisek, 1972) and also allows us to estimate the effective recombination coefficient indispensable for solving the electron continuity equation. It is interesting to note that the time delay also depends on the propagation path and VLF frequency. Note that the perturbation is weaker in the NWC signal. Also in both the cases, the VLF signal decays slower than that observed in GOES data, since the recombination time scale in the ionosphere is larger.

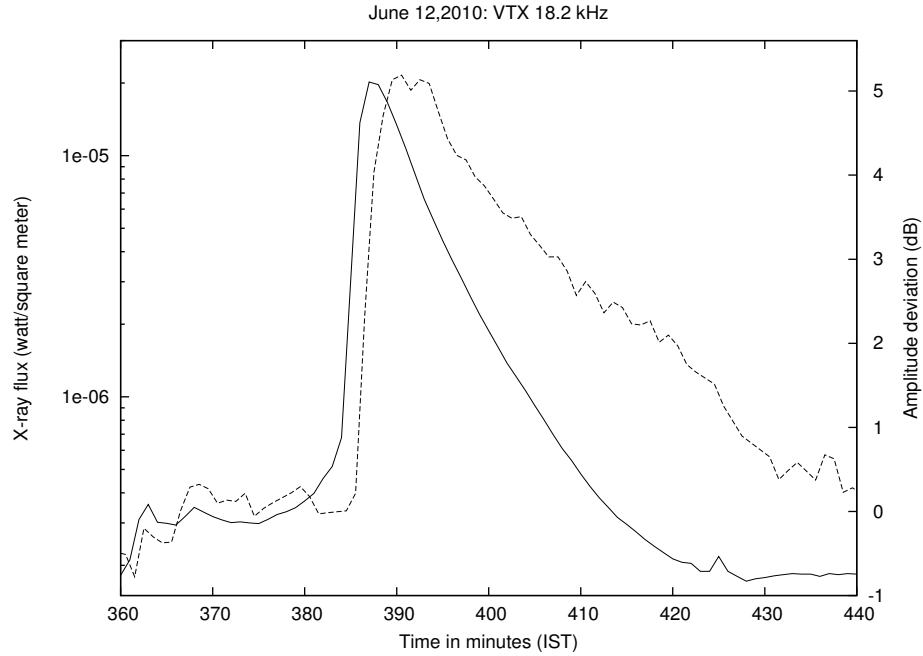

Figure 3.3: Solar flare event (solid line) with a peak X-ray irradiance at 06:27 IST on 12th June, 2010 and the corresponding disturbance of VLF signal amplitude of VTX (18.2 kHz) transmitter (dashed line). The VLF signal amplitude peak occurred at 06:29.5 IST.

Now to deduce the changes in the ionospheric parameters due to this M2.0 flare, we used the LWPC code developed by Ferguson et al. (1998). The Long-Wave Propagation Model (LWPM) in LWPC treats the ionosphere as having exponential increase in conductivity with height. A log-linear slope ($\beta$ in $km^{-1}$) and a reference height ($h'$) define this exponential model. The quiet day values of $\beta$ and $h'$ have been calculated during a flare time in the absence of flare. Then the range-exponential model of LWPC has been used to calculate the set of $\beta$ and $h'$ parameters that lead to the best agreement between the observed and calculated phase and amplitude perturbations during the flare. Thus, using the time varying $\beta$ and $h'$, the time variation of the electron density height profile during solar flare event has been obtained.

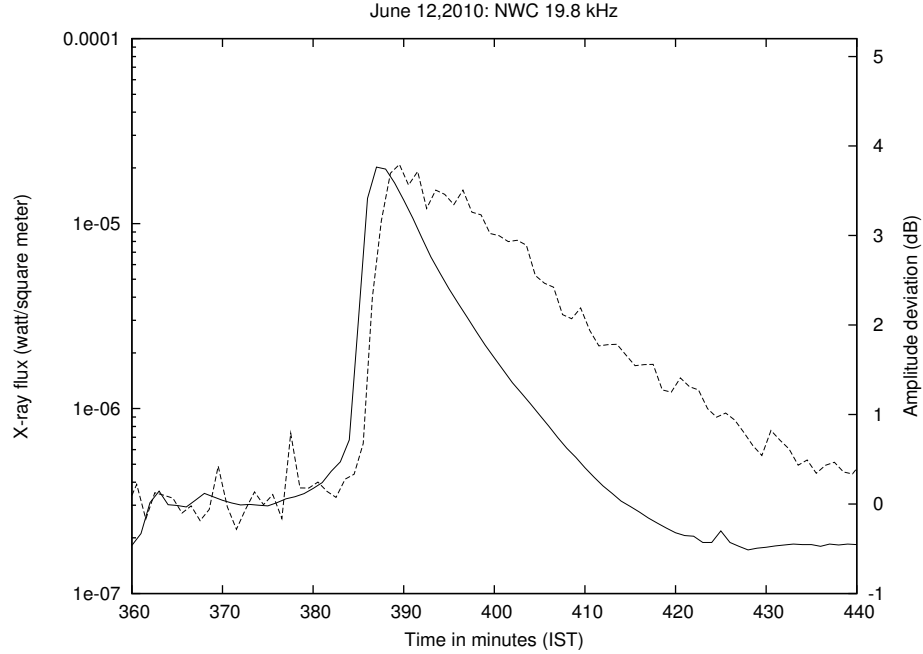

Figure 3.4: Solar flare event (solid line) with a peak X-ray irradiance at 06:27 IST on 12 June, 2010 and the corresponding disturbances of VLF signal amplitude of NWC (19.8 kHz) transmitter (dashed line). The VLF signal amplitude peak occurred at 06:28.5 IST.

### 3.3.1  Determination of $\beta$ and $h'$ Under a Flare Condition

We determine the observed amplitude perturbations $\Delta A$ by subtracting the quiet day data on 11th June from the perturbed data on 12th June as-

$$\Delta A = A_{perturb} - A_{quiet}.$$

These perturbations $\Delta A$ are then added to the simulated unperturbed value which is obtained from the LWPC default program at the receiver site to obtain,

$$A'_{perturb} = A_{lwpc} + \Delta A.$$

These $A'_{perturb}$ are then used for obtaining the $\beta$ and $h'$ parameters under flare conditions. The unperturbed values obtained from LWPC program for the two propagation paths are-

Table 3.1: An example of how the ionospheric parameters vary due to a solar flare.

| Frequency (KHz) | FPT (IST) | OPT (IST) | Time | $\Delta A(dB)$ | $\beta \ (km^{-1})$ | $h'$ (km) | $N_e$ at 74 km $(m^{-3})$ |
|---|---|---|---|---|---|---|---|
| VTX, 18.2 | 06:27 | 06:29 | 06:20 | 0.176 | 0.329 | 73.2 | 0.28117E+09 |
| | | | 06:29 | 5.135 | 0.48 | 68.0 | 0.38498E+10 |
| | | | 06:40 | 3.713 | 0.395 | 69.6 | 0.12288E+10 |
| | | | 07:20 | 0.208 | 0.32 | 73.36 | 0.26522E+09 |
| NWC, 19.8 | 06:27 | 06:28 | 06:20 | 0.244 | 0.32 | 74.1 | 0.20930E+09 |
| | | | 06:29 | 3.794 | 0.42 | 68.3 | 0.23679E+10 |
| | | | 06:40 | 2.996 | 0.388 | 69.5 | 0.12386E+10 |
| | | | 07:20 | 0.465 | 0.322 | 73.5 | 0.25386E+09 |

VTX-Kolkata = 71.56 dB (above $1\mu$V/m),

NWC-Kolkata = 58.82 dB.

Both of these correspond to $\beta$=0.3 $km^{-1}$ and $h' = 74.0$ km.

To calculate the electron density profile for the lower ionosphere the following well-known Wait's formula was used (Wait, 1962; Thomson, 1993; Grubor, 2008):

$$N_e(h, h', \beta) = 1.43 \times 10^{13} exp(-0.15h') exp[(\beta - 0.15)(h - h')],$$

where $N_e$ is in $m^{-3}$.

Table 1 shows the deduced $\beta$ and $h'$ parameters and electron densities at different times during the flare. In the second column we give the flare peak time (FPT) according to GOES data. In the third column, we present the observed peak time (OPT). The fourth column gives the time at which we compute the ionospheric parameters. Other columns are self-explanatory.

### 3.3.2   Results

From Table 1, we see that the VLF reflection height due to the flare decreases approximately by 6 km from the normal value for both the propagation path (VTX-Kolkata - 6 km and VTX-NWC - 6.3 km).

In Figure 3.5, we present the variation of the electron number density with height (in km) at various times during the flare using the VTX data. The dotted line gives the density profile in a quiet day. The profile at 6:20 am, just at the beginning of the rising phase of the flare, the profile is similar. At the peak of the flare (6:29

IST), the variation became steeper – at higher altitude, say, at 90 km, the number density is increased by a factor of a few hundred, while at a lower height, say at about 60 km, the number density went up by the factor of a few. As time passes by and the flare decays, the electron number density distribution tries to go back to its 'normal' quiet value.

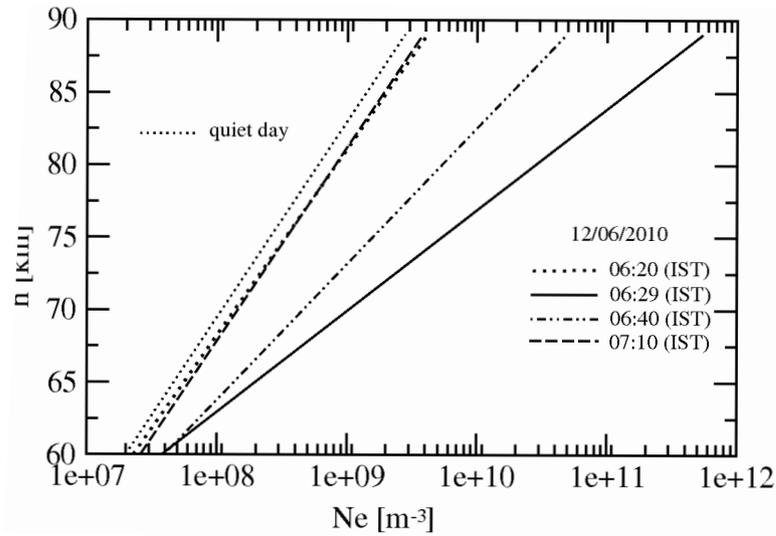

Figure 3.5: Changes in the electron density profile in the lower ionosphere from 60 km-90 km during the solar flare M2.0 as deduced from the VTX (18.2 kHz) data [Pal & Chakrabarti, 2010].

The sharpness in electron distribution ($\beta$) and the height parameter $h'$ were also found to respond to the X-ray flux. Figures 3.6 and 3.7 show the variation of $\beta$ and $h'$ with X-ray flux for this flare. We note that there is roughly a linear correlation between $\beta$ or $h'$ and X-ray flux.

Depending on the recombination time scale at different heights, the number density of the electrons would vary with time. In Figure 3.8, we show in the same plot variations of both the X-ray flux and electron density with time at a height of 74 km. The shape roughly follows the VLF signal amplitude, but not the X-ray flux, owing to the time delay due to recombination.

It is to be noted that the parameters one obtains from the LWPC are average properties along the propagation path. They are not the absolute values at the receiver point or the mid-point of the path (i.e., the place of first reflection). This can be easily verified by comparing the deduced values for two VLF signals which record the same flare. In Figure 3.9, we draw a plot similar to Figure 3.5 for the

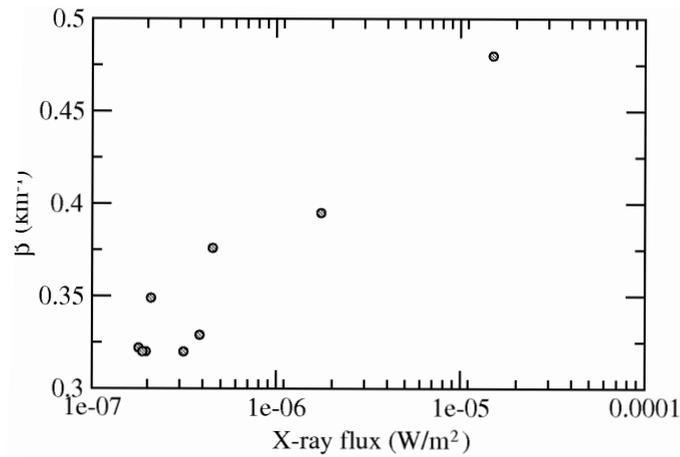

Figure 3.6: The D-region sharpness parameter $\beta$ as a function of the X-ray flux in the band $0.1 - 0.8$ nm.

NWC signal. We note that the highest electron number density is a factor of five or so less than that in Figure 3.5 at a height of 90 km. In this case the path is longer, and thus the average electron density is lower. In Figure 3.10, where the plot of the variation of the electron density (dashed curve) is shown, we note that the peak number density is much lower for NWC path than for the VTX path (Figure 3.8).

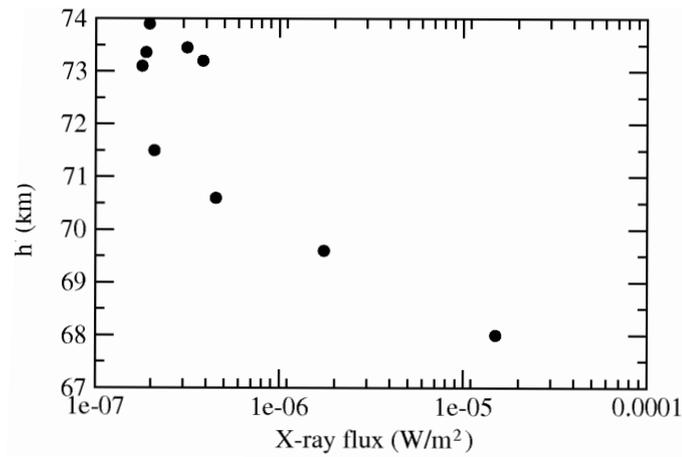

Figure 3.7: The D-region height parameter $h'$ as a function of the X-ray flux in the band $0.1 - 0.8$ nm.

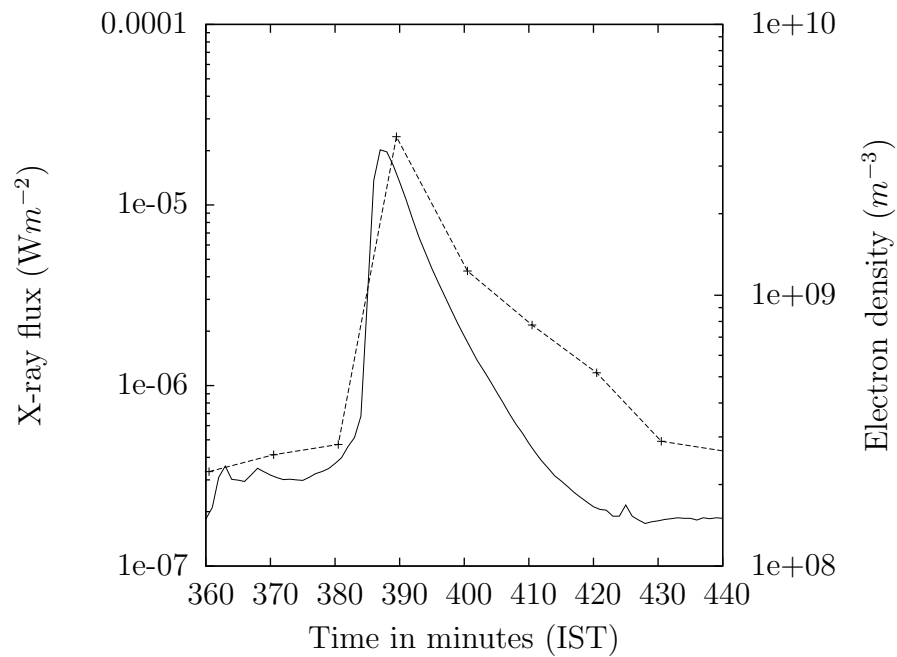

Figure 3.8: Variation of average electron number density along the VTX-Kolkata propagation path at a height of 74 km in the course of solar flare (dashed line). The variation of X-ray flux (solid line) is also shown.

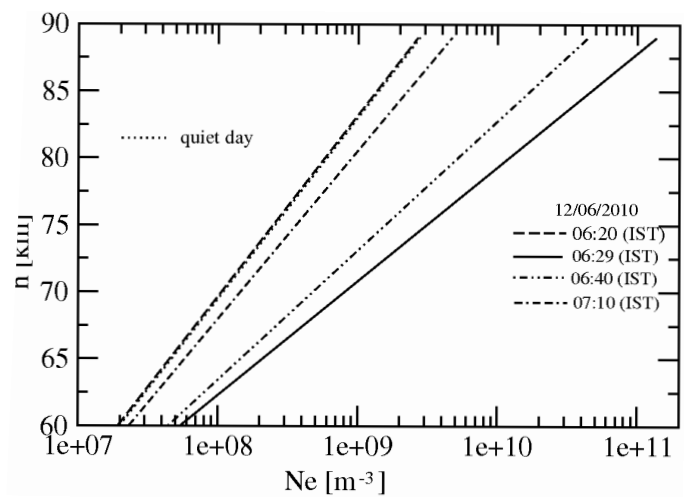

Figure 3.9: Changes in the electron density profile in the lower ionosphere from 60–90 km during the the solar flare M2.0 deduced from NWC (19.8 kHz) data [Pal & Chakrabarti, 2010].

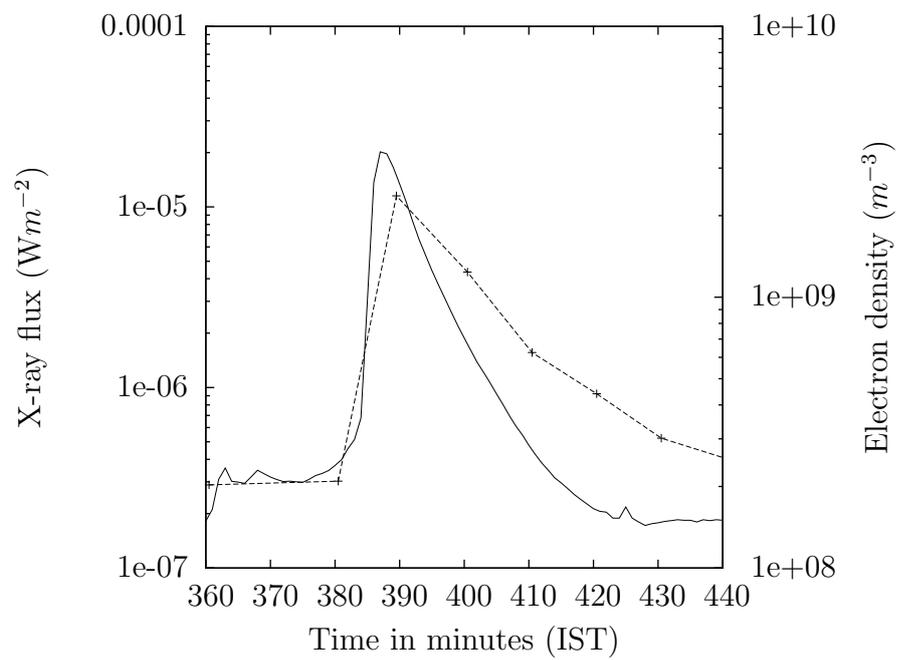

Figure 3.10: Variation of average electron number density along the NWC-Kolkata propagation path at a height of 74 km in the course of solar flare (dashed line). The variation of X-ray flux (solid line) is also shown. The number density achieved is much lower than what was obtained for VTX-Kolkata path.

## 3.4  Dependence of VLF Perturbations on Solar Zenith Angle

The amplitude of VLF perturbations depends not only on the frequency, magnitude of solar flux and propagation path but also on the position of the Sun along the propagation path. That means the solar flares of nearly same magnitude can produce different VLF perturbations depending on the position of the Sun in the path or in other words on the time of occurrence of flares in the day time. In Figure 3.11, maximum VLF perturbation (termed as VLF response) is plotted against the solar zenith angle at the receiver corresponding to the peak time of flares. Here all the flares are between M1.0-M2.0 and produced perturbations on the NWC-Kolkata path in 2011. The zero of solar zenith angle represents the local noon time of the receiver with minus and plus zenith angle corresponding to pre-noon and post-noon periods.

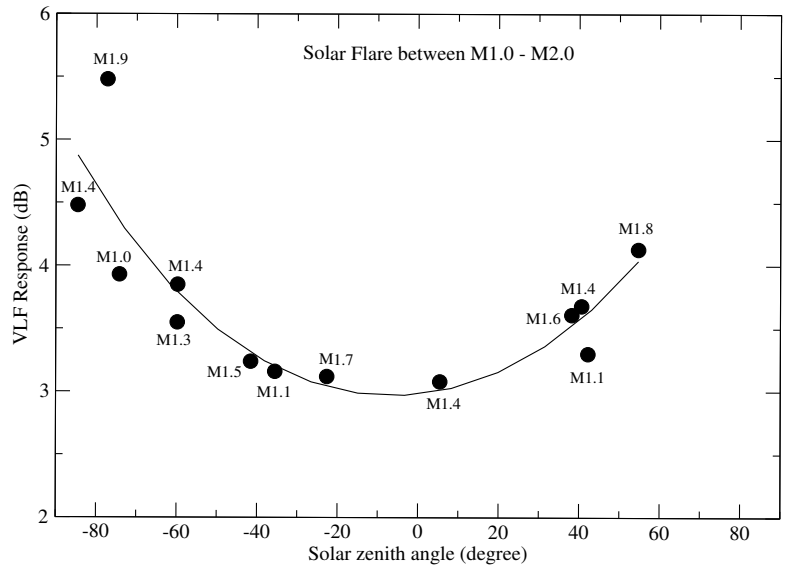

Figure 3.11: Maximum VLF perturbation due to solar flares is plotted against solar zenith angle at the receiver for solar flares between M1.0 to M2.0.

## Chapter 4

# Modeling VLF Signal Perturbations Associated with Total Solar Eclipse

## 4.1 Introduction

A solar eclipse produces disturbances in the ionosphere and provides us with a unique and rare opportunity to verify our theoretical understanding about how the ionosphere reacts to impinging radiation, or, for that matter, withdrawal of it. During a solar eclipse, the ionosphere experiences a virtual sunset and a sunrise in quick succession. The UV and soft X-ray photons from the solar disk are totally or partially blocked for a certain time interval and the precise time of this event is known well ahead of time. Thus it is possible to study the behavior of different atmospheric processes including the ion chemistry.

## 4.2 Earlier Works

There are a large number of papers in the literature related to the solar eclipse effects on the ionosphere. Most of the works are concerned with the ionospheric behavior above 100 km altitudes during eclipse. Theoretical investigations of the effects of the eclipse on ionospheric E and F regions were made by Rishbeth (1968). Variation of total electron content (TEC) of the ionosphere during solar eclipses have been extensively studied by Yeh et al., 1997; Huang et al., 1999; Tsai & Liu, 1999; Afraimovich et al., 2002; Jakowski et al., 2008; Krankowski et al., 2008. Latitude and longitude dependence of the ionospheric response in terms of TEC due to solar eclipses have been reported by Le et al., 2009 and Krankowski et al., 2008. The difficulties in ground based D-region measurements limit the study of the behavior of lower ionosphere (specially D-region) during solar eclipses. Much of the study of the D-region during solar eclipses depend on the use of ELF/VLF wave propagation along the Earth-ionosphere waveguide.

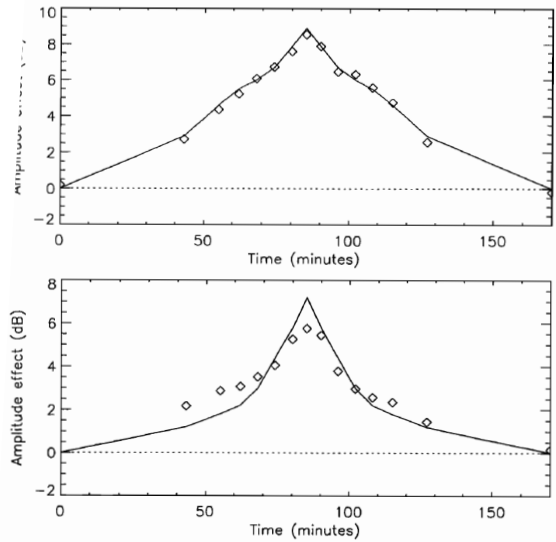

Figure 4.1: The LWPC modeled amplitude effect (solid curve) compared with corresponding observed changes (diamond) for FTA2 (20.9 kHz) to Saint Ives and DHO (23.4 kHz) to Saint Wolfgang path [Clilverd et al., 2001].

The first report of the effect of the solar eclipse on VLF path between GBR (Rugby, England) to Cambridge was reported almost half a century ago (Bracewell, 1952). Subsequently, several authors have presented the results of amplitude and phase variations due to solar eclipses. Crary & Schneible (1965) have found 2 to 3 dB amplitude change due to solar eclipse associated with 6 to 11 km enhancement in the VLF reflection height for short VLF paths (≤1000 km). Lynn (1981) studied the effects of October, 1976 solar eclipse over three long VLF paths. The VLF phase response was found to be a non-linear function of solar obscuration and the effective time constant of ionospheric response was found to be $4.3 \pm 1.5$ min. There are very few papers which discussed the modeling of VLF signals perturbations associated with a solar eclipse. Clilverd et al. (2001) presented the effects of total solar eclipse of August 11, 1999 on VLF signals in Europe. They had five receiving sets which observed several stations each and thus had altogether 17 different paths. While their path lengths varied from 90 km to 14,510 km, majority were $< 2000$ km. They reported positive amplitude change on path lengths $< 2000$ km, negative amplitude changes on paths $> 10,000$ km and negative phase changes on most paths independent of path lengths. These authors also gave an explanation of the nature of the signals using the LWPC waveguide code (Ferguson, 1998). They found an 8 km rise in $h'$ parameter during eclipse conditions over short paths, while over the medium-length paths, the estimation was a rise of 5 km in $h'$ parameter. They also calculated the variation of electron density at 77 km altitude throughout the period of solar eclipse, which showed a linear variation in electron production rate with solar ionizing radiation. Figure 4.1 shows their observed and simulated results for two VLF paths. A similar result was reported by Fleury et al. (2000), who found an increase of about 5 km in the $h'$ parameter at mid-eclipse. The path lengths were within 1000 km and they used a constant $\beta = 0.5$ parameter throughout the eclipse time while changing $h'$ linearly from day time to night time values. They also assumed the disturbance of height to be uniform along the whole radio path.

During the Total Solar Eclipse of July 22, 2009, ICSP conducted a week-long VLF campaign in which data from more than a dozen places in India and Nepal were collected before, during, and after the eclipse [Chakrabarti et al., 2012a]. All the receivers were monitoring VTX (18.2 kHz) and NWC (19.8 kHz) transmitters during this time. Special interests were given to the VTX transmitter since the location of the transmitter near the southern tip of India gives a uniform geophysical condition in Indian context. Also, the magnetic meridian passing through VTX vertically splits the Indian sub-continent into roughly two halves with different signal characteristics in the eastern and western halves (discussed in Chapter 2). Thus spreading around receivers in the sub-continent allowed us to study the signals under varied wave propagation conditions on a short path ($< 3000$ km). In this Chapter, we present the observations and modeling of the signatures of the VTX signal due to the total solar eclipse (TSE) which took place on the 22nd of July 2009 in India.

## 4.3     Observed Effects of Solar Eclipse on VLF Paths

We use ICSP made VLF antenna/receiver systems to monitor VLF signals. Each set consists of a loop type rectangular antenna and Gyrator III type receiver. The system was controlled by PC with a sound card for real time data acquisition.

In Figure 4.2, we show the path of totality during July 22nd, 2009 eclipse. While ICSP obtained data at twelve places (Chakrabarti et al., 2010a, 2012a), we chose six representative propagation paths (VTX-Kolkata, VTX-Khukurdaha, VTX-Malda, VTX-Raignaj, VTX-Pune and VTX-Kathmandu) of varying character and noise-free signals for the modeling purpose. The VTX-Kolkata, VTX-Khukurdaha and VTX-Pune propagation paths are very much below the belt of totality, while the VTX-Malda (2151 km) path nearly touches the southern boundary of totality. The VTX-Raiganj (2207 km) path entered the totality belt while the VTX-Kathmandu propagation path (2296 km) actually crossed the path of totality.

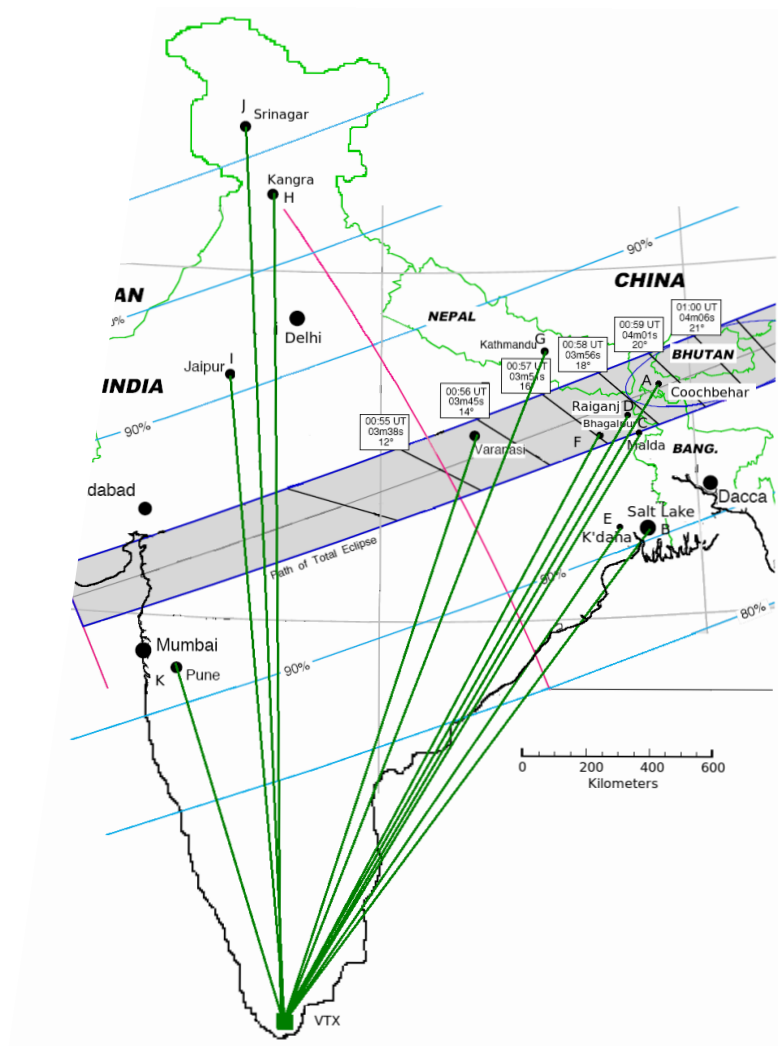

Figure 4.2: The path of totality over the Indian subcontinent during the total solar eclipse on 22nd July, 2009 is shown by shaded region. Our receivers and the VTX station are marked with names of the places and an English alphabet as an identifier. This figure is modified from that provided by F. Espenak & J. Anderson in NASA 2009 Eclipse Bulletin to include our campaign locations [Chakrabarti et al., 2010a].

Table 1 shows the parameters of and at the receiving stations which concern our study. We show the locations with geographic latitude and longitude, bearing angle, distance between the transmitter and the receiver, the percentage of eclipse coverage, the time of the first contact, the time at which the maximum eclipse took place, the last contact and the elevation (E) of the Sun (with respect to the local horizon) in degrees at these times.

Table 4.1: Solar eclipse parameters at different receiving stations in our campaign.

| Symbol | Place & Geog. Lat, Long | Bearing & Dist. (km) | Coverage % | Sunrise | 1st (IST) A(°) | Mid (IST) A(°) | 2nd (IST) A(°) |
|---|---|---|---|---|---|---|---|
| a | Coochbehar 26°19′N, 89°28′E | 30°22′ 2346 | Total 100 | 04:53 | 05:30:16 7.1 | 06:28:43 19.6 | 07:34:06 34 |
| b | Salt Lake 22°34′N, 88°24′E | 34°37′ 1946 | Partial 89.9 | 05:04 | 05:28:46 05 | 06:26:20 17 | 07:30:54 32 |
| c | Malda 25°N , 88°09′E | 29°35′ 2151 | Partial 99.6 | 05:00 | 05:29:34 05 | 06:27:24 18 | 07:32:04 32 |
| d | Raiganj 25°36′N, 88°08′E | 28°36′ 2207 | Total 100 | 04:59 | 05:29:52.3 5.6 | 06:27:43.1 18.1 | 07:32:22 32.4 |
| e | Khukurdaha 22°27′N , 87°45′E | 33°13′ 1894 | Partial 90.7 | 05:07 | 05:28:43.8 04 | 06:26:3.0 17 | 07:30:16.9 31 |
| f | Bhagalpur 25°15′N , 87°01′E | 26°31′ 2116 | Total 100 | 05:04 | 05:29:41 05 | 06:27:02 17 | 07:31:04 31 |
| g | Kathmandu 27°45′N , 88°23′E | 19°20′ 2296 | Partial 96.3 | 05:21 | 05:31:10 04 | 06:27:43 16 | 07:30:30 30 |
| - | Varanasi 25°22′N , 83°E | 15°46′ 1948 | Total 100 | 05:20 | 05:30:04 01 | 06:25:44 13 | 07:27:33 27 |
| h | Kangra 32°58′N , 76°16′E | 357°40′ 2624 | Partial 65.3 | 05:32 | 05:37:19.7 00 | 06:28:45.3 10.1 | 07:24:46.1 21.6 |
| i | Jaipur 26°55′N , 75°52′E | 354°43′ 2070 | Partial 87.6 | 05:46 | 05:32:53 -03 | 06:25:32 07 | 07:23:18 20 |
| j | Srinagar 34°08′N , 74°51′E | 354°28′ 2878 | Partial 59.4 | 05:35 | 05:40:12 00 | 06:30:04 10 | 07:24:04 21 |
| k | Pune 18°31′N , 75°55′E | 339°52′ 1183 | Partial 93.2 | 06:08 | 05:30:00 -09 | 06:21:54 02 | 07:19:03 15 |

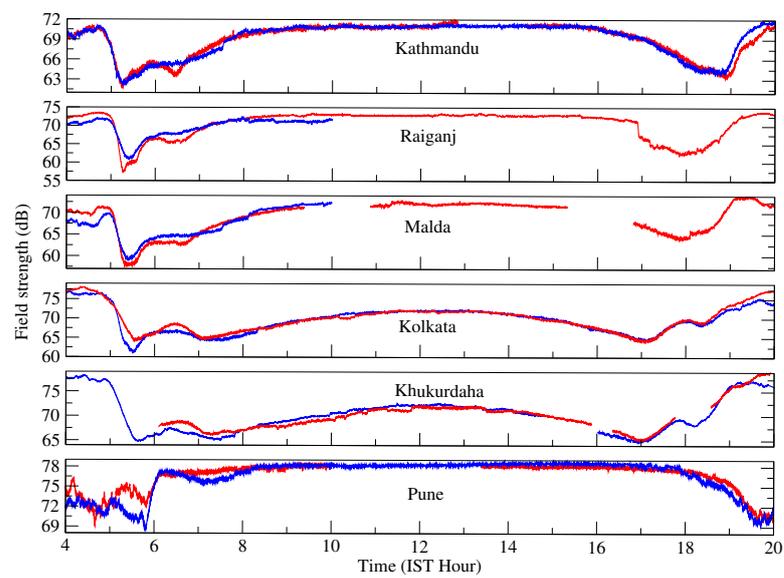

Figure 4.3: The data from 4 h to 20 h (IST) at (a) Kathmandu (b) Malda, (c) Khukurdaha, (d) Raiganj, (e) Salt Lake and (f) Pune. The blue curves represent the normal diurnal behavior (average of the day before and after the eclipse) at different places and the red curves represent the diurnal behavior on solar eclipse day at those places [Chakrabarti et al., 2012a].

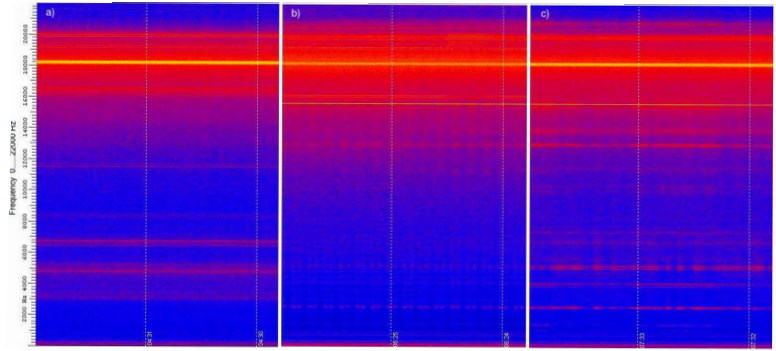

Figure 4.4: Three screen-shots at (a) pre-eclipse, (b) maximum eclipse, (c) post-eclipse of the broadband data from the Raiganj site, which experienced the total solar eclipse. The shots are at pre-eclipse, maximum eclipse and post-eclipse conditions respectively. The VTX signal (yellow color) clearly became weaker during the maximum phase of the eclipse. The monitored signal is much higher than the blue noise floor [Chakrabarti et al., 2012a].

In Figure 4.3, we present the data at six stations from 4 h to 20 h (IST). These stations are: (a) Kathmandu, (b) Malda, (c) Khukurdaha, (d) Raiganj, (e) Salt Lake and (f) Pune. The blue curves represent the normal diurnal behavior (average of the day before and after the eclipse) at different places and the red curves represent the diurnal behavior on solar eclipse day at those places. Since the night time signal is highly variable, we concentrated on the chunk of the signal around the day time. The day time VLF signal amplitude is found to be otherwise highly repeatable. The gap in the data are due to power failure. Note that there are no ionospheric disturbances in the signal on the eclipse day due to geomagnetic activities or solar flares. Indeed, according to NOAA/NGDC data, the Kp index for 22nd July, 2009 was 3 between 5:30 IST (0 UT) and 8:30 IST (3 UT). The Dst index was 5, 4 and −5 for 6:30 IST (1 UT), 7:30 IST (2 UT) and 8:30 IST (3 UT) respectively. So the eclipse period was geomagnetically quiet and the observed deviation of the signal amplitude can be safely assumed to be due to the effects of eclipse alone.

In Figure 4.4, we present three screen-shots of the broadband data from the Raiganj site, which experienced the total solar eclipse. The shots are at pre-eclipse, maximum eclipse and post-eclipse conditions respectively. The yellow color is the VTX signal which clearly became weaker during the maximum phase of the eclipse. The monitored signal is much higher than the natural noise floor, so there is no scope for confusion.

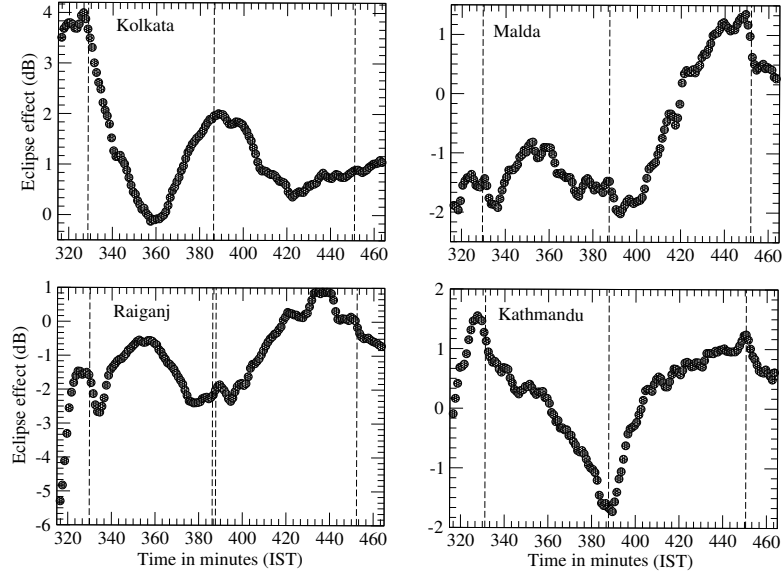

Figure 4.5: The variation of the differential amplitude of the VTX signal at four receiving stations during the eclipse time. These variations of amplitude are obtained by subtracting an average unperturbed data from the eclipsed data on 22nd July, 2009. The time of first contact, eclipse maximum (or, second and third contacts, when available), and the last contact are indicated by vertical dashed lines at each receiver [Pal et al., 2012a].

In Figure 4.5, we show the differential amplitude variations of the VTX signal (18.2 kHz) for the four propagation paths (VTX-Kolkata, VTX-Malda, VTX-Raiganj and VTX-Kathmandu) during the eclipse time only. Along the X-axis, the time given is in Indian Standard time (IST=UT+5:30). These variations of amplitude are obtained by subtracting an unperturbed data (average of the day before and after the eclipse) from the eclipsed data on the 22nd July, 2009. In the top left panel, we show the result of VTX-Kolkata path. The eclipse was partial and the maximum obscuration of the solar disk was ∼ 90% (Figure 4.6). Here, the change in amplitude is positive. The signal is enhanced by +2 dB close to the maximum of the eclipse. The top right panel, the bottom-left and bottom-right panels are for VTX-Kathmandu, VTX-Raiganj and VTX-Malda respectively. The signal amplitude was reduced by about 2 dB in all these places.

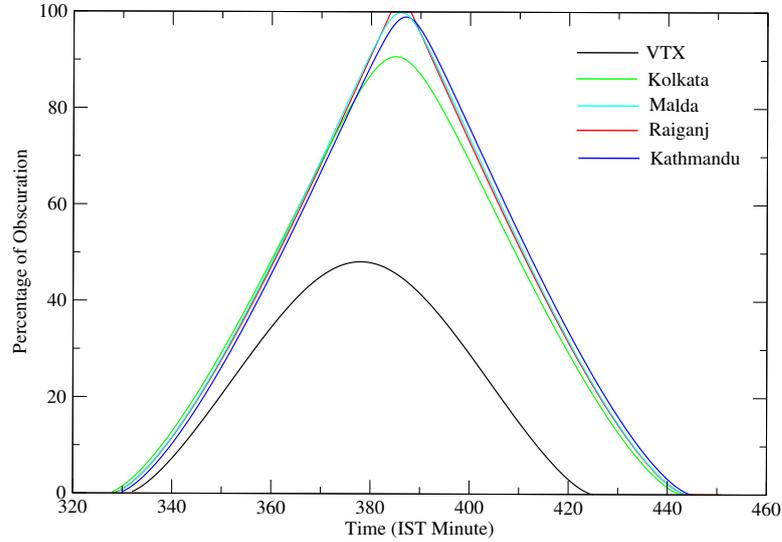

Figure 4.6: Variation of percentage of solar obscuration (S) as a function of time at different places during the TSE 2009.

## 4.4   Modeling VLF Signal Amplitudes Using Wave-hop Code

As far as the VTX is concerned all the propagation paths within India are less than 3000 km and thus the wave-hop model should be able to explain the variation of amplitude with time, provided some assumptions are made about the ionospheric reflection height variation with the degree of obscuration of the Sun. There are many works in the literature which use the wave-hop theory (Wait, 1962; Budden, 1966; Wakai, 2004). Following these, we developed our own code (discussed details in Chapter 2) (Pal & Chakrabarti, 2010) to describe the wave amplitude and phase in the context of VTX transmitter. In our code, the received signal is calculated from the sum of the ground wave propagating along the Earth and the sky-waves reflected by the ionosphere. The phase variations in the sky component can occur if the ionospheric conditions change with time as is the case in the present case. As a result, at a given receiving station, the resultant signal may vary with time due to the occurrences of constructive and destructive interferences between the ground and sky waves. The strength of the ground wave along the propagation path is calculated over sea-water using the permittivity of 70 F/m and the conductivity of 5 S/m. For waves propagating over grounds, the permittivity of 15 F/m and conductivity of 0.002 S/m were chosen.

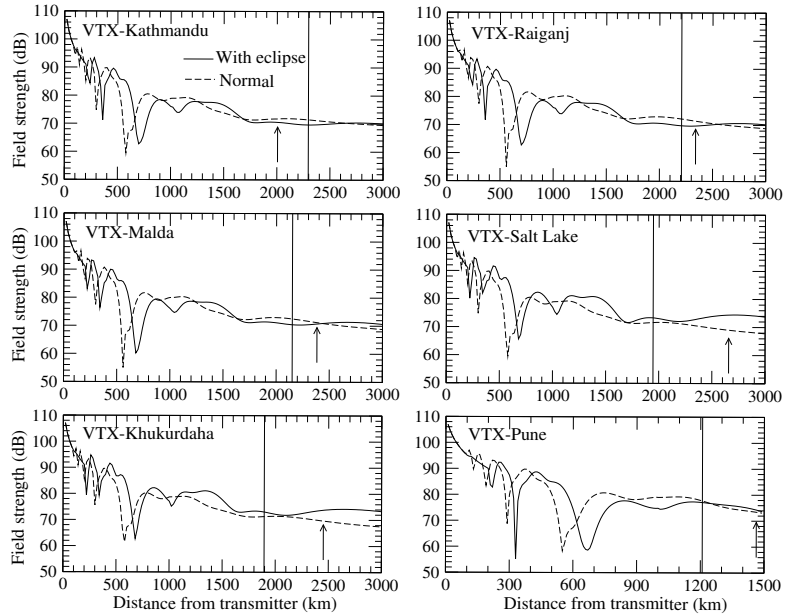

Figure 4.7: Amplitude variations along the propagation paths for the six different places at mid-eclipse condition using the wave-hop code. The dashed curve represents the VTX signal amplitude variation at normal day conditions and the solid curve represents the same variation during the maximum eclipse condition. The vertical line is placed at the distance of the receiver from the transmitter while the vertical arrow represents the location where maximum eclipse occurs. Observed and simulated deviations from the normal agree very well.

For modeling purpose, we assume that the effect of eclipse is to increase the reflection height. We assume that this height is increased by about 10 km at the center-line of the path of totality (Clilverd et al., 2001) and it falls off on both sides following a Gaussian shape along the propagation path. The sharpness of the ionosphere characterized by the reflection coefficient is kept as a constant along the propagation path for convenience (Fleury et al., 2000). It is observed that the resultant signal variation with time shows generally correct behavior if the reflection coefficient is chosen to be $R_c \sim 0.4$ (instead of $\sim 0.3$ for normal days), however a better match requires fine tuning of $R_c$. Thus, for instance, the propagation paths along Kathmandu, Raiganj, Malda, Salt Lake, Pune and Khukurdaha requires $R_c$ = 0.375, 0.376, 0.385, 0.50, 0.31 and 0.48 respectively. In Figure 4.7, we show the variation of the amplitude of the resulting signal from the transmitter along the respective bearing (See, Table 1) of six different receiving stations at maximum
eclipse condition (around 6:30 AM IST = 1:00 UT) on the 22nd of July, 2009 using
our wave-hop code. The dashed curve represents the signal amplitude variation
from the VTX transmitter on the 22nd of July assuming the eclipse is absent. The solid
curve shows the variation when the total eclipse is assumed. The expected deviation
(in dB) due to the eclipse at various distances from the transmitter is the difference
between the solid curve and the dashed curve. The vertical line corresponds to
the location of the receiver, while the vertical arrow points to the location of the
center-line of totality along the respective propagation path. We clearly see that in
some places, the signals are enhanced and in some others, the signals are reduced.
This agrees with what we observed in Figure 4.3. This leads us to believe that we
have been able to capture the basic physics behind the signal properties during the
eclipse.

The observed positive and negative amplitude enhancement due to solar eclipse
depends on the position of the receiver along the propagation path. So it can not
be simply explained by saying that those paths crossing the solar eclipse totality
belt show amplitude reduction and the rest show positive amplitude enhancement
due to eclipse. To show this, we present the Figure 4.8. The first panel shows the
amplitude variation at 6:30 AM (IST; 1:00 UT) on the 22nd of July, 2009 using our
wave-hop code. The dashed curve represents the signal amplitude variation from the
VTX transmitter on a normal day along a direction with a bearing of 30° when there
was no eclipse. The solid curve shows the same variation when the total eclipse is
assumed. Six receiving places (which are roughly within a bearing of 10° interval) in
this general direction are marked in the figure by solid vertical lines. In the second
panel, we show the difference of the two curves. This theoretical difference $\Delta A$
is the expected deviation (in dB) due to the eclipse at various distances from the
transmitter. We see that in some places the signals are enhanced ($\Delta A > 0$) and
in some others the signals are reduced ($\Delta A < 0$). In the inset, we zoom the region
surrounding these six stations to show that the signal deviations as predicted by the
theory roughly agree with our observations (i.e., $\Delta A_{predicted} \sim \Delta A_{observed}$).

## 4.5   Modeling Using LWPC Code

We also use the LWPC code to calculate the amplitudes at receiving sites corre-
sponding to normal and partially eclipsed-ionosphere conditions. LWPC code is a
very versatile code and it uses the waveguide mode theory. One has to provide
the suitable boundary conditions for the lower and upper waveguides. The lower waveguide parameters, i.e., permittivity ($\epsilon$) and conductivity ($\sigma$) of the earth have been automatically selected by the code itself. These parameters are constants corresponding to normal and eclipse conditions. The upper waveguide, i.e., ionospheric parameters are specified by the electron density $N_e(h)$ and the electron-neutral collision frequency $\nu_e(h)$ profiles. We use Wait's exponential ionosphere (Wait & Spies, 1964) for the electron density profile with gradient parameter $\beta$ and reference height $h'$ as given by the equation,

$$N_e(h, h', \beta) = 1.43 \times 10^7 exp(0.15 h') exp[(\beta - 0.15)(h - h')],$$

in $cm^3$. The electron-neutral collision frequency ($s^{-1}$) is given by the equation,

$$\nu_e(h) = 1.816 \times 10^{11} exp(-0.15h).$$

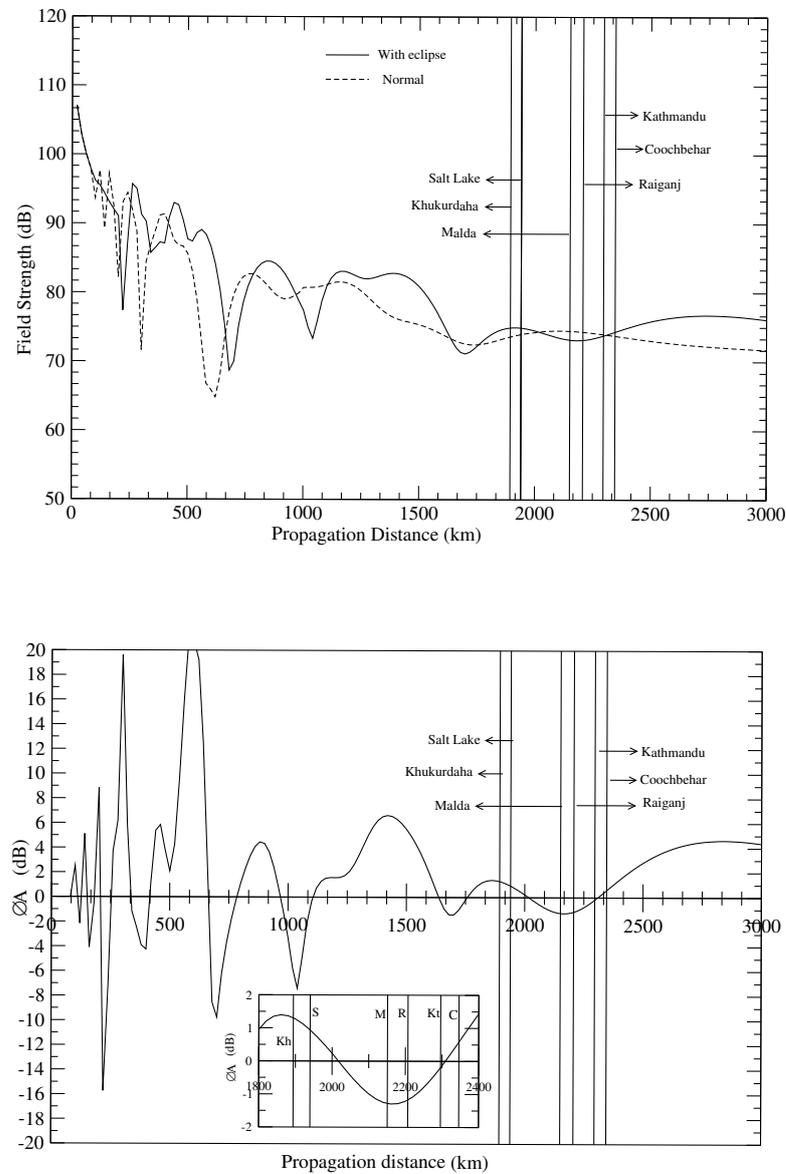

Figure 4.8: Amplitude variation at 6:30 AM using the wave-hop code, dashed curve being the signal amplitude variation from VTX transmitter at normal conditions with bearing at 30 degree. The solid curve shows the same variation during the maximum eclipse condition. The receiving stations roughly along the same bearing angle are marked by solid straight lines. (b) Differential amplitude ($\Delta A$) variation between the eclipse and normal day with the wave-hop code. Within 1750-2000 km of VTX, the fractional change of amplitude is positive, within 2000-2300 km fractional change is negative and beyond 2300 km fractional change is again positive. Six receivers were within this range (1750-2400) and exhibited the same features at the time of the maximum eclipse.

These profiles are quite standard and generally agree with directly observed normal D-region ionospheric profiles (Sechrist, 1974; Cummer et al., 1998).

The optimized set of $\beta$ and $h'$ parameters to describe the normal ionospheric condition at 6:30 AM for each propagation path are chosen by comparing between the electron density from Waits model and IRI-2007 model (Bilitzaa & Reinisch, 2008). The IRI electron density profiles are calculated at the mid-points of the propagation paths. As an example, we show in Figure 4.9, the electron density profile at 6:30 AM from IRI-2007 model compared with the electron density profile for different set of $\beta$ and $h'$ parameters for VTX-Salt Lake path. The parameter set $\beta = 0.35 \pm 0.01$ and $h' = 75.5 \pm 1.0$ km are chosen for this path at 6:30 AM since this set of parameters produce amplitude very close to the observed value of 65.45 dB with respect to 1 $\mu$V m$^{-1}$. The parameters are quiet reasonable at this time since the stable D-region is not completely formed till that time. Similarly, we choose the values $\beta = 0.3$ km$^{-1}$ and $h' = 74.0$ km as appropriate for the VTX-Kathmandu, VTX-Raiganj and VTX-Malda propagation paths as unperturbed ionosphere [Pal et al., 2012a]. Once the unperturbed morning (6:30 AM) parameters ($\beta$ and $h'$) of the ionosphere are chosen for a given path, we calculate the perturbed parameters for that path to know the increase of the height and sharpness parameters of the lower ionosphere.

To model the partially eclipsed D-region ionosphere we need to calculate the degree of solar obscuration as a function of time along each path. We developed a numerical code to obtain the solar obscuration function $S(t)$ at a particular point along a path using the formalism of Mollmann & Vollmer (2006). The input parameters (i.e., time of starting and ending of eclipse, magnitude of eclipse) for the model are taken from the website of solar eclipse calculator

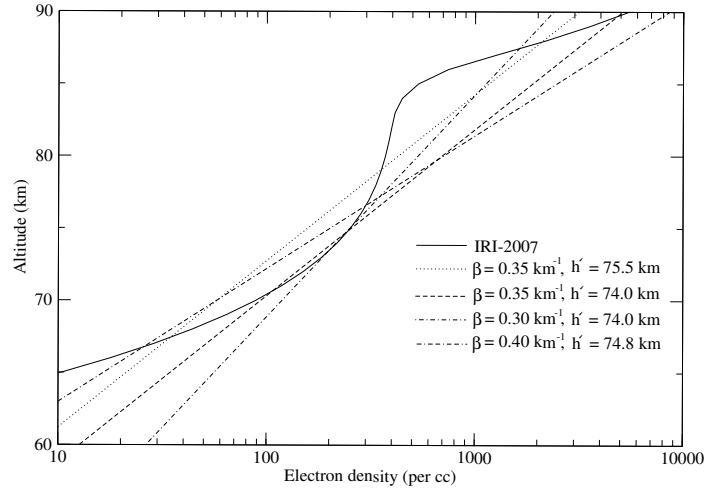

Figure 4.9: The electron density profile at 6:30 AM from IRI-2007 model compared with the electron density profile for different sets of $\beta$ and $h'$ parameters for VTX-Salt Lake path. Parameters for different curves are marked.

($http://www.chris.obyrne.com/Eclipses/calculator.html$) for the solar eclipse of July, 2009. The variation of solar obscuration at some receiving places has been shown in Figure 4.6. Then we divided each propagation path into several ($\sim 15$) segments along the transmitter to receiver great circle path and calculated the degree of solar obscuration as a function of time.

As a first order approximation, we assumed that the ionospheric parameters $\beta$ and $h'$ vary linearly with obscuration function from the unperturbed one, though the ionospheric response to solar eclipse are generally non-linear phenomena (Lynn, 1981; Patel et al., 1986). Thus the parameters vary with time and distance from the transmitter. Therefore, if the maximum deviations for $\beta$ and $h'$ are given at any time, the perturbed parameters will be obtained by summing the unperturbed values of the the parameters with the product of the maximum deviation and solar obscuration function. This assumption automatically brought in the non-uniformity of the ionospheric parameters along the propagation paths.

## 4.6   Results

In Figure 4.10a, we show the amplitude variation (solid curve) obtained by the model fit and compare it with the corresponding observed changes (circle) of VTX signal as received at Kathmandu. The maximum eclipse effect occurs when the effective reflection height $h'$ is increased by +4.0 km and $\beta$ is increased by +0.04 $km^{-1}$ at $t = 383.0$ minute (IST) while the observed maximum eclipse effect occurs at around $t = 389.5$ minute. The variation of the model amplitude with distance from the transmitter (in km) for unperturbed (no eclipse) condition (solid line), maximum eclipse condition (dotted curve) and at $t = 420.0$ minute (dashed curve) are shown in Figure 4.10b. The location of the receiver is indicated by a vertical line.

Figures 4.11(a-b) show similar plots for the VTX-Raiganj propagation path. Here also the maximum eclipse effects occur when the VLF reflection height $h'$ has increased by +4.0 km and $\beta$ has increased by +0.04 $km^{-1}$ at $t = 383.0$ minutes and the observed maximum eclipse effect occurs at around $t = 377.5$ minutes.

In Figures 4.12(a-b), we show similar plots for the VTX-Kolkata propagation path. Here, maximum VLF reflection height $h'$ has increased by +1.8 km and the ionospheric sharpness factor $\beta$ has increased by +0.02 $km^{-1}$ at $t = 390.0$ minute at Kolkata. Figures 4.13(a-b) correspond to the VTX-Malda propagation path where $h'$ has increased by +3.0 km and $\beta$ has increased by +0.02 $km^{-1}$ due to the maximum eclipse at $t = 385.0$ minute at Malda.

In Figure 4.14, we show the variation of the amplitude of the resulting signal from the transmitter along the respective bearing (see Table 1) of six different receiving stations at maximum eclipse condition (around 6:30 AM IST=1:00 UT) on July 22nd, 2009. The black curves represent the signal amplitude variation from the VTX transmitter on the 22nd of July assuming the eclipse is absent. The red curves show the variation when a total eclipse is assumed. We note that the signal amplitude shows several ups and downs on the way. These could be interpreted to be due to occurrences of the constructive and destructive interferences among the sky-wave components and the ground wave component. The expected deviation (in dB) due to the eclipse at various distances from the transmitter is the difference between the red curve and the black curve. This agrees with what we observed in Figure 4.3. The maximum increase in the $\beta$ and $h'$ parameters, denoted as, $\Delta\beta_{max}$ (km) and $\Delta h'_{max}$ ($km^1$) of the lower boundary of the D-region due to the solar eclipse at different propagation paths are given in Table 2.

These results for the VTX signal are quite reasonable. In particular, we find that the path which crosses the whole eclipse belt (i.e., that of VTX-Kathmandu) required the highest shift in height, while the path which is farthest from totality (VTX-Kolkata) required the least height variation of the lower ionosphere.

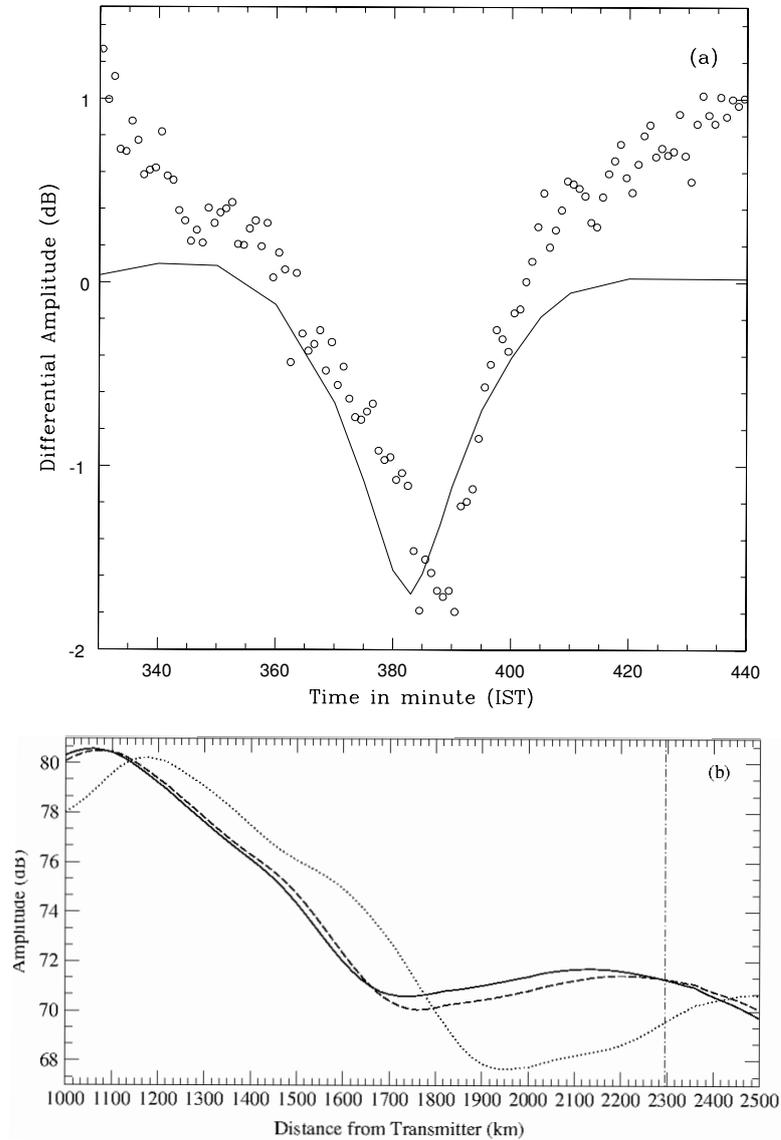

Figure 4.10: a) The model amplitude variation (solid curve) is compared with corresponding observed changes (circle) at Kathmandu. Maximum eclipse effects occur when $h'$ has increased by $+4.0$ km and $\beta$ has increased by $+0.04$ km$^{-1}$ at $t = 383$ minute (IST). b) The variation of the model amplitude with distance from the transmitter for unperturbed conditions (solid line), maximum eclipse conditions (dotted curve) and at $t = 420$ minute (dashed curve). The location of the receiver is indicated by a vertical line [Pal et al., 2012a].

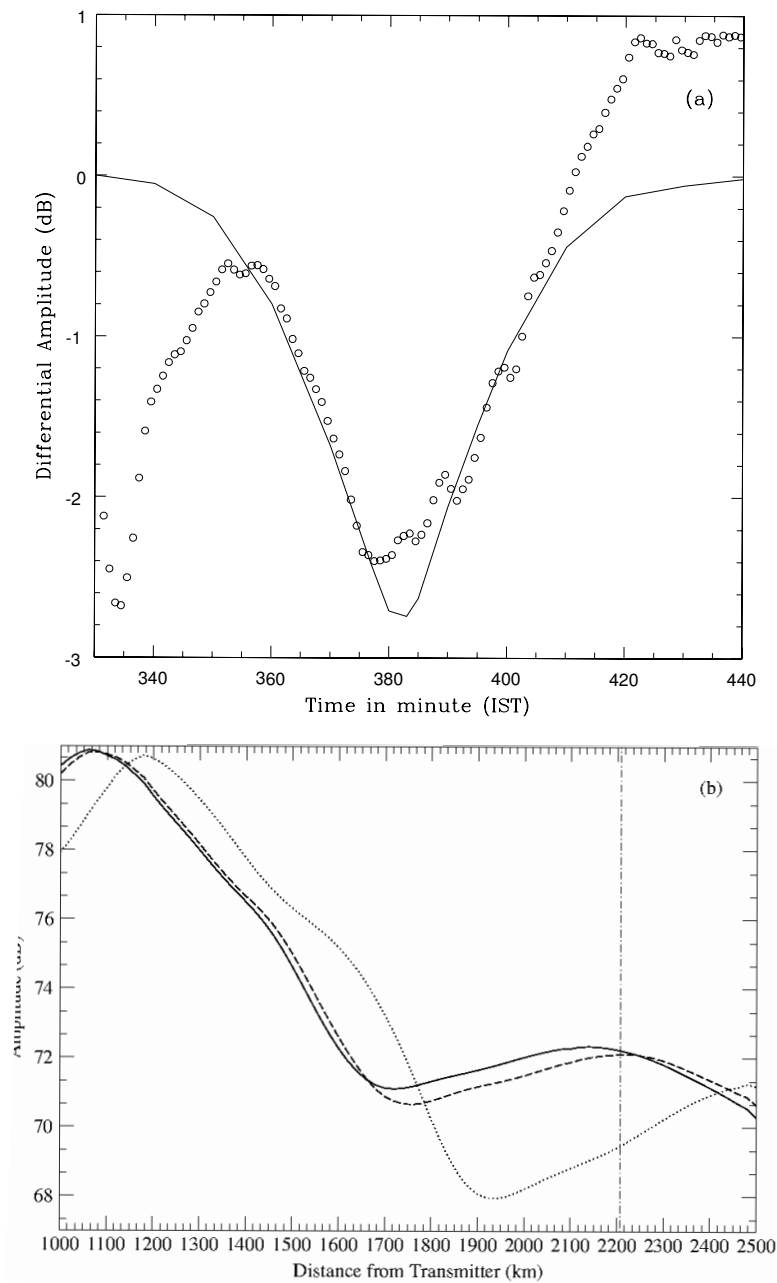

Figure 4.11: A plot similar to Figure 4.10 for the VTX-Raiganj propagation path. Here the reflection height parameter $h'$ has increased by +4.0 km and sharpness factor $\beta$ has increased by +0.04 km$^{-1}$ at $t = 383$ (IST) minute at the region with the totality of eclipse.

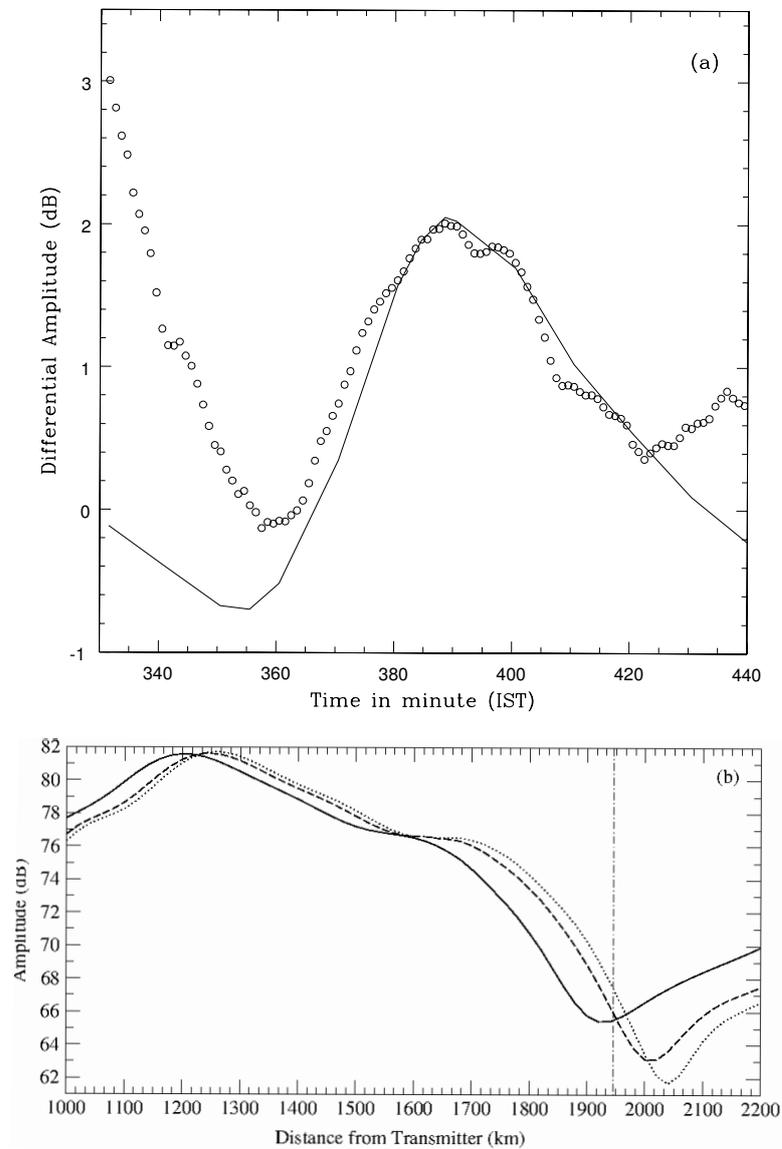

Figure 4.12: a) A plot similar to Figure 4.10 for the VTX-Kolkata propagation path. Maximum eclipse effects occur when $h'$ has increased by +1.8 km and $\beta$ has increased by +0.02 km$^{-1}$ at $t = 390$ (IST) minute at Kolkata. b) The variation of the model amplitude with distance from the transmitter for unperturbed conditions (solid line), maximum eclipse conditions (dotted curve) and at $t = 420$ minute (dashed curve) (7:00 AM IST). The location of the receiver is indicated by a vertical line.

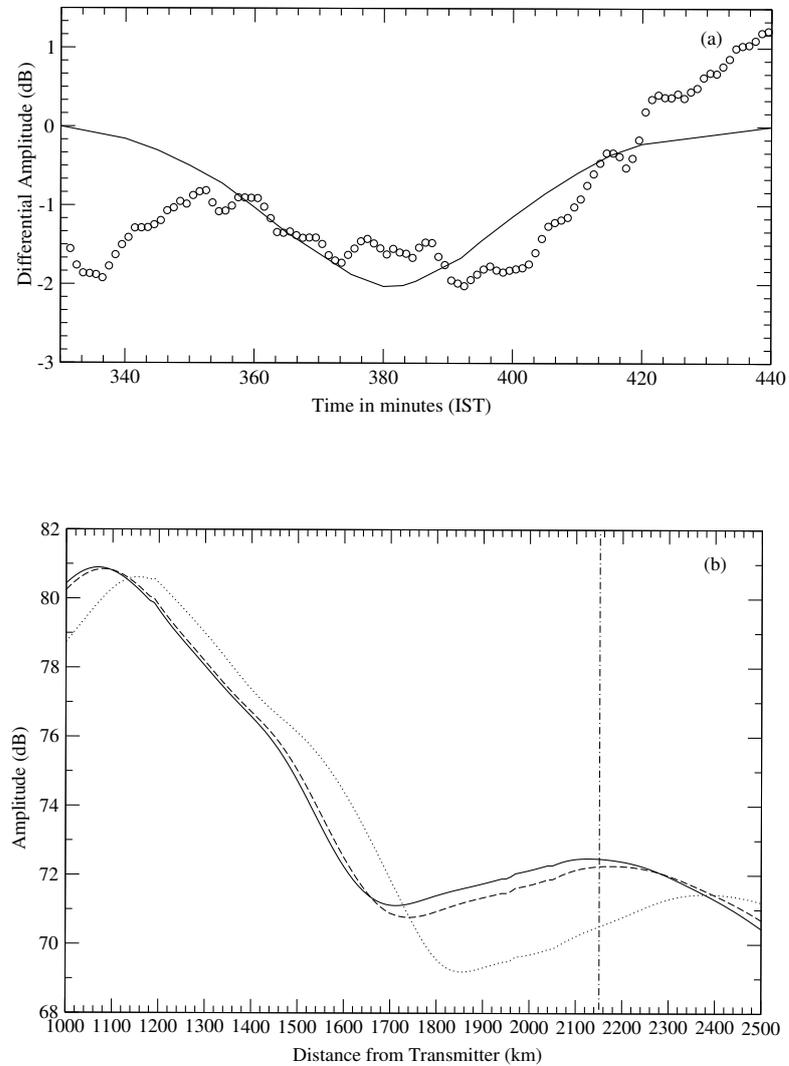

Figure 4.13: A plot similar to Figure 4.10 for VTX-Malda propagation path. Here the maximum eclipse effects occur when $h'$ has increased by +3.0 km and $\beta$ has increased by +0.02 km$^{-1}$ at $t = 385.0$ minute at Malda [Pal et al., 2012a].

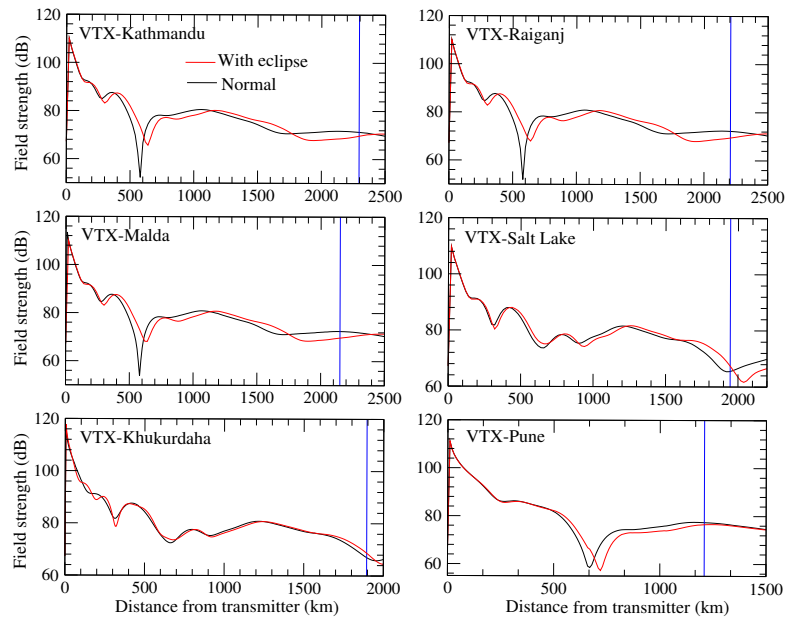

Figure 4.14: Amplitude variation along the propagation paths for the six different places at mid-eclipse condition using the LWPC code. The dashed curve represents the VTX signal amplitude variation at normal day conditions and the solid curve represents the same variation during the maximum eclipse condition. The vertical line is placed at the distance of the receiver from the transmitter. Observed and simulated deviations from the normal behavior agree very well [Chakrabarti et al., 2012a].

Table 4.2:  Maximum increase of $\beta$ and $h'$ parameters due to the solar eclipse

| Propagation path | $\Delta h'_{max}(km)$ | $\Delta\beta_{max}$ $(km^{-1})$ |
|---|---|---|
| VTX - Kathmandu | 4.0 | 0.04 |
| VTX - Raiganj | 4.0 | 0.04 |
| VTX - Malda | 3.0 | 0.02 |
| VTX - Kolkata | 1.8 | 0.02 |
| VTX - Khukurdaha | 1.5 | 0.02 |
| VTX - Pune | 1.9 | 0.02 |

During totality, a near-nighttime condition is expected to prevail in the lower ionosphere and thus the number density of electrons should be reduced as the D-region nearly disappears. To show this, we consider the data at Raiganj which witnessed totality (see, Table 1). In Figure 4.15, we show how the distribution of the electron density (cm$^{-3}$) with altitude changes with time during the eclipse at Raiganj. The electron density at Raiganj is calculated using the so-called Wait's formula using the calculated variation of $\beta$ and $h'$ parameters with time. We note that in lower heights the near neutrality is achieved during the mid eclipse. At a height of $\sim 90$ km, of course, there are still some electrons. Figure 4.16 represents a snapshot of the variation of VLF reflection height ($h'$ parameter) in km along the VTX-Raiganj propagation path at the time of maximum eclipse for which the Figure 4.15 has been drawn. Figure 4.17 shows the electron density variation similar to Figure 4.15. However, it is shown for Kolkata during the solar eclipse period.

## 4.7   Discussions

In this Chapter, we have modeled the effects of the solar eclipse of July 22nd, 2009 on the VLF signals. Our waveguide model is based on the assumption of linear variation of ionospheric parameters with the degree of solar obscuration. There appears to be still some discrepancies between our result and that of the observation, especially close to the sunrise region. This means that our crucial assumption that the variation of ionospheric parameters with the solar obscuration is linear at all points of the propagation path may not be quite accurate. Furthermore, the data close to the sunrise time is contaminated by increasing attenuation due to lowering of the lower ionosphere. A complete agreement could perhaps be reached when the sunrise effects are also included.

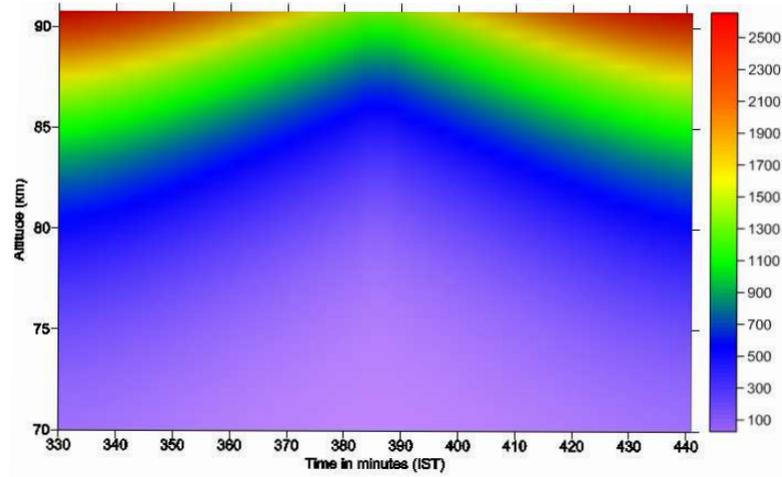

Figure 4.15: Altitude variation of electron number density (cm$^{-3}$) obtained from Wait's formula during eclipse time at Raiganj. Note that at lower height, the number density is ∼0 due to the near night condition when the D-layer completely disappeared. The ionization of the D-region at lower altitudes recover at a faster rate than the ionization at higher altitudes [Pal et al., 2012a].

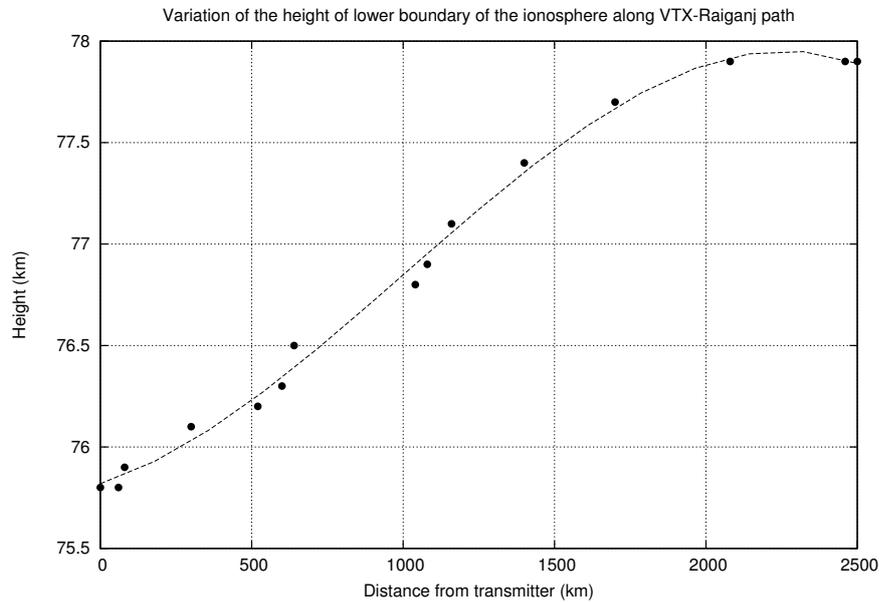

Figure 4.16: Snapshot of the variation of the effective VLF reflection height ($h'$) in km along the VTX-Raiganj propagation path at the time of maximum eclipse at Raiganj.

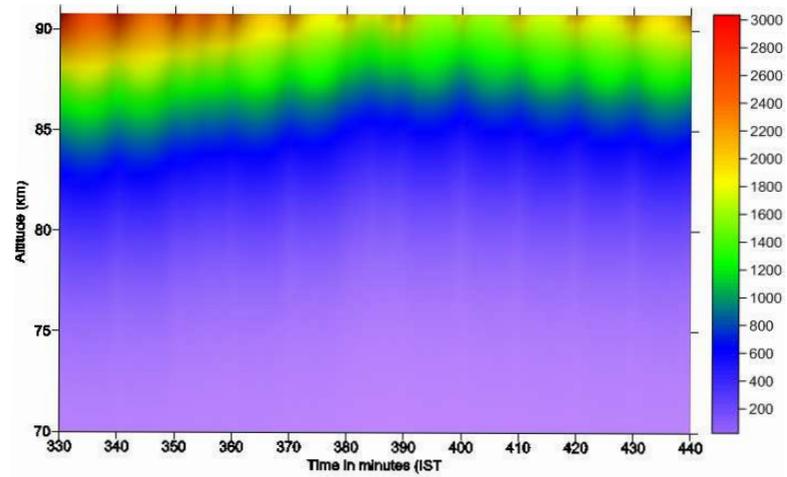

Figure 4.17: Altitude variation of electron number density (cm$^{-3}$) similar to Figure 4.15 but using the Kolkata data. The apparent periodic variation is due to an artifact of the plotting routine.

## Chapter 5

# Ionospheric Effects of an Occulted Solar Flare During a Solar Eclipse

## 5.1 Introduction

On January 15th, 2010, there was an annular solar eclipse as seen from the southern regions of India. At Khukurdaha (22°27′N, 87°45′E) station, the eclipse was partial with a maximum obscuration of about 75%. From this station, we have been monitoring NWC transmitter operating at 19.8 kHz. During the solar eclipse period, a solar flare also occurred which was partly blocked by the lunar disk. Thus the resultant signal was perturbed both by the eclipse and by the flare.

In this Chapter, we describe our results in detail and compare the VLF signals with the light curve of hard and soft X-rays obtained by the GOES-14 and RHESSI satellites. We also present the magnetogram around the flare obtained by the GONG project and the image of the flare from the HINODE X-Ray Telescope (XRT) data. From the relative positions of the Sun, the Moon and the receiving station, we derive the time variation of the lunar occultation of the flare. From these, we obtain the time variation of the electron number density in the lower ionosphere over the entire event.

## 5.2 The Solar Eclipse and the Solar Flare

ICSP made Gyrator-III VLF receiver (1-22 kHz) with a loop antenna was monitoring NWC station from Khukurdaha. The annular solar eclipse of January 15, 2010 started at 12:05 IST (=UT+5:30) and continued up to 15:28 IST. The eclipse was partial (maximum coverage 75%) as seen from Khukurdaha. The maximum obscuration at the receiver occurred at 13:56:26 IST. The distance between transmitter and receiver is about 5700 km (Figure 5.1). Thus the transmitter was far away from the annular solar eclipse belt. During propagation of signal from NWC to Khukurdaha, there was no eclipse in 24.5% of the total path while 75.4% of the total path experienced some eclipse. The whole propagation path experienced an average value of solar obscuration at the most 40% calculated using a numerical integration over the whole path. Figure 5.2 shows the degree of solar obscuration during eclipse at the receiver and over the whole path at 70 km above the ground. The dotted curve represents the average value of solar obscuration integrated over the whole path.

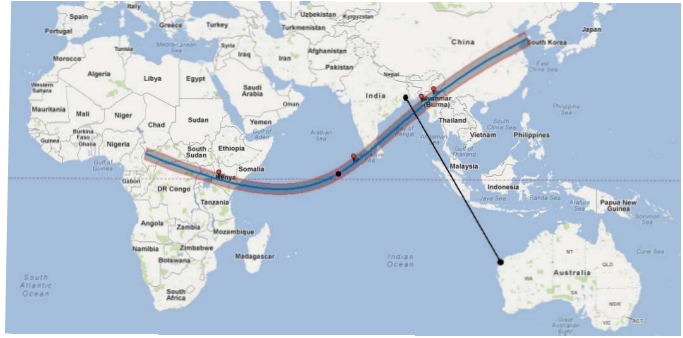

Figure 5.1: Great circle path between the NWC (19.8 kHz) transmitter and the receiver at IERC, Sitapur. The shaded area is the path of total annularity of January 15, 2010 [Pal et al., 2012b].

While the eclipse is in process, a C1.3 solar flare started at 07:22 UT (12:52 IST) and continued till 10:22 UT (15:52 IST) with an extended maximum in soft X-ray from 08:41 UT (14:11 IST) to 08:44 UT (14:14 IST). The hard X-ray peaked at 08:36 UT (14:06 IST). Thus the hard and the soft X-Ray peak within the eclipse period. The soft X-ray peak is about 15 minutes after the maximum obscuration of the solar disk. The peak energy flux in $1.5 - 12.5$ keV was $1.3 \times 10^{-6} ergs/cm^2$ which is well above the detectability limit of our VLF antenna (typically, C1.0 flare).

## 5.3   The Experimental Set-up and the Observational Results

A loop antenna and VLF receiver were used to monitor VLF data at Khukurdaha. Real time narrow band signal amplitudes from several transmitters were recorded automatically in the computer. In Figure 5.3, we give an example of deviation of the NWC signal amplitude obtained from a C4.0 flare which occurred on the 6th February, 2010, three weeks after the solar eclipse event which we report here. For comparison with this, we also present the soft X-Ray flux from the GOES-14 satellite of this flare [Figure 5.3].

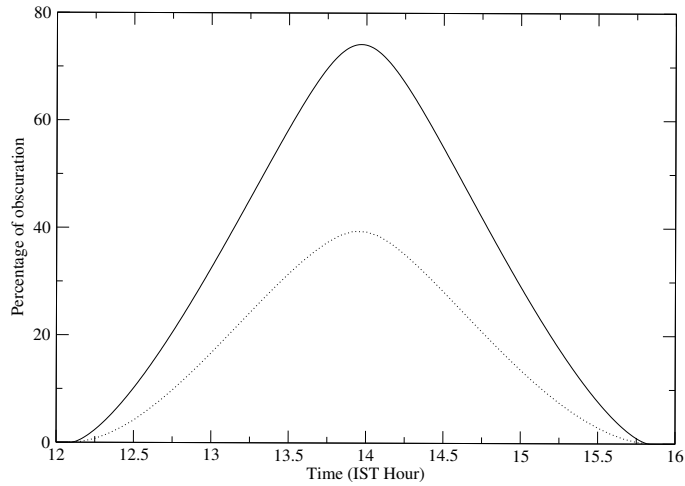

Figure 5.2: Variation of percentage of solar obscuration (S) as a function of time. Solid curve represents the solar obscuration at the receiver and the dotted curve represents the average value of solar obscuration over the whole path [Pal et al., 2012b].

In Figure 5.4, we present the amplitude (dB) variation of the NWC signal as obtained by our receiver on 15th January, 2010, the day of the annular eclipse. The box highlights the region around the eclipse which clearly shows a dip near the eclipse maximum and an upward rising kink near the flare maximum. The time is given in IST (=UT+5:30). In Figure 5.5, we plotted the net deviation of NWC signal from the normal data which is defined as the average of the data of 14th and 16th January for the solar eclipse time only. Here, the upper panel shows the VLF amplitude deviation from the normal data due to the combined effects of solar eclipse and solar flare. Note that the net deviation due to the combined effect is 1.3 dB. In the middle panel, the dashed curve shows hard X-ray flux and the solid curve shows soft X-ray flux of the flare from GOES-14 satellite. The two vertical lines correspond to the time of maximum phase of the eclipse and the time at which the peak of the solar flare occurs. The lower panel shows the RHESSI data of X-ray flux in different energy bands for the same flare. Unfortunately, the satellite was over South Atlantic Anomaly region for much of the time (SAA) and the flare was seen only during the period denoted by F.

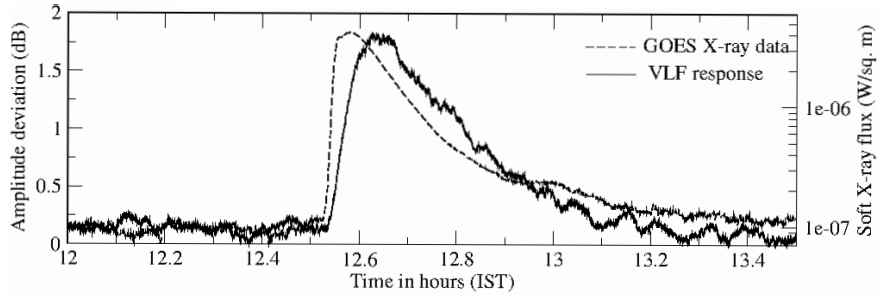

Figure 5.3: Example of a normal (non-occulted) C4.0 solar flare which occurred on February 6, 2010. Dashed curve represents the GOES soft X-ray flux and solid curve represents the response of the NWC signal amplitude to the flare [Pal et al., 2012b].

GOES satellite was observing the solar flare un-obstructed by the lunar occultation. In Figure 5.6, we superimpose the position of the lunar limb on the magnetogram of the X-ray flaring region (AR 11040) at different times. This gives us an indication that the flare was partially occulted by the moon from the VLF receiving station. In Figure 5.7, the exposed solar disks are superimposed on the HINODE-XRT image of the solar flare during the solar eclipse at every five minutes, starting from 13:50:00 IST to 14:25:00 IST (top-left to top-right; bottom-left to bottom-right). This shows that the flare was actually blocked by the moon.

In Figure 5.8, we present the RHESSI image taken in the period marked by F in Figure 5.5. Thus the image was taken just after the flare had the peak emission. However, assuming that the flare peak did not shift, we could identify the region where the peak emissions occurred. The black rectangular and circular regions show the peaks of the soft X-ray (3-12 keV) and hard X-ray (12-25 keV) emitting regions respectively. We find that the peak region of the soft X-ray was fully covered by the moon at 14:10 hrs (IST) and the hard X-ray peak region was covered from 14:20 to 14:25 hrs. It is interesting to note that the HINODE satellite also observed the occultation of the same flare but at a slightly different time and from a different perspective.

For the sake of completeness, in Figure 5.9, we present the snapshots of X-ray flares at different times (14:16:00 IST, 14:17:43 IST, 14:17:47 IST, 14:18:10 IST, 14:18:39 IST, 14:24:58 IST) as seen by HINODE.

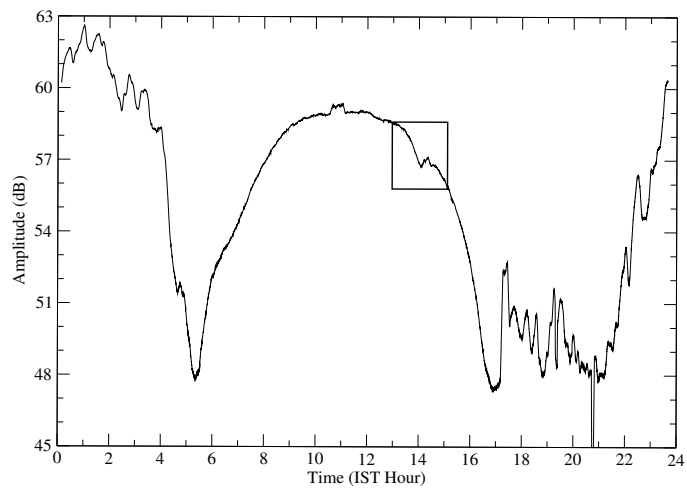

Figure 5.4: The normalized diurnal variation of NWC signal on the eclipse day. The box highlights the ionospheric disturbances observed due to the joint effects of the eclipse and the flare. The time is in IST (UT + 5:30) [Maji et al., 2012; Pal et al., 2012b].

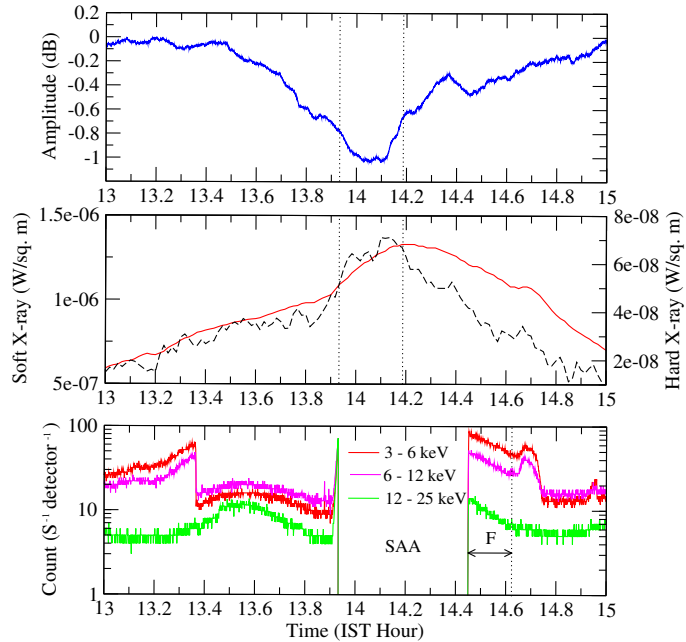

Figure 5.5: Observed VLF data with GOES-14 and RHESSI soft and hard X-ray light curve. Upper panel shows the VLF amplitude deviation due to the combined effect of the solar eclipse and the solar flare obtained by subtracting the data of January 15th from the average of the normal data of 14th and 16th January, 2010. In the middle panel, the dashed curve shows the hard X-ray flux and solid curve shows the soft X-ray flux of the flare as obtained from the GOES satellite. Lower panel shows the RHESSI data of X-ray light curve in different energy bands at the same time. The region SAA and F marks the region over South Atlantic Anomaly and the period when the flare was observed by RHESSI. The curves marked by (I), (II) and (III) are for (3-6) keV, (6-12) keV and (12-25) keV respectively [Pal et al., 2012b].

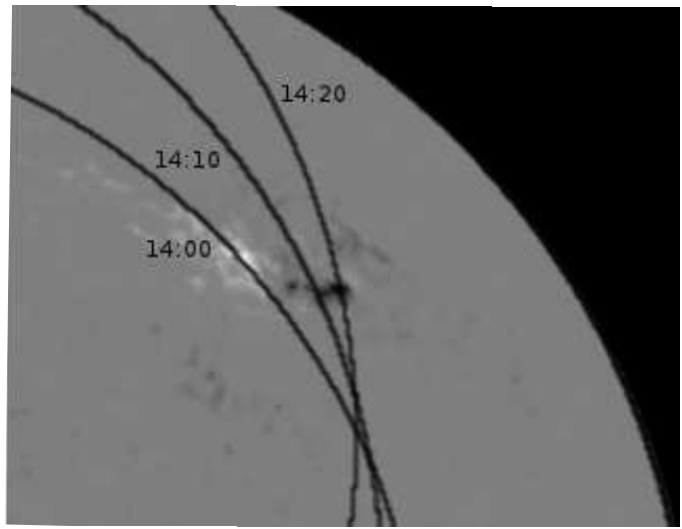

Figure 5.6: Closer view of the positions of the lunar limb superimposed on the magnetogram (taken from GONG project data at 14:34 IST) of X-ray flaring region at different times to show that it was variously occulted by the moon from a direct view [Pal et al., 2012b].

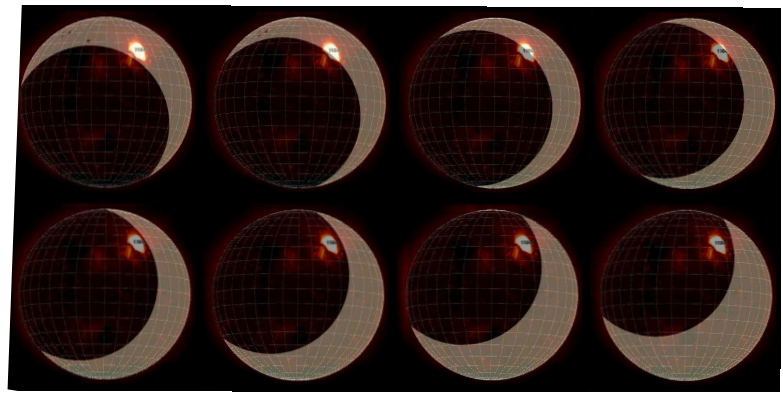

Figure 5.7: Exposed solar disks at every 5 minutes (from top-left to top-right; bottom-left to bottom-right), starting from 13:50:00 IST to 14:25:00 IST, are superimposed on the XRT image of the solar flare during the solar eclipse [Pal et al., 2012b].

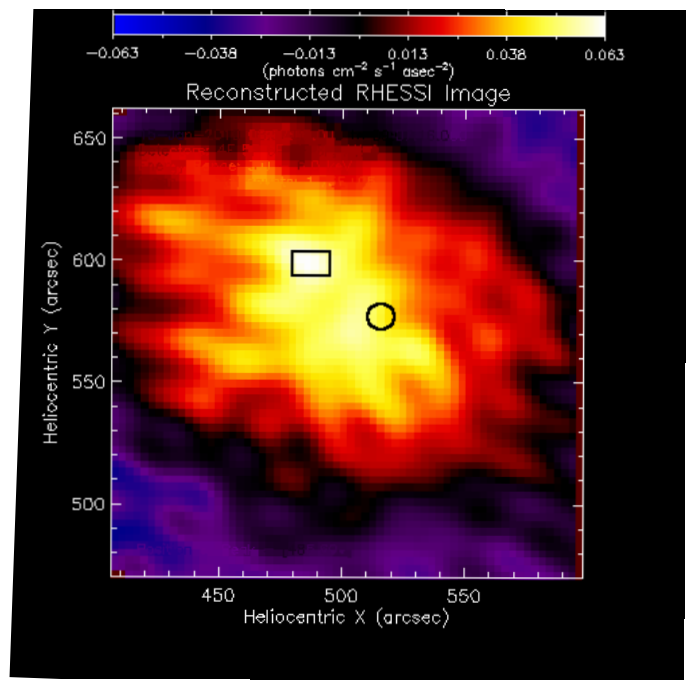

Figure 5.8: In this RHESSI image, the black rectangle and circle show the peaks of soft X-ray (3-12 keV) and hard X-ray (12-25 keV) emitting regions. At 14:10 hrs (IST), the soft X-ray peak and between 14:20 to 14:25 hrs hard X-ray peak were covered by the moon.

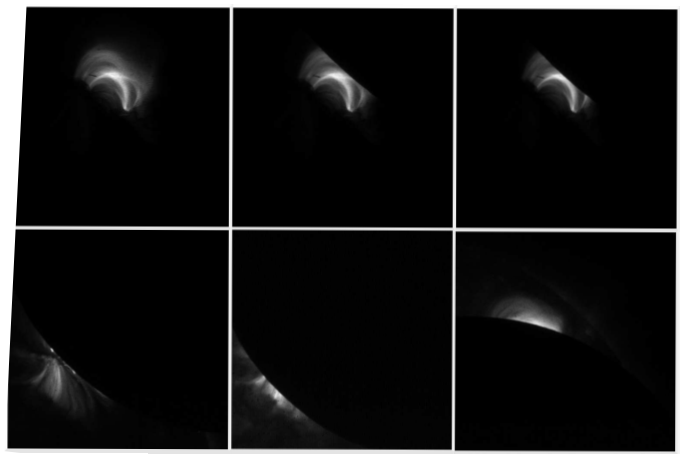

Figure 5.9: Occultation of the X-ray flare by HINODE satellite but at a slightly different time (14:16:00 IST, 14:17:43 IST, 14:17:47 IST, 14:18:10 IST, 14:18:39 IST, 14:24:58 IST – from top-left to top-right and bottom-left to bottom-right respectively) and from a different perspective.

## 5.4   Separation of the Effects of the Solar Eclipse and those of the Flare

Separation of the effects of the eclipse and the flare is tricky, since the flare is occulted partially by the moon during the eclipse. Thus, the observed VLF signal is neither due to the effect of the eclipse alone nor due to the flare alone. It is a composite effect that we observed. However, since the same antenna/receiver system has taken data from a similar flare at a similar time frame, we can separate the effects easily provided we assume that the effect is a linear superposition of the two effects. The details of the separation procedure can be found in Maji et al., (2012). In Figure 5.10a, we present two curves – the solid curve is the deviation of the signal due to the combined effects. This is what we actually observe. The dotted curve is obtained by subtracting the best estimated blocking of the flare by superposing the satellite images of the flaring region and the lunar disk. This would have been the signal deviation in case the flare was totally absent. This separation is thus due to the enhancement of the signal by the occulted flare. The open circles and the asterisks are the times when the LWPC code was used to obtain the electron number densities presented in the next Section. In Figure 5.10b, we derive the variation of the effective VLF reflection height as obtained by the LWPC code. As expected, the reflection height would have been larger if the flare was absent (dashed curve). The presence of the flare reduces the height, but not as much as would have happened if the obscuration were absent.

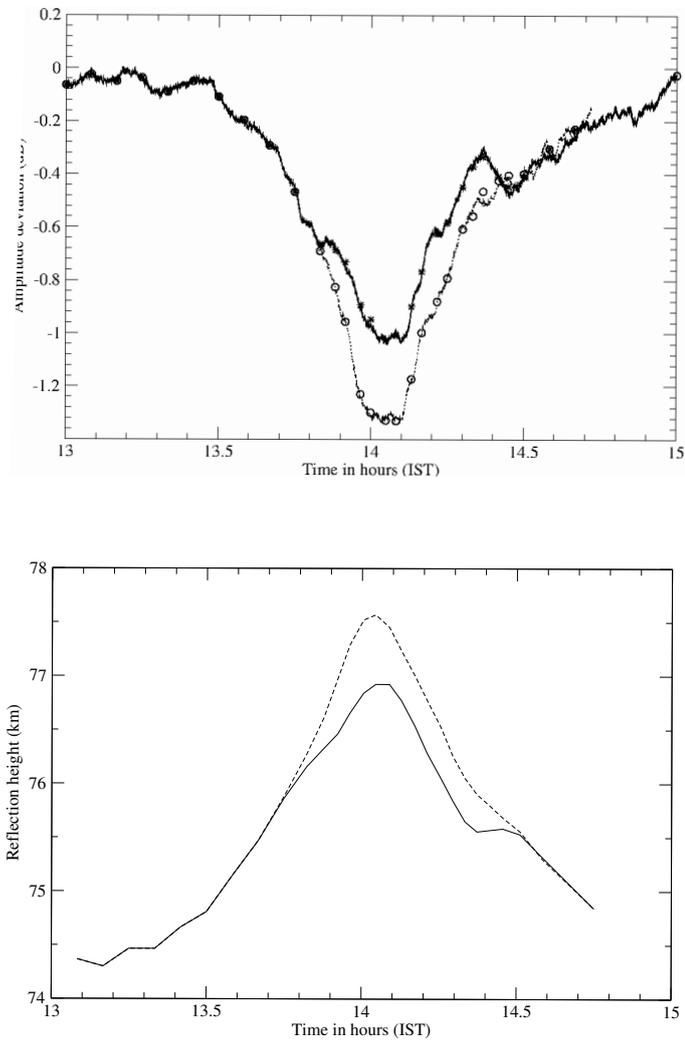

Figure 5.10: (Top panel, a) Variation of differential amplitude of NWC (19.8 kHz) signal with time. The solid curve is the observed signal due to the combined effect of solar eclipse and the occulted X-ray flare. Dotted curve is the model signal if the flare were absent. Open circle and asterisk represent the times used for modeling electron density with LWPC. (Bottom panel, b) The corresponding effective VLF reflection height variation [Pal et al., 2012b].

## 5.5   Modeling VLF data and Derivation of Ionospheric Parameters

Armed with the VLF signals with and without the flare, we can now model the time variation of the Earth's ionosphere. We use the LWPC code (see Chapter 2 & 3) to calculate the amplitudes at receiving sites corresponding to normal and partially eclipsed-ionosphere conditions. User of this model has to provide a suitable lower and an upper boundary conditions for the waveguide. The lower waveguide parameters i.e., permittivity ($\epsilon$) and conductivity ($\sigma$) of the Earth have been automatically selected by the code itself. These parameters are constants corresponding to normal and eclipse conditions. The upper waveguide i.e., ionospheric parameters are specified by the electron density $N_e(h)$ and the electron-neutral collision frequency $\nu_e(h)$ profiles. We use Wait's exponential ionosphere for electron density profile.

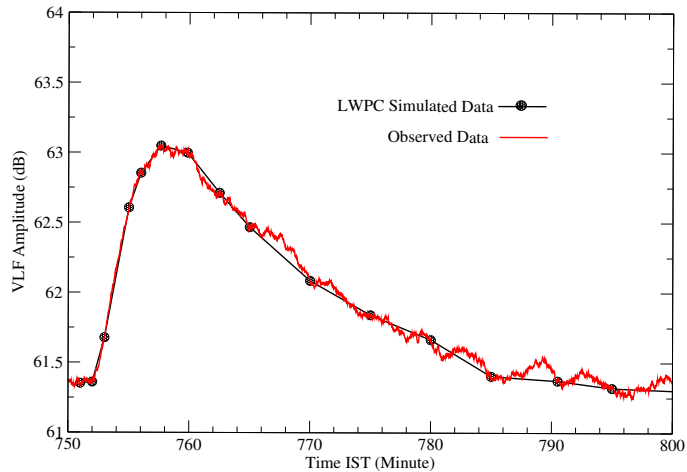

Figure 5.11: Simulated result of VLF amplitude perturbations due to a C4.0 solar flare (black curve) and the corresponding observed VLF data (red line) are plotted with IST (UT + 5.5).

We took $h' = 71.0$ km and $\beta = 0.43$ km$^{-1}$ as the unperturbed ionosphere [Thomson, 1993] while simulating the C4.0 flare, since these parameters reproduce the signal amplitude at the receiver exactly as we observe. To model the partially eclipsed D-region ionosphere we calculated the degree of solar obscuration as a function of time along the path. For this, we developed a numerical code to obtain the solar obscuration function $S(t)$ at a particular point along a path using the formalism of Mollmann & Vollmer (2006). The input parameters (i.e., the time of starting and ending of the eclipse, magnitude of eclipse) for the model are taken from the website of solar eclipse calculator (*http://www.chris.obyrne.com/Eclipses/calculator.html*) for the solar eclipse of 15th January, 2010. We then divide the propagation path into several segments ($\sim 40$) along the great circle path connecting the transmitter and the receiver. We calculate the degree of solar obscuration as a function of time. We find that $\sim 70$ % of the total propagation path experienced partial solar eclipse.

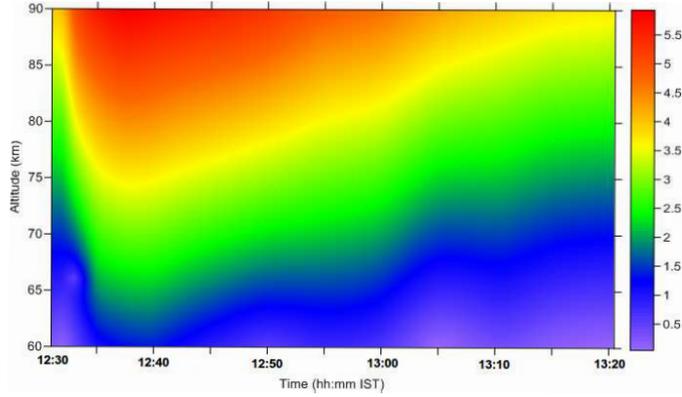

Figure 5.12: Variation of electron number density ($cm^{-3}$) with altitude due to the normal C4.0 flare. The lower ionosphere decreases from 71 km to 68 km while the sharpness factor increases from 0.43 $km^{-1}$ to 0.48 $km^{-1}$.

The ionospheric response to the incoming solar radiation is generally non-linear in the lower ionosphere [Lynn, 1981; Patel et al., 1986]. To model the observed non-linearity between the solar illumination ($S(t)$) and VLF amplitude response, we assumed that the disturbed ionospheric parameters vary according to $S^n$, where $S$ is the solar obscuration function on the points along the path. For this propagation path, comparing with our calculated effect of the solar eclipse, we found that the values of $n$ lie between 1.5 to 1.0 from the start of the observed deviation to the maximum deviation.

We first apply this method to compute the time variation of enhanced electron number density due to the C4.0 flare (Figure 5.3). Figure 5.11 shows the simulation results where the observed deviation due to the flare (red curve) has been reproduced (black curve) by the LWPC code. Figure 5.12 shows the electron density variation during the course of the flare. The color scale shows the logarithmic number density on the right. The numbers are comparable to that reported by Grubor et al. (2008).

We found that the VLF reflection height came down from 71 km to 68 km, a decrease of 3.0 km, while the sharpness factor $\beta$ increases from 0.43 km$^{-1}$ to 0.48 km$^{-1}$.

We now apply the same method to compute how the electron number density changed in the ionosphere on the 15th of January, 2010. Here, we choose $h' = 74$ km, and $\beta = 0.3$ km$^{-1}$ as parameters for the unperturbed ionosphere. In Figure 5.13, we show the results in presence of the eclipse and the occulted flare (top panel). The data was obtained by using LWPC at 'open circles' in the dotted curve of Figure 5.10a. If we subtract the effects of the occulted flare as a function of time, and use the asterisks on the solid curve in Figure 5.10a we get the electron density distribution as shown in Figure 5.13b. In Figure 5.14a, we show the difference of the number densities in the top and the bottom panels of Figure 5.13. Far from having a profile similar to that in Figure 5.12, we find a prominent dip in the electron number density at around 14:07 IST at all the altitudes. If we take cross-sections at different heights we get the time evolution of the electron number density of the ionosphere as a function of the height. We plot this in Figure 5.14b. We find that below $\sim 60$ km, the ionosphere is not perturbed significantly. With an increasing in height, there is a prominent dip in the middle. In fact, at 14:07 IST, the $h'$ parameter is decreased by 0.41 km. This dip is due to the occultation of the flare. It appears that the soft X-ray peak is covered by 14:10 and the hard X-ray peak region is covered by 14:25 IST. The number density rises again when the flare is exposed again though the short-lived flare was already in its decaying phase by that time.

## 5.6    Discussions

In this Chapter, we presented and analyzed a historic event, namely, the lunar occultation of a solar flare during an annular eclipse. This is the first time VLF monitoring was carried out during the event, and the VLF signal contains the effects of all the three events, namely, the solar flare, lunar occultation of it and the eclipse itself. Although a flare was observed during a lunar occultation before [Kreplin & Taylor, 1971], there was no VLF monitoring of that event. We used the data from a multiple number of satellite and ground based observations, such as the GOES-14 (both hard and soft X-Ray flux), GONG Project (Magnetogram data), HINODE (images of the flare) RHESSI (for X-ray light curves and image) which in conjunction with the VLF signal gives a time line of the sequence of events which took place. We reproduced the possible VLF signal for the effects of the partial eclipse only. We accordingly computed the electron density profile of the eclipsed ionosphere. We also computed the enhanced electron number density solely due to the occulted flare and clearly found the effects of the occultations of the soft and hard peak emission regions. The eclipse causes the effective reflection height to rise as the electron number density decreases. On the other hand, the effect of the solar flare, which is present throughout the eclipse period, is to partially counteract the eclipse behavior by increasing the electron density and decreasing the reflection height. The radiation from the solar flare and its ionospheric effect is briefly interrupted when the flare is occulted by the moon.

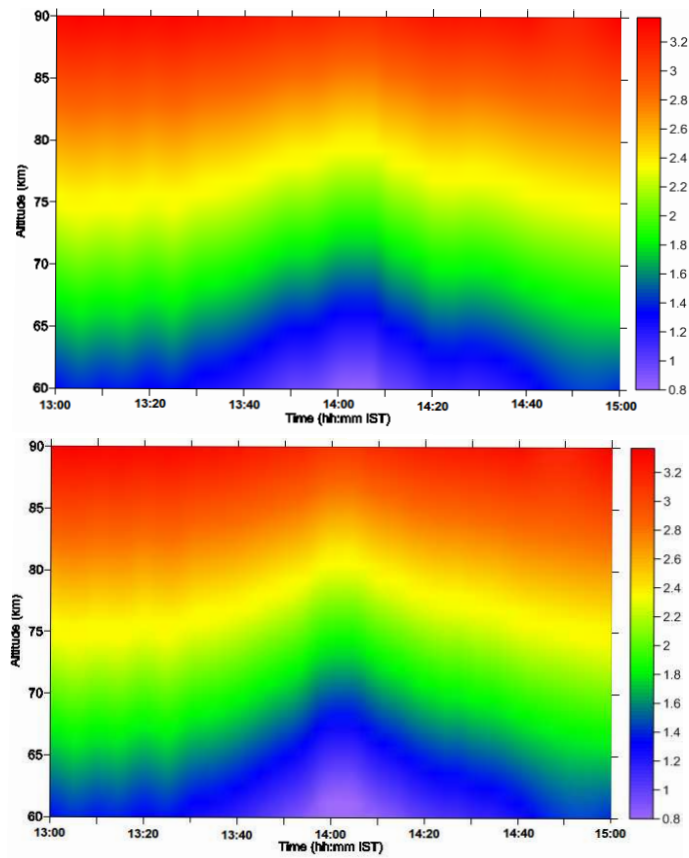

Figure 5.13: Variation of electron density in logarithmic scale ($cm^{-3}$) with altitude for the combined effects (top panel, a) (solid curve in Figure 5.10a) and (bottom panel, b) for the eclipse effect only (dashed curve in Figure 5.10b).

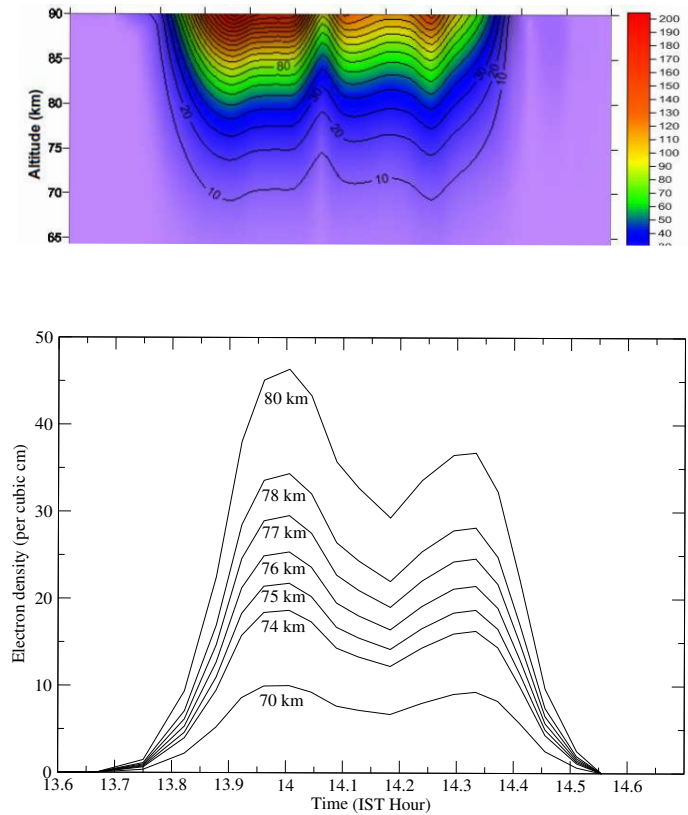

Figure 5.14: (Top panel, a) The electron number density variation induced only by the flare and occultation (difference between Figure 5.13a and Figure 5.13b) in linear scale. The dip between 14:05 and 14:10 is consistent with the obscuration of the soft X-ray peak (see, Figures 5.6-5.8). (Bottom panel, b) Variation of electron density ($cm^{-3}$) at different layers of the ionosphere at receiver location due to the occulted X-ray flare [Pal et al., 2012b].

# Chapter 6

# Summary and Future Works

## 6.1 Summary

The purpose of this dissertation was to study the theoretical and observational characteristics of the Very Low Frequency (VLF) radio wave propagation in the equatorial low latitude regions. Main motivation was to simulate the VLF signal behavior under various realistic conditions of the EIWG boundaries so that the behavior of the lower ionosphere under normal or disturbed conditions can be studied via the effects on the VLF signals.

In Chapter 2, we gave details about the simulation procedures. Two methods namely, the wave-hop and the LWPC codes have been used for simulation of diurnal behavior of VLF signals. First we developed the wave-hop code and applied this code for the propagation of VLF signal from the Indian Navy transmitter VTX over Indian sub-continent. We compared the results of the wave-hop code with that of the LWPC code. The wave-hop and LWPC roughly give equivalent results for the signal strength variation along the VTX-Kolkata baseline. Further, the LWPC code has been used to explain the VTX (18.2 kHz) signal behavior at different places in India obtained during the ICSP-VLF campaign.

In Chapter 3, we gave details about the application of the narrow band VLF measurements to detect and characterize the D-region ionospheric variations due to solar flares. We have shown the effects on the D-region due to a M2.0 solar flare simultaneously detected on two VLF paths (VTX-Kolkata, NWC-Kolkata). We found that short VLF paths act as a good detector for such solar disturbances.

In Chapter 4, we simulated the VLF perturbations due to the Total Solar Eclipse of July 22, 2009, for the VTX (18.2 kHz) signal using the wave-hop and LWPC codes. We assumed that the ionospheric parameters due to eclipse vary according to the degree of solar obscuration on the way to the receivers. On the basis of this assumption we reproduced the observed VLF deviation due to the eclipse for several VLF paths in Indian sub-continent. Both the observations and simulation results match very well.

We also modeled the perturbations in the NWC (19.8 kHz) signal due to combined effects of a solar flare and solar eclipse both occurred in the same time frame. Also, the flare was occulted by the Moon during the time of eclipse. Details have been shown in the Chapter 5. This was a first report of this kind. The distribution of D-region electron density profile due to the solar flare has been computed which showed a distinct dip in its profile. This confirms that the flare was occulted by the lunar disk.

## 6.2   Future Works

### 6.2.1   Improvements of the Wave-hop Model

The study of the amplitude and phase of VLF signals is important to diagnose the ionospheric conditions. The ray theory gives an easier way to calculate the total field for short VLF paths. However, it is necessary to include the effect of the geomagnetic field on VLF propagation as it does not arise explicitly to wave-hop model. The magnetic field affects the ionospheric reflection coefficient for different direction of propagation. Therefore, the reflection coefficient used in the model should be calculated as a function of magnetic field parameter, electron density and electron-ion collision frequency parameter. Also, for inhomogeneous waveguides such as across day-night ionospheric gradients, the geometry of the propagation path needs to be improved. The model needs to be verified for as many as short VLF paths by extensive observations as well.

### 6.2.2   Coupled Ionospheric Chemistry and LWPC code

The Earth's lower ionosphere is a very good detector of high energy (X-ray and Gamma ray) sources. Therefore, the study of its detailed ionization process is of utmost importance to interpret the results of VLF and other radio waves propagation through it. The present LWPC code computes the amplitude and phase of the VLF signal for any arbitrary ionospheric conditions that has been provided by the user along the VLF path. Here the ionospheric parameters are either the Wait's $\beta$ and $h'$ parameters for electron density or user may choose to provide his/her model of the ionosphere. Thus it is necessary to couple a D-region ionospheric chemistry code with that of LWPC code, so that the changes in VLF signals can be determined self-consistently as a function of injected energy in the D-region. Some coupled work has been done recently along the NWC-Kolkata path to simulate the VLF perturbations due to solar flares [Palit et al., 2013]. More work is necessary for such study.

### 6.2.3   Modeling Earthquake Pre-cursors

Earthquakes are reported to produce large perturbations in the ionospheric plasma states which in turn perturb the VLF wave propagation in the EIWG. As a result, the amplitude and phase anomaly are observable in VLF data prior to earthquakes. The modeling procedure as described in Chapter 4 and Chapter 5, can be applied to in disturbances produced in the VLF signals as a result of future seismic activities so that it is possible to explain the anomalous fluctuation of VLF signals mainly at night time which are believed to be associated with earthquakes. These works to be done in future to understand all three types of precursors [Chakrabarti et al., 2010b; Sasmal et al., 2009, 2010; Ray et al., 2011] of earthquakes.

# Appendix A

# ABBREVIATIONS

| | |
|---|---|
| $\dot{A}$ | Angstrom, $1 \times 10^{-8}$ $cm$ |
| EIWG | Earth-Ionosphere Waveguide |
| ELF | Extremely Low Frequency, 300–3000 Hz |
| EUV | Extreme Ultraviolet, 10–120 nm |
| GOES | Geostationary Operational Environmental Satellite |
| GRB | Gamma Ray Burst |
| HF | High Frequency, 3–30 MHz |
| ICSP | Indian Centre for Space Physics |
| IERC | Ionospheric and Earthquake Research Centre |
| IRI | International Reference Ionosphere |
| LF | Low Frequency, 30–300 kHz |
| LWPC | Long Wave Propagation Capability |
| Mm | Mega meter |
| nm | nano meter |
| PCA | Polar Cap Disturbance |
| SID | Sudden Ionospheric Disturbance |
| SNBNCBS | S. N. Bose National Centre for Basic Sciences |
| SPE | Solar Proton Event |
| SRT | Sunrise terminator time |
| SST | Sunset terminator time |
| TEC | Total Electron Content |
| TSE | Total Solar Eclipse |
| UV | Ultraviolet, 100–400 nm |
| VHF | Very High Frequency, 30–300 MHz |
| VLF | Very Low Frequency, 3–30 kHz |

# References

Afraimovich, E.L., Kosogorov, E.A., Lesyuta, O.S., 2002, Effects of the August 11, 1999 total solar eclipse as deduced from total electron content measurements at the GPS network, Journal of Atmospheric and Solar-Terrestrial Physics, 64, 1933–1941.

Afraimovich, E. L., A. T. Altynsev, V. V. Grechnev, and L. A. Leonovich, 2001, Ionospheric effects of the solar flares as deduced from global GPS network data, Advances in Space Research, 27, 1333–1338.

Appleton, E. V. and Barnett, M. A. F., 1925, On some direct evidence for downward atmospheric reflection of electric rays, Proceedings of the Royal Society, vol. 109 no. 752 621-641.

Bailey, D.K., 1964, Polar cap absorption, Planetary and Space Science 12, 495–541.

Barber N. F. and D. D. Crombie, 1959, V.L.F. reflections from the ionosphere in the presence of a transverse magnetic field, Journal of Atmospheric and Solar-Terrestrial Physics, 10, 37–45.

Barr, R., 1987, The effect of the Antarctic icecap on the propagation of Omega navigation signals, 1987, Proceedings of the $12^{th}$ Annual Meeting of the International Omega Association, Hawaii, 27.1–27.6.

Barr, R., Jones, D.L., Rodger, C.J., 2000, ELF and VLF radio waves, Journal of Atmospheric and Solar-Terrestrial Physics, 62, 1689–1718.

Barrington-Leigh, C. P., U. S. Inan, and M. Stanley, 2001, Identification of sprites and elves with intensified video and broadband array photometry, J. Geophys. Res., 106, 1741–1750.

Beech M., P Brown and J Jones, 1995, VLF detection of fireballs, Earth Moon Planets, 68, 181–188.

Bilitzaa, D., Reinisch, B.W., 2008, International reference ionosphere 2007: improvements and new parameters, Advances in Space Research, 42 (4), 599–609.

Bracewell, R.N., 1952, Theory of formation of an ionospheric layer below E layer based on eclipse and solar flare effects at 16 kc/s, Journal of Atmospheric and Terrestrial Physics, 2, 226–235.

Breit, G., and Tuve, M. A., 1926, A test of existance of conducting layer, Physical Review, 28, 554–573.

Budden, K. G., 1966, Radio waves in the ionosphere, Cambridge University Press, Cambridge.

Burgess, B., Jones, T.B., 1967, Solar flare effects and VLF radio-wave observations of the lower ionosphere, Radio Science, 2, 619–626.

Burke, C. P., and D. L. Jones, 1992, An experimental investigation of ELF attenuation rates in the Earth-ionosphere duct, Journal of Atmospheric and Terrestrial Physics, 54, 243–250.

Chakrabarti, S. K., S. Pal, K. Acharya, S. Mandal, S. Chakrabarti, R. Khan, B. Bose, 2002, VLF observation during Leonid Meteor Shower-2002 from Kolkata, Indian Journal Physics, 76B, 693.

Chakrabarti, S. K., Saha, M., Khan, R. et al., 2005, UnusualsunsetterminatorbehaviourofVLFsignalsat17 kHzduringtheEarthquake episodeofDec.,2004, Proceedings of the XXVIIIth URSI General Assembly.

Chakrabarti, S., Sasmal, S., Saha, M., Khan, M., Bhowmik, D., and Chakrabarti, S. K, 2007, Unusual behaviour of D-region ionization time at 18.2 kHz during seismically active days, Indian Journal Physics, 81, 531–538.

Chakrabarti, S. K., Sasmal, S., Pal, S., Mondal, S.K., 2010a, Results of VLF campaigns in summer, winter and during solar eclipse in Indian subcontinent and beyond, AIP Conference Proceedings, 1286, AIP, New York.

Chakrabarti, S. K., Sasmal, S., and Chakrabarti, S, 2010b, Ionospheric anomaly due to seismic activities – Part 2: Evidence from D-layer preparation and disappearance times, Nat. Hazards Earth Syst. Sci., 10, 1751–1757, doi:10.5194/nhess-10-1751-2010.

Chakrabarti, S. K., Pal. S., Sasmal. S., et al., 2012a, VLF campaign during the total eclipse of July 22nd, 2009: observational results and interpretations, Journal of Atmospheric and Solar-Terrestrial Physics, 86, 65–70, http://dx.doi.org/10.1016/j.jastp.2012.06.006.

Chakrabarti, S. K., Mondal S. K., Sasmal S., Pal S. et al., 2012b, VLF signals in summer and winter in the Indian sub-continent using multi-station campaigns, Inian Journal of Physics, 86, 5, 323–334.

Chilton, C. J., 1961, VLF phase perturbation associated with meteor shower ionization, J. Geophys. Res., 66, 2, 379–383.

Clilverd, M. A., Rodger, C. J., and Thomson, N. R., 1999, Investigating seismoionospheric effects on a long subionospheric path, J. Geophys. Res., 104, 28171–28179.

Clilverd, M.A., Rodger, C.J., Thomson, N.R., et al., 2001, Total Solar Eclipse effects on VLF signals: observations and modeling, Radio Science, 36 (4), 773–788.

Crary, J.H., and D.E. Schneible, 1965, Effect of the eclipse of 20 July 1963 on VLF signals propagating over short paths, Radio Science, 69D, 947–957.

Crombie, D.D., 1965, On the use of VLF measurements for obtaining information on the lower ionosphere (especially during solar ares), Proceedings of IEEE 53, 2027–2034.

Cummer, S.A., 1997, Lightning and ionospheric remote sensing using VLF/ELF radio atmospherics, Ph. D. Thesis, Standford University, Department of Electrical Engineering.

Cummer, S.A., Inan, U.S., Bell, T.F., 1998, Ionospheric D-region remote sensing using VLF radio atmospherics, Radio Science, 33, 1781–1792.

Davies, K., 1990, Ionospheric Radio, published by Peter Peregrinus Ltd., London, U. K.

De, S. S., B. Bandyopadhyay, S. Barui et al., 2012, Studies on the Effects of 2009 Leonid Meteor Shower on Subionospheric Transmitted VLF Signals and Vertical Electric Potential Gradient, Earth Moon Planets, 108, 111–121.

Demirkol, M. K., 1999, VLF remote sensing of the ambient and modified lower ionosphere, Ph.D. Thesis, Stanford University.

Donnelly, R. F., 1971. Extreme ultraviolet flashes of solar flares observed via sudden frequency deviations: experimental results, Solar physics, 20, 188–203.

Dowden, R. L., 1996, Comment on "VLF signatures of ionospheric disturbances associated with sprites by Inan et al., Geophys. Res. Lett., 23, 3421–3422.

Ferguson, J. A., and F. P. Snyder, 1980, Approximate VLF/LF mode conversion model, Technical Document, 400, Naval Ocean Systems Center, San Diego, California, USA.

Ferguson, J. A., F. P. Snyder, D. G. Morfitt, and C. H. Shellman, 1989, Long-wave propagation capability and documentation, Technical Document, 1518, Naval Ocean Systems Center, San Diego, California, USA.

Ferguson, J. A., 1998, Computer Programs for Assessment of Long-Wavelength Radio Communications, version 2.0, Technical document, 3030, Space and Naval Warfare Systems Center, San Diego, California, USA.

Fishman, G., and Inan, U.S., 1988, Observation of an ionospheric disturbance caused by a gamma-ray burst, Nature, 331, 418–420.

Fishman, G., Woods, P.M., Hossfield, C., Anderson, L., 2002, XRF 020427: Sudden Ionospheric Disturbance (SID), GCN Circular, No 1394.

Fleury, R., Lassudrie-Duchesne, P., 2000, VLF-LF Propagation Measurements during the 11 August 1999 Solar Eclipse, IEEE Conference Publication.

Galejs, J., 1972, Terrestrial Propagation of Long Electromagnetic Waves, Pergamon Press, Oxford.

Grubor, D., D. Sulic, and V. Zigman, 2008, Classification of X-ray solar flares regarding their effects on the lower ionosphere electron density profile, Ann. Geophys., 26, 1731–1740.

Hayakawa, M., and Fujinawa, Y., 1994, Electromagnetic Phenomena Related to Earthquake Prediction, Terra Science, Tokyo.

Hayakawa, M., Molchanov, O. A., Ondoh, T., and Kawai, E., 1996, The precursory signature effect of the Kobe earthquake on VLF subionospheric signals, J. Comm. Res. Lab., Tokyo, 43, 169–180.

Hayakawa M., (Editor), 1999, Atmospheric and Ionospheric Electromagnetic Phenomena Associated with Eathquakes, TERRAPUB, Tokyo.

Hayakawa, M. and Molchanov, O. A., 2000, Effect of earthquakes on lower ionosphere as found by subionospheric VLF propagation, Advances in Space Research, 26, 8, 1273–1276.

Hayakawa, M. and O. A. Molchanov (Editors), 2002, Seismo-Electromagnetics: Lithosphere-Atmosphere-Ionosphere Coupling, TERRAPUB, Tokyo.

Hayakawa, M., Molchanov, O. A., Shima, N., Shvets, A. V., and Yamamoto, N., 2003, Seismo Electromagnetics: Lithosphere Atmosphere-Ionosphere Couplings, edited by: Hayakawa, M. and Molchanov, O. A., TERRAPUB, Tokyo, Japan.

Huang, C.R., Liu, C.H., Yeh, K.C., Lin, K.H., Tsai, W.H., Yeh, H.C., Liu, J.Y., 1999, A study of tomographically reconstructed ionospheric images during a solar eclipse, J. Geophys. Res., 104, 79–94.

Inan, U. S., D. C. Shafer, W. Y. Yip, and R. E. Orville, 1988, Subionospheric VLF signatures of nighttime D region perturbations in the vicinity of lightning discharges, J. Geophys. Res., 93(A10), 11455–11472.

Inan, U.S., Bell, T.F., Pasko, V.P., Sentman, D.D., Wescott, E.M., Lyons, W.A., 1995, VLF signatures of ionospheric disturbances associated with sprites, Geophys. Res. Lett., 22, 3461–3464.

Inan, U.S., Lehtinen, N.G., LEV-Tov, S.J., Bell, T.F., et al., 1999, Ionization of the lower ionosphere by $\gamma$–rays from a magnetar: detection of a low energy (3–10 keV) component, Geophys. Res. Lett., 26, 3357–3360.

Inan, U.S., Lehtinen, N.G., Moores, R.C., Hurley, K., et al., 2007, Massive disturbance of the day time ionosphere by the giant gamma-ray flare from magnetar SGR 1806-20, Geophys. Res. Lett., 34, 8103.

Jackson, J.D., 1999, Classical Electrodynamics (3rd ed.), New York, Wiley.

Jakowski, N., Stankov, S.M., Wilken, V., Borries, C., Altadill, D., Chum, J., Buresova, D., Boska, J., Sauli, P., Hruska, F., Cander, Lj.R., 2008, Ionospheric behavior over Europe during the solar eclipse of 3 October 2005, Journal of Atmospheric and Solar-Terrestrial Physics, 70, 836–853.

Johnson, M. P., and Inan, US., 2000, Sferic clusters associated with early/fast VLF events, Geophys. Res. Lett., 27, 1391–1394.

Kasahara, Y., Muto, F., Hobara, Y., and Hayakawa, M., 2010, The ionospheric perturbations associated with Asian earthquakes as seen from the subionospheric propagation from NWC to Japanese stations, Nat. Hazards Earth Syst. Sci., 10, 581–588.

Keay C. S. L., 1995, Continued Progress in Electrophonic Fireball Investigations, Earth Moon and Planets, 68, 361–368.

Kelley, M. C., 2009, The Earths Ionosphere, PLasma Physics and Electrodynamics, Chapter 2.2, Academic Press, second edition.

Krankowski, A., Shagimuratov, I.I., Baran, L.W., and Yakimova, G.A., 2008, The effect of total solar eclipse of October 3, 2005, on the total electron content over Europe, Advances in Space Research, 41, 628–638.

Kreplin R. W and Taylor R. G., 1971, Localization of the source of flare X-ray during the eclipse of 7 March, 1970, Solar Physics, 21, 452–459.

Larsen, T. R., 1979, Solar-terrestrial predictions proceedings II, US GPO, Washington, DC 20402.

Le, H., Liu, L., Yue, X., Wan, W., Ning, B., 2009, Latitudinal dependence of the ionospheric response to solar eclipses, J. Geophys. Res., 114, doi:10.1029/2009JA014072.

Liu J. Y., Chiu C.S., and Lin C. H., 1996, The solar flare radiation responsible for sudden frequency deviation and geomagnetic fluctuation, J. Geophys. Res., 101, 10855–10862.

Liu, J. Y., and C. H. Lin, 2004, Ionospheric solar flare effects monitored by the ground-based GPS receivers: Theory and observation, J. Geophys. Res., 109, A01307, doi:10.1029/2003JA009931.

Lynn, K. J. W., 1981, The total solar eclipse of 23 October, 1976 observed at VLF", Journal of Atmospheric and Terrestrial Physics, 43, 1309–1316.

Lynn, K. J. W., 2010, VLF Waveguide Propagation: The Basics, AIP Conference Proceedings, 1286, 3–41.

Maekawa, S., Horie, T., Yamauchi, T., Sawaya, T., Ishikawa, M., Hayakawa, M., and Sasaki, H., 2006, A statistical study on the effect of earthquakes on the ionosphere based on the subionospheric LF propagation data in Japan, Ann. Geophys., 24, 2219–2225, doi:10.5194/angeo-24-2219-2006.

Maji, S. K., Chakrabarti, S. K. and Mondal, S. K., 2012, Unique observation of a solar flare by Lunar occultation during the 2010 Annular Solar Eclipse through ionospheric disturbances of VLF signal, Earth Moon and Planets, 108, 243–251.

Manju G. , K.G. Simi and S.R. P. Nayar, 2012, Analysis of solar EUV and X- ray flux enhancements during intense solar flare events and the concomitant response of equatorial and low latitude upper atmosphere, Journal of Atmospheric and Solar-Terrestrial Physics, http://dx.doi.org/10.1016/j.jastp.2012.05.008.

Marshall, R. A., U. S. Inan, and W. A. Lyons, 2006, On the association of early/fast very low frequency perturbations with sprites and rare examples of VLF backscatter, J. Geophys. Res., 111, D19108, doi:10.1029/2006JD007219.

Masuda, S., Kosugi, T., Hara, H., Tsuneta, S., Ogawara, Y., 1994, A loop-top hard X-ray source in a compact solar flare as evidence for magnetic reconnection, Nature, 371, 6497, 495–497.

Matsushita, S., 1967, Physics of Geomagnetic Phenomena I, Academic Press, New York.

Mika, A., and C. Haldoupis, 2008, VLF studies during TLE observations in Europe: A summary of new findings, Space Sci. Rev., 137, 489–510, doi:10.1007/s11214-008-9382-8.

Mitra, A. P., 1951, The D-layer of the ionosphere, J. Geophys. Res., 50, N0. 3, 373–402.

Mitra, A. P. and Shain, C. A., 1953, Journal of Atmospheric and Terrestrial Physics, 4, 204–218.

Mitra, A. P., and J. N. Rowe, 1972, Ionospheric effects of solar flares – VI. Changes in D region ion chemistry during solar flares, Journal of Atmospheric and Terrestrial Physics, 34, 795–806.

Mitra, A. P., 1974, Ionospheric Effects of Solar Flares, Astrophysics ans Space Science library, 46, D. Reidel Publishing Company, Boston.

Molchanov, O. A. and Hayakawa, M., 1998, Subionospheric VLF signal perturbations possibly related to earthquakes, J. Geophys. Res., 103, 17489–17510.

Molchanov, O.A., A. V. Shvets, and M. Hayakawa, 1998, Analysis of lightning induced ionization from VLF Trimpi events, J. Geophys. Res., vol. 103, 23443–23458.

Mollmann, K.P., Vollmer, M., 2006, Measurements and predictions of the illuminance during a solar eclipse, Eur. J. Phys., 27, 1299–1314.

Moore, C. R., C. P. Barrington-Leigh, U. S. Inan, and T. F. Bell, 2003, Early/fast VLF events produced by electron density changes associated with sprite halos, J. Geophys. Res., 108(A10), 1363, doi:10.1029/2002JA009816.

Morfitt, D. G., and C. H. Shellman, 1976, MODESRCH: An improved computer program for obtaining ELF/VLF/LF propagation data, Technical Report NOSC/TR 141, Naval Ocean System Centre, San Diego, California.

Nicolls, M.J. and M.C. Kelley, 2005, Strong evidence for gravity wave seeding of an ionospheric plasma instability, Geophys. Res. Lett., 32, 679.

Ohshio M., 1971, Negative sudden phase anomaly, Nature, 229, 239–240.

Pal, S., Chakrabarti, S. K., 2010, Theoretical models for Computing VLF wave amplitude and phase and their applications, AIP Conference Proceedings., 1286, 42–60.

Pal, S., Basak, T. and Chakrabarti, S. K., 2011, Results of computing amplitude and phase of the VLF wave using wave-hop theory, Advances in Geosciences, Solar Terrestrial (ST), 27, World Scientific, 111.

Pal, S., Chakrabarti, S.K., Mondal, S.K., 2012a, Modeling of sub-ionospheric VLF signal perturbations associated with total solar eclipse, 2009 in Indian subcontinent, Advances in Space Research, 50, 196–204.

Pal S., S. K. Maji and S. K. Chakrabarti, 2012b, First ever VLF monitoring of Lunar occultation of a solar flare during the 2010 Annular Solar Eclipse and its effects on the D-region electron density profile, Planetary and Space Science, 73, 310-317, doi:10.1016/j.pss.2012.08.016.

Pappert, R. A., Gossard, E. E. and Rothmuller, I.J., 1967, A numerical investigation of classical approximations used in VLF propagation, Radio Science, 2, 387–400.

Pappert, R.A., Snyder, F.P., 1972, Some results of a mode conversion program for VLF, Radio Science, 7, 913–923.

Pappert, R.A., Morfitt, D.G., 1975, Theoretical and experimental sunrise mode conversion results at VLF, Radio Science, 510, 537–546.

Parrot, M., Achache, J., Berthelier, J. J., Blanc, E., et al., 1993, High-frequency seismo-electromagnetic effects, Physics of the Earth and Planetary Interiors, 77, 65–83.

Parrot, M., 1995, Electromagnetic noise due to earthquakes, In: Volland, H. (Ed.), Handbook of Atmospheric Electrodynamics, 2nd Edition, 2, CRC Press, Boca Raton, FL, 95–116.

Parrot, M., J.J. Berthelier, J.P. Lebreton, J.A. Sauvaud, O. Santolk, J. Blecki, 2006, Examples of unusual ionospheric observations made by the DEMETER satellite over seismic regions, Physics and Chemistry of the Earth, 31, 486–495.

Patel, D. B., K. M. Kotadia, P. D. Lele, and K. G. Jani, 1986, Absorption of radio waves during a solar eclipse, Earth Planet Science, 95, 193–200.

Poulsen, W. L., 1991, Modeling of very low frequency wave propagation and scattering within the Earth-ionosphere waveguide in the presence of lower ionospheric disturbances, Ph.D. Thesis, Stanford University.

Ray, S., S. K. Chakrabarti, S. K. Mondal and and S. Sasmal, 2011, Ionospheric anomaly due to seismic activities-III: correlation between night time VLF amplitude fluctuations and effective magnitudes of earthquakes in Indian sub-continent, Nat. Hazards Earth Syst. Sci., 11, 2699–2704, doi:10.5194/nhess-11-2699-2011.

Raulin, J. P., Fernando C. P. Bertoni, Hernan R. Gavilan, and Jorge C. Samanes, 2010, Longterm and transient forcing of the low ionosphere monitored by SAVNET, AIP Conference Proceedings, 1286, 103–125, doi:http://dx.doi.org/10.1063/1.3512872.

Recommendation, ITU-R, 1992, Ground-wave propagation curves for frequencies between 10 kHz and 30 MHz, ITU, 368-7.

Recommendation ITU-R, 2002, Prediction of field strength at frequencies below about 150 kHz, ITU, 684-3.

Reder, F.H., and Westerlund, S., 1970, VLF signal phase instabilities produced by propagation medium – Experimental results, In: Davies, K. (Ed.), Phase and Frequency Instabilities in Electromagnetic Wave Propagation, AGARD Conference Proceedings, 33, Technovision Services, Slough, England, 103–136.

Rishbeth H., 1968, Solar eclipses and ionospheric theory, Space Science Reviews, 8, 543–554.

Rishbeth, H., Garriott, O.K., 1969, Introduction to Ionospheric Physics, Academic Press, New York, London.

Rodger, C. J., 2003, Subionospheric VLF perturbations associated with lightning discharges, Journal of Atmospheric and Solar-Terrestrial Physics, 65, 5, 591-606.

Round H. J., Eckersley T. L., Tremellen K. and Lunnon F. C., 1925, Report on measurements made on signal strength at great distances during 1922 and 1923 by an expedition sent to Australia, Jour. Inst. Elect. Engs., 63, 933–1011.

Russell, D.A., 1979, The enigma of the extinction of the dinosaurs, Annual Reviews of Earth & Planetary Science, 7, 163–182.

Sampath H. T., Inan U. S. and Johnson M. P., 2000, Recovery signatures and occurrence properties of lightning-associated subionospheric VLF perturbations, J.Geophys. Res., 105, A1, 183–191.

Sao K., Yamashita M., Tanahashi S., Jindoh H., and Ohta K., 1970, Sudden enhancements (SEA) and decreases (SDA) of atmospherics, Journal of Atmospheric and Terrestrial Physics, 32, 9, 1567–1576.

Sasmal S., and Chakrabarti S. K., 2009, Ionosperic anomaly due to seismic activities – Part 1: Calibration of the VLF signal of VTX 18.2 KHz station from Kolkata and deviation during seismic events, Nat. Hazards Earth Syst. Sci., 9, 1403–1408.

Sasmal S., Chakrabarti S. K., and Chakrabarti S., 2010, Studies of the correlation between ionospheric anomalies and seismic activities in the Indian subcontinent, AIP Conference Proceedings, 1286, 270–290.

Schmitter, E. D., 2011, Remote sensing planetary waves in the midlatitude mesosphere using low frequency transmitter signals Ann. Geophys., 29, 1287–1293, doi:10.5194/angeo-29-1287-2011.

Schnoor, P.W., Welch, D.L., Fishman, G., and Price, A., 2003, GRB030329 observed as a sudden ionospheric disturbance (SID), GCN Circular No. 2176.

Schunk, R. W. and Nagy, A. F., 2009, Ionospheres, Physics, Plasma Physics, and Chemistry, Second Edition, Cambridge University Press.

Sechrist Jr. C.F., 1974, Comparisons of techniques for measurement of D-region electron densities. Radio Science, 9, 137–149.

Sivjee G., D. McEwen, and R. Walterscheid, 2003, Polar Cap Disturbances: Mesosphere and Thermosphere –Ionosphere Response to Solar-Terrestrial Interactions, Sodankyla Geophysical Observatory Publications, 92, 69–72.

Smith, R., 1974. Approximate mode conversion coe cients in the Earth-ionosphere waveguide for VLF propagation below an anisotropic ionosphere, Journal of Atmospheric and Terrestrial Physics, 36, 1683–1688.

Smith, R., 1977, Mode conversion coefficients in the Earth-ionosphere waveguide for VLF propagation below a horizontally stratified, anisotropic ionosphere, Journal of Atmospheric and Terrestrial Physics, 39, 539–543.

Stonehocker G. H., 1970, Advanced telecommunication forecasting technique in AGY, 5th., Ionospheric forecasting, AGARD Conference Proceedings, 29, 27.

Swift, D.W., 1961, The effect of solar X-rays on the ionosphere, Journal of Atmospheric and Solar-Terrestrial Physics, 23, 29–56.

Tanaka, Y.T., Terasawa, T., Yoshida, M., Horie, T., Hayakawa, M., 2008, Ionospheric disturbances caused by SGR 1900+14 giant gamma ray flare in 1998: Constraints on the energy spectrum of the flare, J. Geophys. Res., A07307, doi:10.1029/2008JA013119.

Thomson, N. R., 1993, Experimental daytime VLF ionospheric parameters, Journal of Atmospheric and Solar-Terrestrial Physics, 55, 2, 173–184.

Thomson, N. R. and Clilverd, M.A., 2001, Solar flare induced ionospheric D-region enhancements from VLF amplitude observations, Journal of Atmospheric and Solar-Terrestrial Physics, 63, 16, 1729–1737.

Todoroki, Y., S. Maekawa, T. Yamauchi, T. Horie, and M. Hayakawa, 2007, Solar flare induced D-region perturbation in the ionosphere, as revealed from a short-distance VLF propagation path, Geophys. Res. Lett., vol. 34, L03103, doi:10.1029/2006GL028087.

Thorsett, S.E., 1995, Terrestrial implications of cosmological gamma-ray burst models, Astrophysical Journal Letters 444: L53, doi:10.1086/187858.

Tsai, H.F., and Liu, J.Y., 1999, Ionospheric total electron content response to solar eclipses, J. Geophys. Res., 104, A6, 12, 657–668.

Tsurutani, B. T. et al., 2005, The October 28, 2003 extreme EUV solar flare and resultant extreme ionospheric effects: Comparison to other Halloween events and the Bastille Day event, Geophys. Res. Lett., 32, L03S09, doi:10.1029/2004GL021475.

Valnicek B., and Ranzinger P., 1972, X-ray emission and D-region sluggishness, Bulletin of the Astronomical Institute of Czechoslovakia, 23, 318–322.

Van Allen, J. A., C. E. McIlwain, and G. H. Ludwig, 1959, Radiation observations with satellite 1958, J. Geophys. Res., 64, 271–286.

Wait, R. J., 1962, Electromagnetic Waves in Stratified Media, Pergamon Press, Oxford.

Wait, J. R., and K. P. Spies, 1964, Characteristics of the earth-ionosphere waveguide for VLF radio waves, NBS Technical Note, U.S. 300.

Wait, J. R., and K. P. Spies, 1965, Influence of finite ground conductivity on the propagation of VLF waves, J. Res. Nat. Bureau Stand., 69D, 1359.

Wait, J. R., 1998, The Ancient and Modern History of EM Ground-Wave Propagation, IEEE Antennas and Propagation Magazine, 40, 5, 7–24.

Wakai, N., Kurihara, N. and Otsuka, A., 2004, Numerical method for calculating LF sky-wave and their resultant wave field strengths, Electronics Letter, 40, 5, 288–290.

Walt, M., 1994, Introduction to Geomagnetically Trapped Radiation, New York, Cambridge University Press.

Wan, W., L. Liu, H. Yuan, B. Ning, and S. Zhang, 2005, The GPS measured SITEC caused by the very intense solar flare on July 14, 2000, Advances Space Research, 36, 2465–2469.

Weidman, C. D., and E. P. Krider, 1986, The amplitude spectra of lightning radiation fields in the interval from 1 to 20 MHz, Radio Science, 21, 964–970.

Westerlund, S., Reder, F.H., and Abom, C.J., 1969, Effects of polar cap absorption events on VLF transmissions, Planetary and Space Science, 17, 1329–1374.

Westerlund, S., and Reder, F., 1973, VLF propagation at auroral latitudes, Journal of Atmospheric and Terrestrial Physics, 35, 8, 1453–1474.

Winch, D.E., Ivers, D.J., Turner, J.P.R. and Stening, R.J., 2005. Geomagnetism and Schmidt quasi-normalization, Geophys. J. Int., 160, 487–504.

Wolf, T. G., 1990, Remote sensing of ionospheric effects associated with lightning very low frequency radio signals, Ph.D. Thesis, Stanford University.

Yeh, K.C., Yu, D.C., Lin, K.H., Huang, C.R., Tsai, W.H., Liu, J.Y., Xu, J.S., Igarashi, K., Xu, C., and Nam, V.H., 1997, Ionospheric response to a solar eclipse in the equatorial anomaly region, Terrestrial Atmospheric and Oceanic Science, 8, 165–178.

Yoshida, Y., Yamauchi, T., Horie, T., Hayakawa, M., 2008, On the generation mechanism of terminator times in subionospheric VLF/ELF propagation and its possible application to seismogenic effects, Nat. Hazards Earth Syst. Sci., 8, 129–134.

Zhang, D. H., and Z. Xiao, 2003, Study of the ionospheric total electron content response to the great flare on 15 April 2001 using the International GPS Service network for the whole sunlit hemisphere, J. Geophys. Res., 108, (A8), 1330–1341.

Zgrablic, G., Vinkovic, D., Gradecak, S., et al., 2002, Instrumental recording of electrophonic sounds from Leonid reballs, J. Geophys. Res., 107, A7, 10.1029/2001JA000310.

Zigman, Z., Grubor, A. D. and Sulic D., 2007, D-region electron density evaluated from VLF amplitude time delay during X-ray solar flares, Journal of Atmospheric and Solar-Terrestrial Physics, 69, 775–792.